\documentclass[12pt,a4paper]{article}

\usepackage{amsmath}
\usepackage{amsfonts}
\usepackage{amssymb}
\usepackage{graphicx}
\usepackage{color}
\usepackage{booktabs}
\usepackage{inputenc}
\usepackage[T1]{fontenc}
\usepackage{mathrsfs}
\usepackage{enumerate}
\usepackage{xcolor,url}
\usepackage{cancel,dsfont} 
\usepackage{multirow}
\usepackage{soul}
\usepackage{braket}

\newcommand{\eea}{\end{eqnarray}}
\newcommand{\bea}{\begin{eqnarray}}
\newcommand{\be}{\begin{equation}}
\newcommand{\ee}{\end{equation}}

\def\be{\begin{equation}}
\def\ee{\end{equation}}

\def\knl{k_{\rm NL}}
\def\km{k_{\rm M}}
\def\kr{k_{\rm R}}

\font\BF=cmmib10
\def\k{{\hbox{\BF k}}}
\def\q{{\hbox{\BF q}}}
\def\z{{\hbox{\BF z}}}
\def\s{{\hbox{\BF s}}}
\def\m{{\hbox{\BF m}}}

\def\x{{\hbox{\BF x}}}

\def\v{{\hbox{\BF v}}}

\def\hinvMpc{h\,{\rm Mpc}^{-1}}
\def\Mpcinvh{{\rm Mpc}\,h^{-1}}

\newcommand{\kmax}{k_{\rm max }}

\newcommand{\code}[1]{\texttt{#1}}

\usepackage[colorlinks,bookmarks]{hyperref}
\definecolor{linkblue}{rgb}{0,0,0.8}
\definecolor{linkgreen}{rgb}{0,0.5,0}

\hypersetup{pdfpagemode=UseNone, pdfstartview=FitH, linkcolor=linkblue,
            citecolor=linkgreen, urlcolor=linkblue}

\font\BF=cmmib10
\def\v{{\hbox{\BF v}}}
\def\k{{\hbox{\BF k}}}
\def\q{{\hbox{\BF q}}}
\def\z{{\hbox{\BF z}}}

\begin{document}

\begin{center}

{\Large \bf Cosmological inference from the EFTofLSS:  \\[0.3cm]
the eBOSS QSO full-shape analysis}  \\[0.7cm]

{\large  Th\'eo Simon${}^{1}$, Pierre Zhang${}^{2,3,4,5}$, Vivian Poulin${}^{1}$\\[0.7cm]}

\end{center}

\begin{center}

\vspace{.0cm}

{\normalsize { \sl $^{1}$ Laboratoire Univers \& Particules de Montpellier (LUPM), CNRS \& Universit\'e de Montpellier (UMR-5299), Place Eug\`ene Bataillon, F-34095 Montpellier Cedex 05, France}}\\
\vspace{.3cm}

{\normalsize { \sl $^{2}$ Department of Astronomy, School of Physical Sciences, \\
University of Science and Technology of China, Hefei, Anhui 230026, China}}
\vspace{.3cm}

{\normalsize { \sl $^{3}$ CAS Key Laboratory for Research in Galaxies and Cosmology, \\
University of Science and Technology of China, Hefei, Anhui 230026, China}}
\vspace{.3cm}

{\normalsize { \sl $^{4}$ School of Astronomy and Space Science, \\
University of Science and Technology of China, Hefei, Anhui 230026, China}}
\vspace{.3cm}

{\normalsize { \sl $^{5}$  Institut fur Theoretische Physik, ETH Zurich,
 8093 Zurich, Switzerland}}\\
\vspace{.3cm}

\end{center}

\hrule \vspace{0.3cm}
{\small  \noindent \textbf{Abstract} \vspace{.1in}

\noindent We present cosmological results inferred from the effective-field theory (EFT) analysis of the full-shape of eBOSS quasars (QSO) power spectrum. 
We validate our analysis pipeline against simulations, and find overall good agreement between the analyses in Fourier and configuration space.
Keeping the baryon abundance and the spectral tilt fixed, we reconstruct at $68\%$ CL the fractional matter abundance $\Omega_m$, the reduced Hubble constant $h$, and the clustering amplitude $\sigma_8$, to respectively $\Omega_m=0.327   \pm 0.035$, $h=0.655   \pm 0.034$, and $\sigma_8 = 0.880   \pm 0.083$ from eBOSS QSO alone.
These constraints are consistent at $\lesssim1.8\sigma$ with the ones from {\it Planck} and from the EFT analysis of BOSS full-shape. Interestingly $S_8$ reconstructed from eBOSS QSO is slightly higher than that deduced from {\it Planck} and BOSS, although statistically consistent.
In combination with the EFT likelihood of BOSS, supernovae from Pantheon, and BAO from lyman-$\alpha$ and 6dF/MGS, constraints improve to $\Omega_m = 0.2985\pm 0.0069$ and $h = 0.6803 \pm 0.0075$, in agreement with {\it Planck} and with similar precision.
We also explore one-parameter extensions to $\Lambda$CDM and find that results are consistent with flat $\Lambda$CDM at $\lesssim 1.3\sigma$.
We obtain competitive constraints on the curvature density fraction $\Omega_k = -0.039 \pm 0.029$, the dark energy equation of state $w_0 = -1.038  \pm 0.041$, the effective number of relativistic species $N_{\rm eff} = 3.44^{+0.44}_{-0.91}$ at $68\%$ CL, and the sum of neutrino masses $\sum m_\nu< 0.274 e$V at $95\%$ CL, without {\it Planck} data. 
Including {\it Planck} data, contraints significantly improve thanks to the large lever arm in redshift between LSS and CMB measurements. In particular, we obtain the stringent constraint $\sum m_\nu < 0.093 e$V, competitive with recent lyman-$\alpha$ forest power spectrum bound. 
\vspace{0.3cm}}
\hrule

\vspace{0.3cm}
\newpage

\tableofcontents

\section{Introduction\label{sec:intro}}

The distribution of matter at large scales contains a wealth of cosmological information, from the initial conditions of the Universe to the gravitational collapse of late-time objects. 
The program of cosmic microwave background (CMB) experiments has now matured to a state where $\Lambda$CDM parameters have been measured to percent level with the {\it Planck} satellite~\cite{Planck:2018vyg}, and with similar precision by subsequent experiments, \emph{e.g.}, ACT~\cite{ACT:2020gnv} and SPT~\cite{SPT:2017sjt,SPT:2019fqo}. 
At the same time, the data volume gathered by large-scale structure (LSS) surveys has been continuously growing. 
As those surveys probe vastly different epochs in the history of the Universe, they allow for a crucial consistency test of the $\Lambda$CDM model and have delivered independent cosmological determinations at precision comparable to CMB measurements, see \emph{e.g.}, the recent results from the photometric surveys DES~\cite{DES:2021wwk} and KIDS~\cite{KiDS:2020suj}, or from the spectroscopic surveys BOSS~\cite{BOSS:2016wmc}. In addition, LSS data have become paramount to break degeneracies of the $\Lambda$CDM model and extensions when combined with CMB. 

Recently, as the accuracy of observations has improved, various cosmological probes have delivered cosmological parameter measurements with a growing level of inconsistency.
The most statistically significant cosmological discrepancy is the ``Hubble tension''~\cite{Verde:2019ivm}, corresponding to a difference of $\sim 5\sigma$ between the determination of the Hubble constant from {\it Planck} data analyzed under $\Lambda$CDM~\cite{Planck:2018vyg} and its local determination from the cosmic distance ladder based on cepheid-calibrated SNIa by the SH0ES team~\cite{Riess:2021jrx}. 
Another intriguing cosmological puzzle, the ``$S_8$ tension'' ($\sim 2-3\sigma$), has emerged between weak lensing measurements~\cite{KiDS:2020suj,DES:2021wwk,HSC:2018mrq,Amon:2022ycy} and CMB~\cite{Planck:2018vyg,ACT:2020gnv} determinations of the local matter fluctuations, parameterized as $S_8 = \sigma_8 \sqrt{\Omega_m/0.3}$, where $\sigma_8$ is the root mean square of matter fluctuations on a $8 \Mpcinvh$ scale and $\Omega_m$ the fractional matter density today. 
Spectroscopic surveys probe the distribution of matter at similar redshifts as the photometric ones, but rely essentially on scales that in average are larger than the one probed in weak lensing.  
Thus, spectroscopic surveys have the potential to play a key role in shedding light on these tensions. 
In particular, an agreement between clustering and CMB data would have, under the assumption that there is no systematic error, significant impact on the interpretation of these tensions.
Regarding the $S_8$ tension, this would hint that the origin lies in the scales beyond the (large) scales included in clustering or CMB analyses (see \emph{e.g.},  Ref.~\cite{Amon:2022ycy}). 
As for the $H_0$ tension, a resolution would then require modifications to the concordance model that can lift both the values measured in the CMB and in the LSS.
\\

The large amount of LSS data available provides us with new opportunities to extract additional cosmological information, by making use of the full-shape of summary statistics built from clustering data. 
Among the spectrocopic surveys, the Extended Baryon Oscillation Spectroscopic Survey (eBOSS), combined with previous phases of the Sloan Digital Sky Survey (SDSS), has mapped more than 11 billion years of cosmic history, providing an unprecedented map of the matter clustering in the Universe~\cite{eBOSS:2020yzd} through different tracers of the underlying matter density distribution, \emph{e.g.}, galaxies, quasars or the lyman-$\alpha$ forest. 
To extract cosmological information from these surveys, the (e)BOSS collaboration follows the convention of compressing information from these surveys into simple parameters that can be easily compared with cosmological models. 
These are usually expressed in the form of the Alcock-Paczynski (AP) parameters measured from the BAO angles \cite{Alcock:1979mp} and the $f\sigma_8$ parameter, where $f$ is the growth factor, measured from redshift space distortions (RSD) \cite{Kaiser:1987qv}. 
However, the large amount of LSS data available provides us with new opportunities to extract additional cosmological information, by making use of the full-shape of summary statistics built from clustering data. 
Given the increasing data volume and the variety of tracers probed, new methods to make reliable predictions for the full-shape are necessary to extract the cosmological parameters in a robust and systematic ways. \\

Thankfully, the underlying density and velocity fields of any tracer respect a set of symmetries in the long-wavelength limit known as Galilean invariance~\cite{Kehagias:2013yd,Peloso:2013zw,Creminelli:2013mca}. 
Moreover, we are interested in objects that are non-relativistic, allowing us to define a \emph{nonlinear scale} as the average distance travelled by the objects during the age of the Universe, under which the underlying fields and their dynamics can be smoothed out~\cite{Baumann:2010tm}.  
Building on those considerations, the Effective Field Theory of Large-scale Structure (EFTofLSS) has emerged as a systematic way to organize the expansions in fluctuations and derivatives of the density and velocity fields of the observed tracers at long wavelengths~\cite{Baumann:2010tm,Carrasco:2012cv,Senatore:2014via,Senatore:2014eva,Senatore:2014vja}.~\footnote{See also the introduction footnote in, \emph{e.g.},  Ref.~\cite{DAmico:2022osl} for relevant related
works on the EFTofLSS.} 
The prediction at the one-loop order for the power spectrum of biased tracers in redshift space from the EFTofLSS~\cite{Perko:2016puo} (see also  Ref.~\cite{Desjacques:2018pfv}) has been used to analyze the full-shape of BOSS clustering data in Refs.~\cite{DAmico:2019fhj,Ivanov:2019pdj}. 
These works have shown that: \emph{i)} higher wavenumbers beyond the linear regime in good theoretical control can be accessed, bringing additional cosmological information (see also Ref.~\cite{Nishimichi:2020tvu}), and \emph{ii)} with reliable predictions, as the cosmological parameters (together with the nuisance parameters) are scanned the template can be varied instead of being held fixed, exploiting the full information of the full-shape beyond the one from geometrical distortions (see Refs.~\cite{Grieb:2016uuo,Sanchez:2016sas} for earlier works where the full-shape predictions, yet not from the EFTofLSS, were varied at each point in parameter space). 
EFT analyses of BOSS data have provided precise and robust determination of $\Lambda$CDM parameters~\cite{DAmico:2019fhj,Ivanov:2019pdj,Colas:2019ret,Philcox:2020vvt,Chen:2021wdi,Zhang:2021yna,Chen:2022jzq,Simon:2022lde}, and pushed down limits on extensions, such as neutrino masses and effective number of relativistic species~\cite{Colas:2019ret,Ivanov:2019hqk,Kumar:2022vee,Allali:2023zbi,Schoneberg:2023rnx}, dark energy~\cite{DAmico:2020kxu,DAmico:2020tty,Carrilho:2022mon}, curvature~\cite{Chudaykin:2020ghx,Glanville:2022xes}, early dark energy~\cite{DAmico:2020ods,Ivanov:2020ril,Niedermann:2020qbw,Simon:2022adh}, non-cold dark matter~\cite{Simon:2022ftd,Rubira:2022xhb}, interacting dark energy~\cite{Nunes:2022bhn}, and more~\cite{Gonzalez:2020fdy,Braglia:2020auw,Lague:2021frh,Allali:2021azp}. 
Some EFT analyses of BOSS data have also included the bispectrum at tree-level~\cite{DAmico:2019fhj,Philcox:2021kcw} and at one-loop~\cite{DAmico:2022osl} (see also  Ref.~\cite{Philcox:2022frc}), pushing down uncertainties on $\Lambda$CDM parameters and setting new bounds from the LSS on non-Gaussianities~\cite{Cabass:2022wjy,DAmico:2022gki,Cabass:2022ymb}.
See also, \emph{e.g.}, Refs.~\cite{Troster:2019ean,Kobayashi:2021oud,Semenaite:2021aen,Neveux:2022tuk,Brieden:2022lsd,Semenaite:2022unt} for results from BOSS and/or eBOSS full-shape analyses using methods different from the EFTofLSS.
In addition, the EFTofLSS has made possible the development of a new consistency test of the $\Lambda$CDM and alternative models based on a sound horizon-free analysis~\cite{Philcox:2020xbv,Farren:2021grl,Philcox:2022sgj}, providing a new way to probe beyond $\Lambda$CDM models \cite{Smith:2022iax}.\\ 

In this paper, we analyze the eBOSS quasars (QSO) full-shape using the prediction from the EFTofLSS. 
There are two main motivations behind this work. 
First, the EFTofLSS has only been used to analyze BOSS luminous red galaxies (LRG) and more recently eBOSS emission line galaxies (ELG) \cite{Ivanov:2021zmi}. 
As QSO are different tracers than LRG, and selected by SDSS at an overall higher redshift than LRG, the eBOSS QSO full-shape analysis complements previous BOSS full-shape analysis, providing yet another important consistency test of $\Lambda$CDM  at a different epoch and for another tracer (while also allowing us to test the assumptions behind the EFTofLSS such as the aforementioned Galilean invariance symmetries).
  
Second, the eBOSS QSO full-shape once combined with other cosmological probes can shed light on extensions to $\Lambda$CDM model. Here, we explore four one-parameter extensions to the flat $\Lambda$CDM model, namely the curvature
density fraction $\Omega_k$, the dark energy equation of state $w_0$, neutrino masses $\sum m_{\nu}$, and the effective number of relativistic species $N_{\rm eff}$.
We compare the obtained limits with the ones from {\it Planck} and with the ones from the standard BAO/$f\sigma_8$ technique, in order to assess both the consistency of the results and the potential improvements brought by the EFT analysis.
\\

Our paper is organized as follow. 
In section~\ref{sec:pipeline}, we describe the EFT analysis pipeline for eBOSS QSO that we built. 
In particular, we review the theoretical prediction of the EFTofLSS in~\ref{sec:model}, and present the dataset, likelihood, and prior chosen for our analysis in~\ref{sec:inference}. 
In~\ref{sec:scalecut}, we assess the highest wavenumbers $\kmax$ that can be included in the analysis of eBOSS QSO full-shape data by making use of a general method that consists in evaluating the size of the theoretical error through the insertion of the dominant next-to-next leading order terms in the EFTofLSS prediction at one-loop. 
In~\ref{sec:systematics}, we address known observational systematic effects and provide tests against simulations. 
In section~\ref{sec:LCDM}, we present and discuss the constraints on flat $\Lambda$CDM from the EFT analysis of the eBOSS QSO full-shape, both in Fourier and configuration space, and in combination with other cosmological probes. 
Results on extensions to $\Lambda$CDM are presented and discussed in section~\ref{sec:ext_LCDM}. 
A summary of our results and concluding remarks are given in section~\ref{sec:conclusion}. 
Additional material can be found in the appendices.
Appendix~\ref{app:extensions} is dedicated to exploring the impact of fixing the spectral tilt $n_s$ and the baryons abundance $\omega_b$  in the base-$\Lambda$CDM analysis of the eBOSS QSO full-shape. 
In appendix~\ref{app:redshift_error}, we discuss uncertainties in the redshift determination of quasars, and argue that the main correction happens to be degenerate with some EFT counterterms, justifying that our analysis is free from those potential systematics.

\section{Analysis pipeline}\label{sec:pipeline}

\subsection{Two-point function at the one loop} \label{sec:model}

\paragraph{Power spectrum}

At linear order, the power spectrum of galaxies in redshift space is given by the famous Kaiser formula \cite{Kaiser:1987qv}:
\begin{equation}
    P_{g}(z, k,\mu) =  \left[ b_1(z) + f\mu^2 \right]^2 P_{11}(z, k) \, ,
    \label{eq:linear_pk}
\end{equation}
where $P_{11}(z, k)$ corresponds to the linear matter power spectrum that can be calculated with a Boltzmann code such as \code{CLASS} \cite{Blas_2011} or \code{CAMB} \cite{Lewis:1999bs}, $f$ is the growth factor, $b_1(z)$ is the linear galaxy bias parameter, and $\mu= \hat{z}\cdot\hat{k}$ is the cosine of the angle between the line-of-sight $\vec{z}$ and the wavevector of the Fourier mode $\vec{k}$. 
At one-loop order, the formula is improved to~\cite{Perko:2016puo}: 
\begin{align}\label{eq:powerspectrum}
& P_{g}(k,\mu) =  Z_1(\mu)^2 P_{11}(k)  + 2 Z_1(\mu) P_{11}(k)\left( c_\text{ct}\frac{k^2}{{ k^2_\textsc{m}}} + c_{r,1}\mu^2 \frac{k^2}{k^2_\textsc{r}} + c_{r,2}\mu^4 \frac{k^2}{k^2_\textsc{r}} \right) \\  
& + 2 \int \frac{d^3q}{(2\pi)^3}\; Z_2(\q,\k-\q,\mu)^2 P_{11}(|\k-\q|)P_{11}(q) + 6 Z_1(\mu) P_{11}(k) \int\, \frac{d^3 q}{(2\pi)^3}\; Z_3(\q,-\q,\k,\mu) P_{11}(q) \nonumber \\
& + \frac{1}{\Bar{n}_g}\left( c_{\epsilon,0} +  c_{\epsilon}^{\textrm{mono}} \frac{k^2}{k^2_\textsc{m}} + 3c_{\epsilon}^{\textrm{quad}} \left(\mu^2-\frac{1}{3}\right) \frac{k^2}{k^2_\textsc{m}}  \right), \nonumber
\end{align}
where $k^{-1}_\textsc{m}$ is the scale controlling the spatial derivative expansion, with size given by the host halo typical extension~\cite{Senatore:2014eva}, while $k^{-1}_\textsc{r}$ is the renormalization scale of the velocity products appearing in the redshift-space expansion~\cite{Senatore:2014vja}. 
We discuss these scales more in details in section~\ref{sec:scalecut}. 
In this equation, the first term corresponds to the linear contribution, which is equivalent to eq.~\eqref{eq:linear_pk}. 
This is followed by the relevant one loop order counterterms, where $c_\text{ct}$ is a linear combination of the dark matter sound speed~\cite{Baumann:2010tm,Carrasco:2012cv} and a higher-derivative bias~\cite{Senatore:2014eva}, while $c_{r,1}$ and $c_{r,2}$ represent the redshift-space counterterms~\cite{Senatore:2014vja}. 
The next two terms, on the second line, are the one-loop perturbation contributions which involves 4 galaxy bias parameters $b_{i\in\{1,\dots,4\}}$. 
Finally, the last terms are the stochastic terms, where $\Bar{n}_g$ is the mean galaxy number density. 

In eq.~\eqref{eq:powerspectrum}, $Z_n$ corresponds to the redshift-space galaxy density kernels of order $n$ (see \emph{e.g.},~\cite{Perko:2016puo}). The equations of $Z_1$, $Z_2$, and $Z_3$ are given by:
\begin{align}\label{eq:redshift_kernels}\nonumber
    Z_1(\q_1) & = K_1(\q_1) +f\mu_1^2 G_1(\q_1) = b_1 + f\mu_1^2  \, ,\\ \nonumber
    Z_2(\q_1,\q_2,\mu) & = K_2(\q_1,\q_2) +f\mu_{12}^2 G_2(\q_1,\q_2)+ \, \frac{1}{2}f \mu q \left( \frac{\mu_2}{q_2}G_1(\q_2) Z_1(\q_1) + \text{perm.} \right) \, ,\\ \nonumber
    Z_3(\q_1,\q_2,\q_3,\mu) & = K_3(\q_1,\q_2,\q_3) + f\mu_{123}^2 G_3(\q_1,\q_2,\q_3) \nonumber \\ 
    &+ \frac{1}{3}f\mu q \left(\frac{\mu_3}{q_3} G_1(\q_3) Z_2(\q_1,\q_2,\mu_{123}) +\frac{\mu_{23}}{q_{23}}G_2(\q_2,\q_3)Z_1(\q_1)+ \text{cyc.}\right)  \, ,
\end{align}
where $\mu= \q \cdot \hat{\z}/q$, $\q = \q_1 + \dots +\q_n$, and $\mu_{i_1\ldots  i_n} = \q_{i_1\ldots  i_n} \cdot \hat{\z}/q_{i_1\ldots  i_n}$, $\q_{i_1 \dots i_m}=\q_{i_1} + \dots +\q_{i_m}$. 
$G_i$ are the standard perturbation theory velocity kernels, while $K_i$ are the galaxy density kernels, reading in the basis of descendants as~\cite{Senatore:2014eva,Angulo:2015eqa,Fujita:2016dne}: 
 \begin{align}
     K_1 & = b_1 \, , \\
     K_2(\q_1,\q_2) & = b_1 \frac{\q_1\cdot \q_2 (q_1^2 + q_2^2)}{2 q_1^2 q_2^2}+ b_2\left( F_2(\q_1,\q_2) -  \frac{\q_1\cdot \q_2 (q_1^2 + q_2^2)}{2 q_1^2 q_2^2} \right) + b_4 \, , \\
     K_3(\q,-\q,\k) & = \frac{b_1}{504 k^3 q^3}\left( -38 k^5q + 48 k^3 q^3 - 18 kq^5 + 9 (k^2-q^2)^3\log \left[\frac{k-q}{k+q}\right] \right) \nonumber \\
    &+ \frac{b_3}{756 k^3 q^5} \left( 2kq(k^2+q^2)(3k^4-14k^2q^2+3q^4)+3(k^2-q^2)^4 \log \left[\frac{k-q}{k+q}\right]  \right) \nonumber \\
    & +\frac{b_1}{36 k^3 q^3} \left( 6k^5 q + 16 k^3 q^3 - 6 k q^5 + 3 (k^2 - q^2)^3 \log \left[\frac{k-q}{k+q}\right] \right) \, ,
 \end{align}
 where $F_2$ is the symmetrized standard perturbation theory second-order density kernel (for explicit expressions see~\emph{e.g.},~\cite{Bernardeau:2001qr}), and the third-order kernel is written in its UV-subtracted version and is integrated over $k \cdot \hat q$.
 
In the following, we consider the multipoles of the galaxy power spectrum, obtained through a Legendre polynomials decomposition of the total galaxy power spectrum: 
\begin{equation}
    P_g(z,k,\mu)=\sum_{\ell=0} \mathcal{L}_\ell(\mu) P_\ell(z,k) \, ,
\end{equation}
where $\mathcal{L}_\ell$ represents the Legendre polynomial of order $\ell$, and $P_\ell(z,k)$ are the multipoles of the galaxy power spectrum defined as:
\begin{equation}
    P_\ell(z,k)=\frac{2\ell+1}{2}\int^1_{-1}d\mu \, \mathcal{L}_\ell(\mu)P_{g}(z,k,\mu) \, .
\end{equation}
Given that most of the signal-to-noise ratio resides in the monopole ($\ell=0$) and the quadrupole ($\ell=2$), we consider only these two moments in our analyses. 
In practice, we evaluate the loop corrections using the FFTLog method~\cite{Simonovic:2017mhp}.

\paragraph{Correlation function}

In this paper, we also compare results obtained with the redshift-space galaxy correlation function instead of the power spectrum.
The correlation function at one loop order corresponds to the inverse Fourier transform of the galaxy power spectrum given in eq.~\eqref{eq:powerspectrum}:
\begin{equation}
\xi_g(z, s, \mu_s) =  \int \, \frac{d^3 \k}{(2\pi)^3} \, e^{i \k \cdot \q} P_g(z, k, \mu_k) \, ,
\end{equation}
and, similarly to the galaxy power spectrum, one considers a Legendre polynomials decomposition:
\begin{equation}
    \xi_g(z, s, \mu_s) = \sum_{\ell=0}\mathcal{L}_\ell(\mu_s)\xi_\ell(z, s) \, .
\end{equation}
We can relate the correlation function multipoles to the power spectrum multipoles through a spherical-Bessel transform:
\begin{equation}
\xi_\ell(s) = i^\ell \int \frac{dk}{2\pi^2} k^2 P_\ell(k) j_\ell(ks) \, ,
\end{equation}
where $j_\ell$ is the spherical-Bessel function of order $\ell$. 
As with the power spectrum, we make use of the monopole ($\ell=0$) and the quadrupole ($\ell=2$) of the correlation function in the following. 
As the Fourier transform of power-laws are simply Dirac-$\delta$ function in configuration space, the stochastic contributions in eq.~\eqref{eq:powerspectrum} drop out from the correlation function prediction~\cite{Zhang:2021yna}. 
In practice, we evaluate the correlation function using the FFTLog method~\cite{Lewandowski:2018ywf,Zhang:2021yna}. 

\paragraph{IR-resummation}

As the long-wavelength displacements are non-perturbative in our Universe, we need to resum them to all orders to accurately describe the scales around the BAO scale~\cite{Senatore:2014via}. 
The IR-resummation of the galaxy power spectrum up to the $N$-loop order is defined as~\cite{Senatore:2014vja, Lewandowski:2015ziq,DAmico:2020kxu}: 

\begin{equation}
P^\ell(k)_{|N} = \sum_{j=0}^N \sum_{\ell'}  4\pi (-i)^{\ell'} \int dq \, q^ 2 \, Q_{||N-j}^{\ell \ell'}(k,q) \, \xi^{\ell'}_j (q) \, , 
\end{equation}
where $P^\ell(k)_{|N}$ corresponds to the resummed power spectrum, and $\xi^{\ell}_j (k)$ are the $j$-loop order pieces of the Eulerian (\emph{i.e.}, non-resummed) correlation function, respectively.

The effects from the bulk displacements are encoded in $Q_{||N-j}^{\ell \ell'}(k,q)$, given by:
\begin{align}
&Q_{||N-j}^{\ell \ell'}(k,q) = \frac{2\ell+1}{2} \int_{-1}^{1}d\mu_k \,\frac{i^{\ell'}}{4 \pi} \int d^2 \hat{q} \, e^{-i\q \cdot \k} \, F_{||N-j}(\k,\q) \mathcal{L}_\ell(\mu_k) \mathcal{L}_{\ell'}(\mu_q) \, , \label{eq:resumQ}\\
&F_{||N-j}(\k,\q) = T_{0,r}(\k,\q)\times T_{0,r}^{-1}{}_{||N-j}(\k,\q) \, , \nonumber\\
&T_{0,r}(\k,\q) = \exp \left\lbrace -\frac{k^2}{2} \left[ \Xi_0(q) (1+2 f \mu_k^2 + f^2 \mu_k^2) + \Xi_2(q) \left( (\hat k \cdot \hat q)^2 + 2 f \mu_k \mu_q (\hat k \cdot \hat q) + f^2 \mu_k^2 \mu_q^2 \right) \right] \right\rbrace \,  , \nonumber
\end{align}
where $\Xi_0(q)$ and $\Xi_2(q)$ are defined as: 
\begin{align}\label{eq:Xi_02}
\Xi_0(q) & = \frac{2}{3} \int \frac{dp}{2\pi^2} \, \exp \left(-\frac{p^2}{\Lambda_{\rm IR}^2} \right)  P_{11}(p) \, \left[1 - j_0(pq) - j_2(pq) \right] \, , \\
\Xi_2(q) & = 2 \int \frac{dp}{2\pi^2}\, \exp \left( -\frac{p^2}{\Lambda_{\rm IR}^2} \right)  P_{11}(p) \, j_2(pq) \, .
\end{align}
In practice, we evaluate the IR-resummation using the FFTLog method~\cite{DAmico:2020kxu}. 

\paragraph{Additional modeling effects}

On top of the description of the biased tracers in redshift space, we account for a number of observational effects, as described in Ref.~\cite{DAmico:2019fhj}: the Alcock-Paszynski effect~\cite{Alcock:1979mp}, the window functions as implemented in Ref.~\cite{Beutler:2018vpe} (see also details in appendix~A of Ref.~\cite{Simon:2022adh}), and binning. 
For a given redshift data slice, we evaluate our predictions at one effective redshift rather than accounting for the redshift evolution. 
In particular, we take the EFT parameters as constant within the redshift slice. 
The accuracy of this approximation has been checked in Ref.~\cite{Zhang:2021uyp} in the context of the BOSS survey. 
We have checked that this approximation leads to negligible shifts in the determined cosmological parameters from eBOSS, as expected from the size of the survey compared to the one of BOSS. 
Tests and further considerations on observational effects are given in section~\ref{sec:systematics}. 
Moreover, in appendix~\ref{app:redshift_error}, we show that the correction for uncertainties in the redshift determination of eBOSS QSOs is degenerate with some EFT counterterms, and therefore that our formalism naturally accounts for it.

\subsection{Cosmological inference setup} \label{sec:inference}

\paragraph{Data}
In this work, we use various sets of cosmological observations, comparing results of the EFTofLSS applied to (e)BOSS data with those from {\it Planck} CMB data, and their combination. We make use of the following datasets: 
\begin{itemize}
\item {\bf eBOSS DR16 QSO}: The main novelty of this work is the full-shape analysis of the quasars (QSO) from the extended Baryon Oscillation Spectroscopic Survey (eBOSS)~\cite{eBOSS:2020yzd}. 
The QSO catalogs are described in Ref.~\cite{Ross:2020lqz}. 
The covariances are built from the EZmocks described in Ref.~\cite{Chuang:2014vfa}. 
There are about $343 \, 708$ quasars selected in the redshift range $0.8 < z < 2.2$, making for a sample of about $0.6$Gpc${}^{3}$ at an effective redshift of $z_{\rm eff}=1.52$, cut into two skies, NGC and SGC. 
We analyze the full-shape of the eBOSS QSO power spectrum multipoles, $\ell=0,2$, namely the monopole and the quadrupole, measured in Ref.~\cite{Beutler:2021eqq}.~\footnote{Publicly available at: \url{https://fbeutler.github.io/hub/deconv_paper.html}.}
The covariances and the window functions we use are also from~\cite{Beutler:2021eqq}.  
We use data (and associated covariance matrices) deconvolved from the window functions, such that one does need to apply them to the prediction~\cite{DAmico:2019fhj,Beutler:2021eqq}. 
We analyze the correlation function multipoles measured in Ref.~\cite{Hou:2020rse}.~\footnote{Publicly available at: \url{https://svn.sdss.org/public/data/eboss/DR16cosmo/tags/v1_0_1/dataveccov/lrg_elg_qso/QSO_xi/}.}
When not explicitly mentioned, our eBOSS results are obtained with the power spectrum.

\item {\bf BOSS DR12 LRG}: We compare and combine the eBOSS QSO with BOSS luminous red galaxies (LRG)~\cite{BOSS:2016wmc}. 
The BOSS catalogs are described in Ref.~\cite{Reid:2015gra}. 
The covariances are built from the patchy mocks described in Ref.~\cite{Kitaura:2015uqa}. 
The BOSS data are cut into two redshift bins, LOWZ and CMASS, spanning ranges $0.2<z<0.43 \  (z_{\rm eff}=0.32)$, $0.43<z<0.7  \ (z_{\rm eff}=0.57)$, respectively, with north and south galactic skies for each, respectively denoted NGC and SGC. 
We use the EFT likelihood of the full-shape of the BOSS LRG power spectrum pre-reconstructed multipoles, $\ell=0,2$ (namely the monopole and the quadrupole), measured and described in Ref.~\cite{Zhang:2021yna}, together in cross-correlation with post-reconstruction BAO compressed parameters obtained in Ref.~\cite{DAmico:2020kxu} on the post-reconstructed power spectrum measurements of Ref.~\cite{Gil-Marin:2015nqa}. 

\item {\bf ext-BAO}: We also combine the data from eBOSS and BOSS with external BAO (ext-BAO) measurements, namely data from 6dFGS at $z = 0.106$ and SDSS DR7 at $z = 0.15$ \cite{Beutler:2011hx,Ross:2014qpa,BOSS:2016wmc}, and the joint constraints from eBOSS DR14~\footnote{These data were recently updated in Ref.~\cite{Cuceu:2022wbd} and are consistent with those used in this work.} Ly-$\alpha$ absorption auto-correlation at $z = 2.34$ and cross-correlation with quasars at $z = 2.35$ \cite{Agathe:2019vsu, Blomqvist:2019rah}.

\item {\bf Pantheon}: We also include the Pantheon18 SNIa catalogue,~\footnote{We note that, as this work was completed, the new Pantheon+ data became available \cite{Brout:2022vxf}. Given that the datasets are broadly consistent we do not expect major changes in our conclusions.} spanning redshifts $0.01 < z < 2.3$ \cite{Scolnic:2017caz}.
We stress that here we are only using the uncalibrated luminosity distance to Pantheon18 SNIa.

\item {\bf Planck}: Finally, we compare constraints obtained from different LSS surveys combinations with the {\it Planck} results obtained from analyzing the high-$\ell$ TTTEEE + lowE + lensing~\cite{Planck:2018vyg}. We use the nuisance parameters marginalized {\it Planck} \code{lite} likelihood when performing the MCMC, but switch to the full likelihood to derive the best-fit. 

\end{itemize} 
We dub ``LSS'' the combination of eBOSS + BOSS + ext-BAO + Pantheon to refer to an analysis that is independent of {\it Planck} (or any CMB data). 

\paragraph{Likelihood and prior}
To describe the eBOSS QSO full-shape data, we use the following likelihood $\mathcal{L}$:
\begin{equation}
-2 \log( \mathcal{L} ) = (D-T(\theta)) \cdot C^{-1} \cdot (D-T(\theta)) - 2 \log p(\theta)\, .
\end{equation}
Here $D$ is the data vector, constructed from the measurements of the monopole and quadrupole of the power spectrum or the correlation function; $T(\theta)$ is the corresponding EFTofLSS prediction, containing also additional modeling effects, as described in section~\ref{sec:model}, where $\theta$ designates generically all parameters-- cosmological and EFT ones-- entering in $T$; 
$C^{-1}$ is the inverse covariance built from the mocks mentioned in previous paragraph; $p(\theta)$ is the prior that we describe next. 

For our baseline $\Lambda$CDM analysis of the LSS data, we vary three cosmological parameters within uninformative large flat prior: $\lbrace \omega_{cdm}, \, h, \, \log (10^{10} A_s)\rbrace$, corresponding respectively to the physical cold dark matter abundance, the reduced Hubble constant, and the $\log$-amplitude of the primordial fluctuations.  
We fix the physical baryons abundance $\omega_b = 0.02233$, as motivated by big-bang nucleosynthesis estimates~\cite{Mossa:2020gjc}, and the spectral tilt $n_s = 0.965$ to {\it Planck} preferred value~\cite{Planck:2018vyg}. 
The impact on the posteriors of letting these parameters free to vary is discussed in appendix~\ref{app:extensions}. 
Let us stress that every time {\it Planck} data are used, we free $\omega_b$ and $n_s$, and co-vary the optical depth to reionization $\tau_{\rm reio}$.

For better comparison with the literature, we present most of our cosmological results on the fractional matter abundance, the reduced Hubble constant, and the clustering amplitude, respectively $\lbrace \Omega_m, \, h, \, \sigma_8 \rbrace$. We also explore one-parameter extensions to this baseline $\Lambda$CDM analysis, freeing either the fractional curvature density $\Omega_k$, the equation of state parameter of a smooth dark energy field $w_0$, the sum of the neutrino masses $\sum m_{\nu}$, or the effective number of relativistic species $N_{\rm eff}$. 
For all runs performed, unless specified, we use {\it Planck} prescription for the neutrinos, taking two massless and one massive species with $m_{\nu} = 0.06e$V~\cite{Planck:2018vyg}.

For the EFT parameters, we analytically marginalized over the parameters appearing only linearly in our predictions with a Gaussian prior centered on $0$ of width $\sim \mathcal{O}(b_1)$ in order to keep them within physical range~\cite{DAmico:2019fhj}. 
As for the remaining ones, $b_1$ and $c_2$, we use flat prior $[0,4]$ and $[-4, 4]$ while scanning them. 
We refer to Ref.~\cite{Simon:2022lde} for a detailed description of our choice of priors, that is dubbed ``West-coast'' prior therein.  
For eBOSS, we use $\bar n_g = 2 \cdot 10^{-5} (\hinvMpc)^3$ as estimated from Ref.~\cite{Lyke:2020tag}, while for BOSS, we use $\bar n_g  = 4 \cdot 10^{-4} (\hinvMpc)^3$ as estimated from Ref.~\cite{BOSS:2016wmc}. 
The values chosen for $k_{\rm M}$ and $k_{\rm R}$ is discussed in section~\ref{sec:scalecut}. 
We assign one set of EFT parameters per skycut (NGC or SGC) and per redshift bin for BOSS and eBOSS. 
To obtain the best-fits, we follow the appendix~C of Ref.~\cite{DAmico:2020kxu} and minimize the full likelihood, but without scanning over the parameters appearing only linearly in our predictions, as described therein. 

When analyzing both BOSS and eBOSS full-shapes, we simply add their corresponding likelihoods, as there is no overlap. 
We also simply add the likelihoods when combining with ext-BAO, Pantheon, or {\it Planck}, neglecting potential small correlation.

\paragraph{Posterior sampling} 

We sample the posteriors from our likelihood using the Metropolis-Hasting algorithm from \code{MontePython}~\cite{Brinckmann:2018cvx, Audren:2012wb}.~\footnote{ \url{https://github.com/brinckmann/montepython\_public}}
The linear power spectra are computed with the \code{CLASS} Boltzmann code~\cite{Blas_2011}.~\footnote{ \url{http://class-code.net}}
The full-shape prediction from the EFTofLSS with additional modeling effects are computed using \code{PyBird}~\cite{DAmico:2020kxu}.~\footnote{\url{https://github.com/pierrexyz/pybird}}

\paragraph{Credible interval and best-fit}
As discussed in Ref.~\cite{Simon:2022lde} in the context of the full-shape analysis of BOSS data with the EFTofLSS, shifts between the means of the 1D marginalized posteriors with respect to the most-likely values can arise due to prior volume effects. 
Therefore, when presenting the cosmological results, on top of providing the $68\%$-credible intervals, we also systematically provide the corresponding most-likely values. Those latter are sensitive to the prior weight but are not affected by the prior volume projection effect, as discussed throughout in Ref.~\cite{Simon:2022lde}. 
Let us also caution about the determination of the best-fit values quoted in this work, as they can be subject to some uncertainty, given the flatness of the likelihood of eBOSS QSO full-shape around its maximum in some particular directions. 
For example, we find for eBOSS alone $\Delta \chi^2 \lesssim 0.2$ when moving by $\sim 1\sigma$ in the direction of $\Omega_m$ or $\sigma_8$, where $\sigma$ is the error bars read from the $68\%$-credible intervals. As a complete profile likelihood analysis is beyond the scope of this work, we leave this to future investigation. See, \emph{e.g.}, Ref.~\cite{Planck:2013nga} for discussions on how to mitigate those uncertainties. 

\subsection{Scale cut from governing scales} \label{sec:scalecut}

In this section, we determine the scale cut of the full-shape analysis of eBOSS QSOs directly from the data by considering the impact of higher-order corrections to our one-loop prediction. 
We additionally validate our likelihood against simulations in order to cross-check the value of the scale cut, as well as test for additional modeling uncertainties independent of the EFTofLSS formalism. 
We summarize here the scale cut values that we use, and for which we find that the determination of the cosmological parameters is safe from significant systematic shifts due to the theory error:
\begin{itemize}
\item When analyzing the eBOSS QSO power spectrum, we include $k \in [0.01, 0.24] \hinvMpc$. 
\item When analyzing the eBOSS QSO correlation function, we include $s \in [20, 160] \Mpcinvh$ (which are all the scales provided in the public data). 
\item For completeness, we recall that in the analysis of BOSS LRG power spectrum, we include $k \in [0.01, 0.23] \, \hinvMpc$ for CMASS, while we include $k \in [0.01, 0.20] \, \hinvMpc$ for LOWZ~\cite{Colas:2019ret,DAmico:2020kxu}. 
\end{itemize}
Let us remark that it may come as a surprise that the maximal wavenumber of the analysis for eBOSS QSO are so close to the one of BOSS. 
Indeed, naively, given that eBOSS data are at higher redshift, therefore with smaller nonlinearities, and are of smaller volume than BOSS data, one may expect that the EFTofLSS prediction at one loop could allow to include (in a controlled manner) deeper scales in the eBOSS full-shape analysis, \emph{i.e.}, $k_{\rm max}^{\rm eBOSS} > k_{\rm max}^{\rm BOSS}$. 
In fact, the scales at which the theory error starts to become important is similar in both analyses. This is because, although eBOSS error bars are larger, the theory error is dominated by terms that renormalize products of velocities in the redshift-space expansion and that happen to be larger for eBOSS QSOs than BOSS LRGs. 
We expand over this issue in the following.

\paragraph{Governing scales in eBOSS QSOs} 
In section~\ref{sec:model} appear three scales governing the EFT expansions: the nonlinear scale, $\knl^{-1}$, the spatial extension of the observed objects, $\km^{-1}$, and the ``dispersion'' scale, $\kr^{-1}$. 
The nonlinear scale $\knl^{-1}$ can be understood as the typical distance matter particles travel during the (finite) age of the Universe $\sim \mathcal{H}^{-1}$. 
It is thus about $\knl^{-1} \sim v / \mathcal{H}$, where $v \ll 1$, as we are dealing with non-relativistic matter. 
For the long-wavelength modes of interest, $k \sim \mathcal{H}$, this implies $k / \knl \ll 1$. 
Therefore, this allows us to organize the expansion of the galaxy density field $\delta_g$ in long-wavelength ``smoothed'' fluctuations, \emph{e.g.}, $\delta_g \sim \partial^2 \phi / \mathcal{H}^{2} \sim (k / \knl)^2 \ll 1$~\cite{Baumann:2010tm}.
In particular, at next-to-leading order, this scale appears explicitly in the counterterm $\propto (\partial^2/\knl^2) \partial^2 \phi / \mathcal{H}^{2} \sim c_s^2 (k^2/\knl^2) \delta^{(1)}$ renormalizing the dark matter field $\delta$ at short distances~\cite{Carrasco:2012cv}.
The spatial extension of the observed objects $\km^{-1}$ is instead controlling the spatial derivative expansion~\cite{Senatore:2014eva}. 
At leading order in derivatives, $\delta_g$ receives a contribution of the form of $\delta_{hd} \sim \ (\partial^2 / \km^2  ) \partial^2 \phi / \mathcal{H}^{2} ~ \sim c_{hd} (k^2/\km^2) \delta^{(1)}$.
For halos, $\km^{-1}$ is typically of a few Mpc's, and similarly for LRGs residing in halos, or QSOs residing in LRGs. 
Therefore, in the galaxy power spectrum, eq.~\eqref{eq:powerspectrum}, given $\knl \sim \km$, the counterterm proportional to $c_{\rm ct} \sim c_s^2 + c_{hd}$ is then the linear combination of the two aforementioned contributions, arising from the renormalization of the dark matter field at short distances and from the spatial derivatives expansion.  
Finally, the products of velocity operators, appearing in the redshift space expansion, are renormalized by counterterms entering with yet another scale, $\kr^{-1}$~\cite{Lewandowski:2015ziq,DAmico:2021ymi}, \emph{e.g.}, $\delta_g \supset \tfrac{1}{2} \mu^2 k^2 (\v \cdot \hat z)^2 \supset \tfrac{1}{2} c_{r,1} \mu^2 (k/ \kr)^2 \delta^{(1)}$, where $c_{r,1}$ is defined in eq.~\eqref{eq:powerspectrum}.

In practice, these scales can be measured directly from the data with associated EFT parameters $\sim \mathcal{O}(1)$. 
As explained above, since for biased tracers, the dark matter counterterm is degenerate with the higher-derivative term with $\knl^{-1}
 \sim \km^{-1}$, we only measure $\km$ and $\kr$.
For BOSS, it was found that $\km \sim 0.7 \hinvMpc$ and $\kr \sim 0.35 \hinvMpc$~\cite{DAmico:2019fhj,DAmico:2021ymi}.~\footnote{Note that in this paper we have adjusted the value of $k_{\rm R}$, redefining accordingly the associated EFT parameters. } 
We find for eBOSS similar value for the scale $\km \sim 0.7 \hinvMpc$, which is expected since QSOs are residing in LRGs. 
In contrast, we measure that $\kr^{\rm eBOSS} \sim 0.7 \kr^{\rm BOSS}$, such that $c_{r,1} \sim \mathcal{O}(1)$. 
Such measurement indicates that the ``velocity dispersion'' along the line-of-sight is larger in eBOSS QSOs than in BOSS LRGs, which could point to the possibility that QSOs populate preferentially satellite galaxies rather than central galaxies (see also Ref.~\cite{2021MNRAS.504..857A} suggesting likewise from the perspective of halo occupation distribution).

For all analyses in this paper, we choose the following values for the renormalization scales, ensuring that $c_{\rm ct}$, $c_{r,1}$  are measured well within their Gaussian prior $\mathcal{N}(0,2)$:
\begin{align}
\km^{\rm BOSS} = 0.7 \hinvMpc, & \qquad  \kr^{\rm BOSS} = 0.35 \hinvMpc , \\
\km^{\rm eBOSS} = 0.7 \hinvMpc, & \qquad \kr^{\rm eBOSS} = 0.25 \hinvMpc.  \nonumber
\end{align}

\paragraph{Next-to-next-to-leading order}
Given the scale estimates above, the theory error associated to the higher order terms not included in our baseline analysis is dominated by the terms associated to $\kr^{-1} > \knl^{-1}, \km^{-1}$. 
Thus, the size of the theory error can be estimated by the size of the largest contributions at next-to-next-to-leading order (NNLO), which are given by the counterterms~\cite{DAmico:2021ymi}: 
\begin{equation}\label{eq:nnlo}
P_{\rm NNLO}(k, \mu) = \frac{1}{4} b_1 \left( c_{r,4} b_1   + c_{r,6} \mu^2 \right) \mu^4 \frac{k^4}{k_{\rm R}^4} P_{11}(k) \, ,
\end{equation}
where $c_{r,4}$ and $c_{r,6}$ are the $\mathcal{O}(1)$-parameters controlling the size of the NNLO counterterms. 
Following Ref.~\cite{Zhang:2021yna}, we measure the shift in each 1D posterior upon adding of the NNLO term given by eq.~\eqref{eq:nnlo} as a function of $\kmax$. 
The scale cut is determined as the highest scale $\kmax$ included in the analysis such that the theory error is safely small.

\paragraph{$\kmax$ vs. theory error} 
In table~\ref{tab:nnlo}, we show as a function of $\kmax$ the shifts upon adding the NNLO term, eq.~\eqref{eq:nnlo}, to the one-loop prediction, eq.~\eqref{eq:powerspectrum}, on the 1D posteriors of the three baseline cosmological parameters, $\Omega_m$, $h$, and $\sigma_8$, and two EFT parameters, $b_1$ and $c_2$. 
Given the scale estimates above, we put a Gaussian prior $\mathcal{N}(0,4)$ on the NNLO parameters $c_{r,4}$ and $c_{r,6}$ to keep their size within physical range with a conservative choice. 
The posteriors are obtained as described in previous subsection, additionally marginalizing over $c_{r,4}$ and $c_{r,6}$ when including the NNLO term in the prediction.

\begin{table*}
\footnotesize
\center
\begin{tabular}{|l|c|c|c|c|}
\hline
\multicolumn{5}{|c|}{  $\Delta^{\rm shift}_{\rm NNLO} / \sigma_{\rm stat}^{\rm eBOSS}$ }\\
 \hline
$\kmax$ [$\hinvMpc$] & 0.21 & 0.24  & 0.27 & 0.30 \\
 \hline
 $\Omega_{\rm m}$ & 0.00 & -0.04 & 0.05 & -0.47\\
 $h$ & -0.02 & -0.07 & -0.06 & 0.09  \\
 $\sigma_8$ & -0.04 & -0.15 & -0.03 & -0.20  \\
\hline
 $b_1^{\rm N}$ & 0.02 & 0.09 & 0.02 & -0.09 \\
 $c_2^{\rm N}$ & 0.02 & 0.06 & 0.00 & -0.18\\
 $b_1^{\rm S}$ & 0.02 & 0.08 & 0.06 & 0.01\\
 $c_2^{\rm S}$ & -0.02 & -0.07 & 0.04 & 0.00\\
\hline
\end{tabular}
\caption{\footnotesize Relative shifts of the cosmological and EFT parameters from the addition of the NNLO to the base-$\Lambda$CDM fit to eBOSS for different values of $\kmax$ in $\hinvMpc$.}
\label{tab:nnlo}
\end{table*}

At $\kmax \leq 0.27\hinvMpc$, the shift in all parameters is small.
At $\kmax = 0.30\hinvMpc$, the shift starts to become appreciable in $\Omega_m$, to about $0.5\sigma$, where $\sigma$ is the $68\%$ CL. 
Therefore, staying on the conservative side and within the EFT regime of validity, we take as our final choice for the scale cut $\kmax = 0.24 \hinvMpc$, where no appreciable shift is observed. Let us note that the same conclusion is reached when considering the shift in the best-fit values upon inclusion of the NNLO: at $\kmax = 0.24\hinvMpc$, we find $\lesssim 0.25\sigma$ on all cosmological parameters.

Formally the EFT expansions are convergent only for $k < \min(\knl, \km, \kr) = \kr$.
Therefore, for eBOSS, this reinforce our choice of $\kmax = 0.24 \hinvMpc$, as it is smaller than the EFT breakdown scale, $\kr \sim 0.25 \hinvMpc$. However, we note that this $\kmax$ is rather close to $\kr$. Consequently, the associated parameter controlling this expansion is close to unity, and the whole tower of counterterms (at all order) proportional to powers of $\kr^{-1}$ can in principle be of the same order as the linear power spectrum around $\kmax$. 
Nevertheless, the theory error should ultimately be compared with the observational error to gauge whether the computation is sufficiently under control. In practice, we find that despite being very close the EFT breakdown scale, for our truncated expansion at one-loop, the shift in parameters presented in table~\ref{tab:nnlo} is much smaller than the observational error up to $\kmax \lesssim \kr$, and it is safe to take $\kmax=0.24\hinvMpc$. 
In addition, an analysis to higher $\kmax$ than the one found here is not precluded by the size of $\kr^{-1}$. Indeed, the multipoles can be rotated to form new linear combinations where the terms associated to $\kr^{-1}$ are suppressed~\cite{DAmico:2021ymi}. 
As shown in Ref.~\cite{DAmico:2021ymi}, not much cosmological constraint is gained by such analysis for BOSS data volume (see also~\cite{Ivanov:2021fbu}). 
Similarly, we find that the addition to higher wavenumbers in the full-shape analysis of eBOSS QSOs does not improve significantly the constraints. 
For simplicity, we thus present the results from the multipoles instead of the rotated ones.

\subsection{Assessing systematics beyond the EFT reach} \label{sec:systematics}

\begin{figure*}
\center
\includegraphics[width=0.5\textwidth]{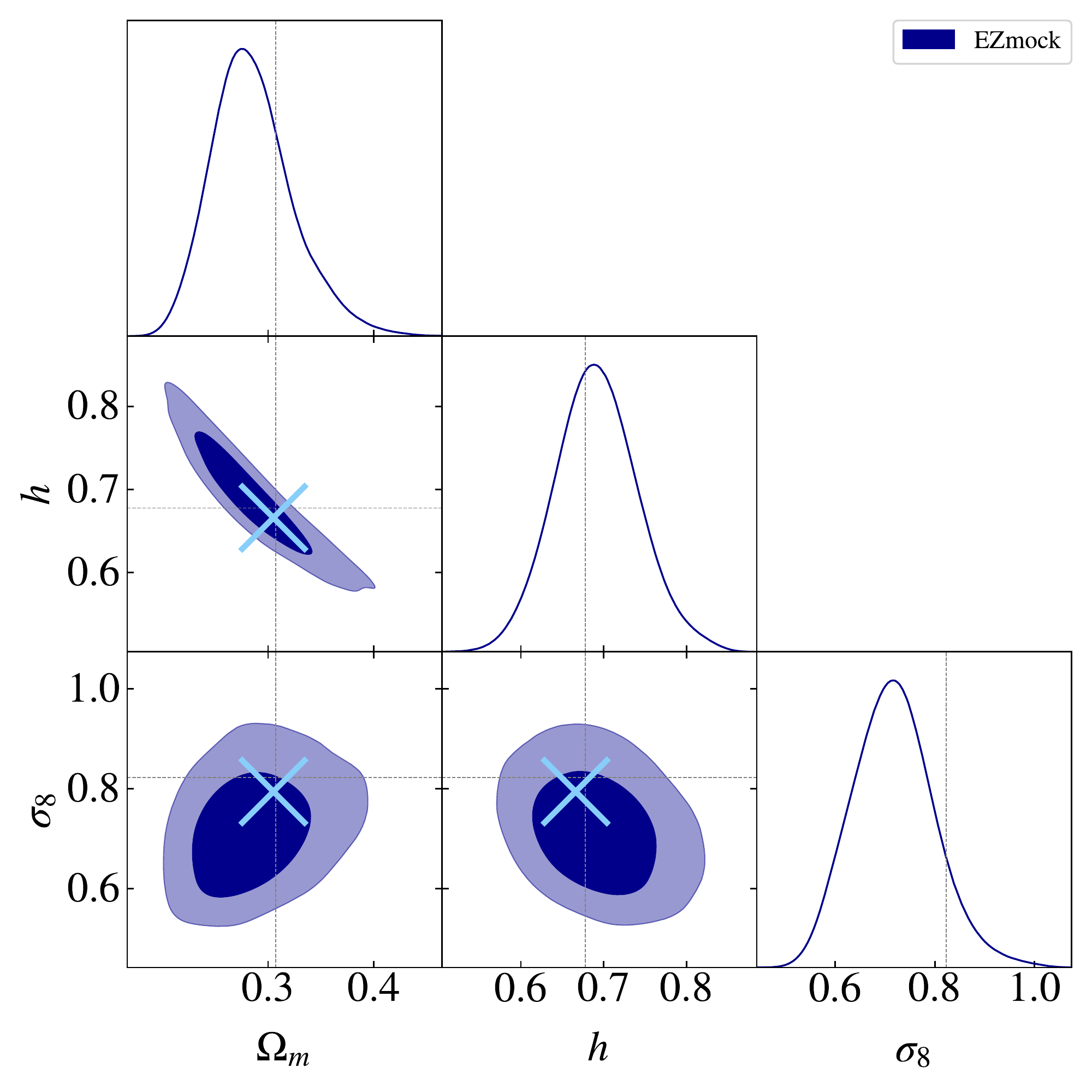}
\caption{\footnotesize Triangle plot (1D and 2D posterior distributions) of the cosmological parameters reconstructed when fitting the mean of all EZmock realizations, analyzed up to $\kmax = 0.24\hinvMpc$. The crosses correspond to the best-fit values while the dashed lines correspond to the truth of the simulation.}
\label{fig:Patchy}
\end{figure*}

Our method to determine the scale cut is convenient as it does not rely on simulations (except the ones used to built the covariance). 
We have simply taken the scale cut as being the $\kmax$ for which the theory error controlled by the largest NNLO contribution, eq.~\eqref{eq:nnlo}, is negligible at the level of the posteriors. 
This is a well-defined procedure relying only on estimates of the scales entering in the EFT expansions. 
However, it does not allow for testing the modeling aspects beyond the EFTofLSS prediction, as \emph{e.g.}, the additional observational effects described at the end of section~\ref{sec:model}. 
In order to assess the accuracy of these extra modeling aspects, in particular the window functions, we perform the following two tests. 

\paragraph{Test against simulations.} We fit the full-shape of the mean over all EZmock realizations. 
As described in section~\ref{sec:inference}, those mocks are built to simulate eBOSS observational characteristics such as the sky mask, redshift selection, systematics weights and veto, etc. 
Results from this fit are shown in figure~\ref{fig:Patchy}. 
We find that up to $\kmax = 0.24 \hinvMpc$, the best-fit values of the cosmological parameters of interest ($\Omega_m, h$, and $\sigma_8$) are shifted with respect to the truth of the simulations by $\lesssim 1/3 \cdot \sigma$.~\footnote{Hereinafter, when comparing results from the \emph{same} dataset, $\sigma$ is taken as the error bar read from the $68\%$-credible interval, regardless of whether we compare the posterior means or best-fits.}

This shows that our modeling of the observational effects in eBOSS are under good control, and our previous determination of the scale cut $\kmax = 0.24 \hinvMpc$ is corroborated by our fit to simulations. 
Before moving on, a comment on the 1D posteriors of the reconstructed cosmological parameters is in order. Although the best-fit parameters agree well with the fiducial model, we  see that the mean of the posterior is shifted with respect to the fiducial by up to $ 1.3\sigma$ (on $\sigma_8$).
The fact that the mean is shifted with respect to the best-fit represents a clear illustration of the {\it prior volume projection effects}, as mentioned in section~\ref{sec:inference} and discussed throughout in Ref.~\cite{Simon:2022lde}: this motivates us to consider the $68\%$-credible intervals \emph{together} with the corresponding best-fit values, especially when comparing results from two different experiments. 

\paragraph{Fourier vs. configuration space.} In the following section, we compare the results obtained when fitting the correlation function or the power spectrum of eBOSS QSO. 
This allows us to assess the potential systematic error associated to the estimators and the differences in the modeling discussed in section~\ref{sec:model}. 
In particular, the effect of the mask cancels in the correlation function estimator, which makes it free from potential inaccuracies associated to the modeling of the window function.  
In the following, we will demonstrate that the posteriors of $\Omega_m, \ h$, and $\sigma_8$ are consistent at $< 0.4 \sigma$ between the Fourier and the configuration space analyses (see table~\ref{tab:parameter_BOSS_eBOSS} and figure~\ref{fig:main_results} for details). 
Bearing in mind that the information content in the power spectrum and the correlation is not exactly the same in the BAO part and due to effectively different scale cuts, such consistency provides a conservative bound on the level of the aforementioned systematics. 
For more discussions on the comparison between the analyses of the power spectrum or the correlation function, see Ref.~\cite{Zhang:2021yna}. 

\paragraph{Additional systematic errors?} Last, we caution that the tests conducted in this work only assess the systematics arising at the level of the summary statistics: the cosmological results are also dependent on the choices made at the level of map-making and catalog selection, which are beyond the scope of the present study. 
We also mention that there are sub-leading contributions that we have not included in our predictions, in particular wide-angle and relativistic effects, line-of-sight dependent effects, and corrections to fiber collisions. 
The wide-angle and relativistic effects have been shown to be negligible for current surveys at $k_{\rm min } = 0.01\hinvMpc$ in, {\it e.g.}, Refs.~\cite{Castorina:2017inr,Beutler:2018vpe,Castorina:2021xzs,Noorikuhani:2022bwc}. 
Similarly, the line-of-sight dependent effects have been shown to lead to a relatively small impact on the determination of the cosmological parameters in the context of the EFT analysis of BOSS in the appendix D of Ref.~\cite{Zhang:2021yna} (see also Ref.~\cite{Zwetsloot:2022chu}). 
However, these conclusions are dependent on the choice of prior on the size of the tidal alignment biases. 
While for LRGs selected by BOSS, such estimate can be found in Ref.~\cite{Hirata:2009qz}, QSOs selected by eBOSS lack, to our knowledge, an estimate of the size of tidal alignments. 
If no estimate can be derived for QSOs, the bispectrum can mitigate those effects (see {\it e.g.}, Ref.~\cite{Agarwal:2020lov}). 
We leave this to future work. 
Finally, the fiber collisions can be treated using the effective window method put forward in Ref.~\cite{Hahn:2016kiy}. 
This correction has been implemented in the EFT analysis of BOSS in Ref.~\cite{DAmico:2019fhj}. 
The largest corrections are degenerate with the EFT counterterms~\cite{Hahn:2016kiy}, and are therefore automatically accounted for in our analysis. 
The remaining corrections, the so-called uncorrelated part~\cite{Hahn:2016kiy}, can also be straightforwardly included. 
However, they were shown to be negligible for BOSS volume~\cite{DAmico:2019fhj}: we therefore also neglect them in the current analysis. 
Finally, we show in appendix~\ref{app:redshift_error} that our analysis is free from errors in the redshift determination of eBOSS QSOs as the corrections to the prediction happen to be degenerate with some EFT counterterms.


\section{Constraints on flat $\Lambda$CDM }\label{sec:LCDM}

In this section, we present results on the flat $\Lambda$CDM model from the EFT analysis of the full-shape of eBOSS QSOs. 
We perform combined analyses with different sets of LSS surveys, namely BOSS full-shape, ext-BAO, Pantheon, as well as with {\it Planck} data, as described in section~\ref{sec:inference}. 
We remind that for our base-$\Lambda$CDM analysis, we fix the baryon abundances to the mean value measured by BBN experiments, the spectral tilt $n_s$ to {\it Planck} preferred value, and the neutrino total mass to its minimal value, as explained in section~\ref{sec:inference}. 
In appendix~\ref{app:extensions}, we show the impact on the cosmological results letting the baryon abundance to vary within a Gaussian prior motivated by BBN experiments and freeing $n_s$. 
The free neutrino mass case, together with other one-parameter extensions to our base-$\Lambda$CDM model, are presented in the next section.

\begin{figure*}
    \centering
    \includegraphics[width=0.49\columnwidth]{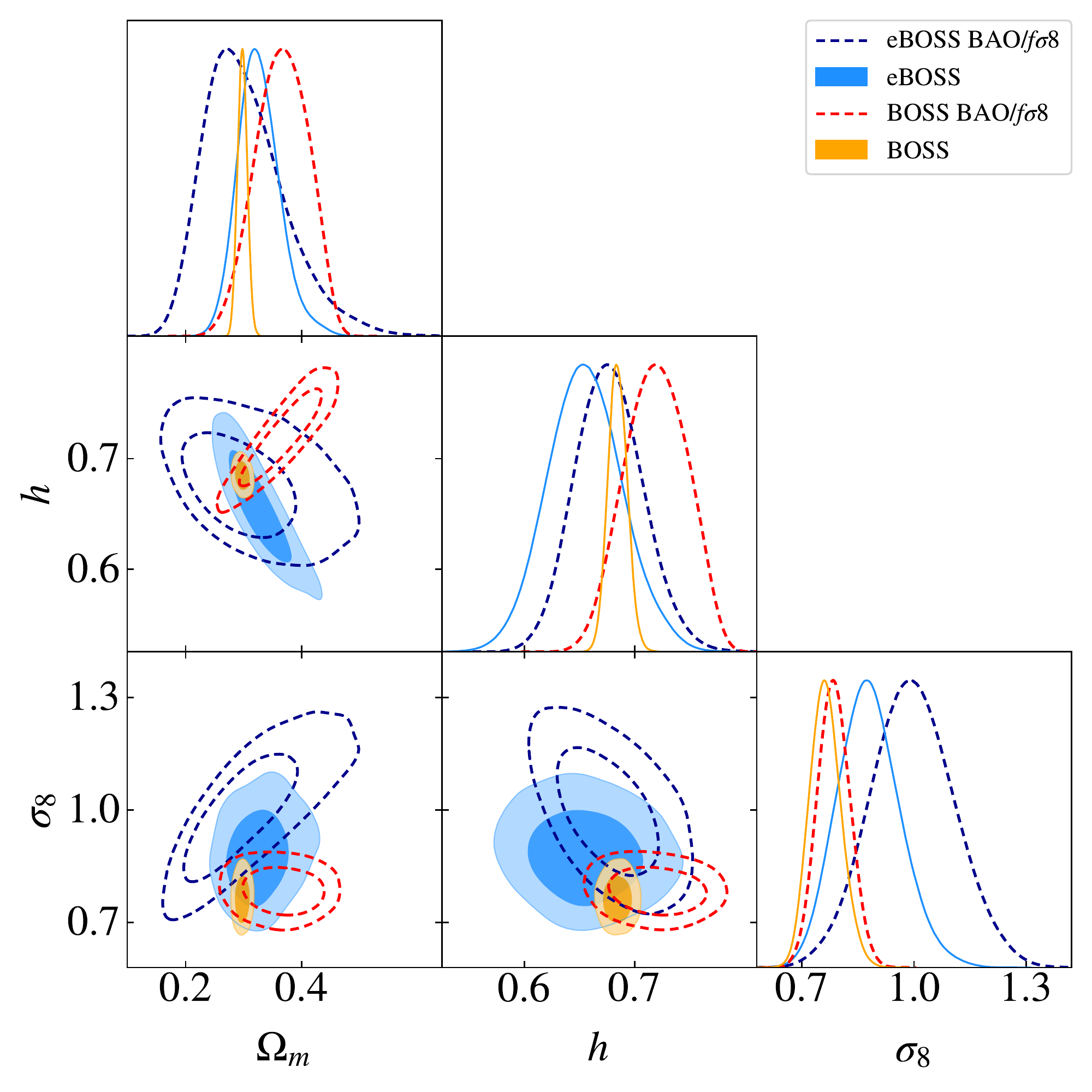}
    \includegraphics[width=0.49\columnwidth]{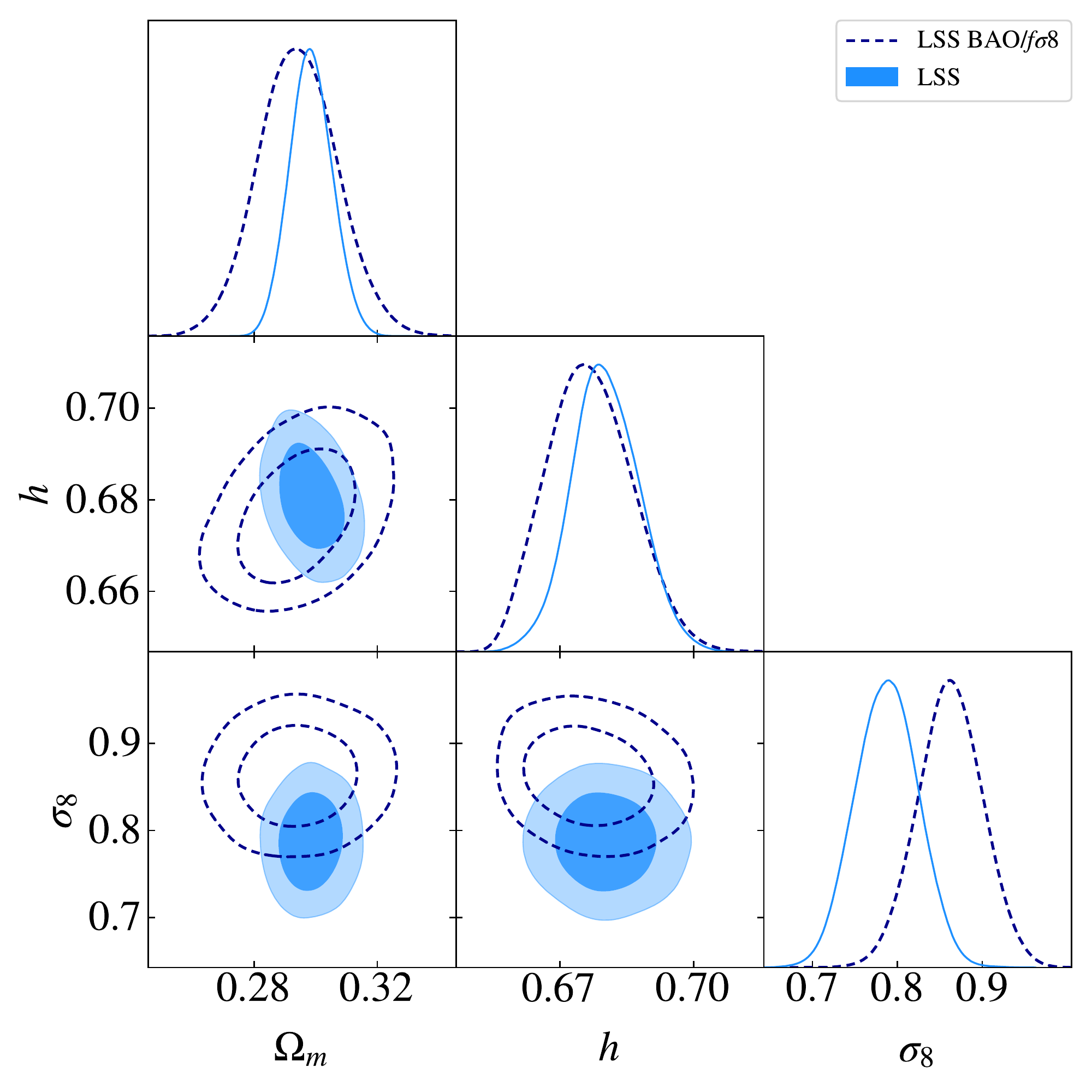}
    \caption{\footnotesize Triangle plots (1D and 2D posterior distributions) of the cosmological parameters reconstructed either from the EFT full-shape analysis or from the compressed  BAO/$f\sigma_8$ parameters. In the {\it left panel}, we have represented these two analyses for eBOSS and BOSS, while in the {\it right pannel} we have combined these two datasets and added the ext-BAO and Pantheon data.
    In this figure and for the rest of the paper, when not explicitly mentioned, eBOSS and BOSS refer to the EFT full-shape analysis of the power spectrum multipoles. }
    \label{fig:fsigma8}
\end{figure*}

As a preliminary analysis and to gauge the impact of the EFT analysis of eBOSS, we show in figure~\ref{fig:fsigma8} the results from the analyses of eBOSS, BOSS and their combination (LSS, referring to eBOSS + BOSS + ext-BAO + Pantheon), using either the combination of BAO and redshift space distortion information (BAO/$f\sigma_8$) as measured by the eBOSS collaboration \cite{Neveux:2020voa} or the EFT full-shape likelihood built in this work. 
One can see that, as expected, the EFT analysis allows us to gain significant constraining power over the conventional BAO/$f\sigma_8$ information. For eBOSS, the error bars of $\Omega_m$ and $\sigma_8$ are reduced by a factor $\sim 2.0$ and $\sim 1.3$, respectively. For BOSS, the error bars of $\Omega_m$ and $h$ are reduced by $\sim 5.4$ and $\sim 3.2$, respectively. Finally, for their combination, we find that the error bars of $\Omega_m$ and $h$ are reduced by $\sim 2.0$ and $\sim 1.25$, respectively. We also note that the EFT likelihood leads to a lower mean for $\sigma_8$ than the BAO/$f\sigma_8$ analysis: 
$\sim1\sigma$ for eBOSS and $\sim2\sigma$ for the full combination of LSS datasets. 
Note that, as can be seen in $\Omega_m - \sigma_8$ plane, the results obtained with BOSS and eBOSS are in  better agreement in the EFT analysis than in the template-based BAO/$f\sigma_8$ analysis (which shows a difference of low statistical significance). 
This may be traced in part to the lower mean value of $\sigma_8$ inferred in the EFT analysis of eBOSS, which in addition may be subject to some prior volume projection effects as we comment on later.

\begin{figure*}[p]
    \centering
    \includegraphics[width=0.7\columnwidth]{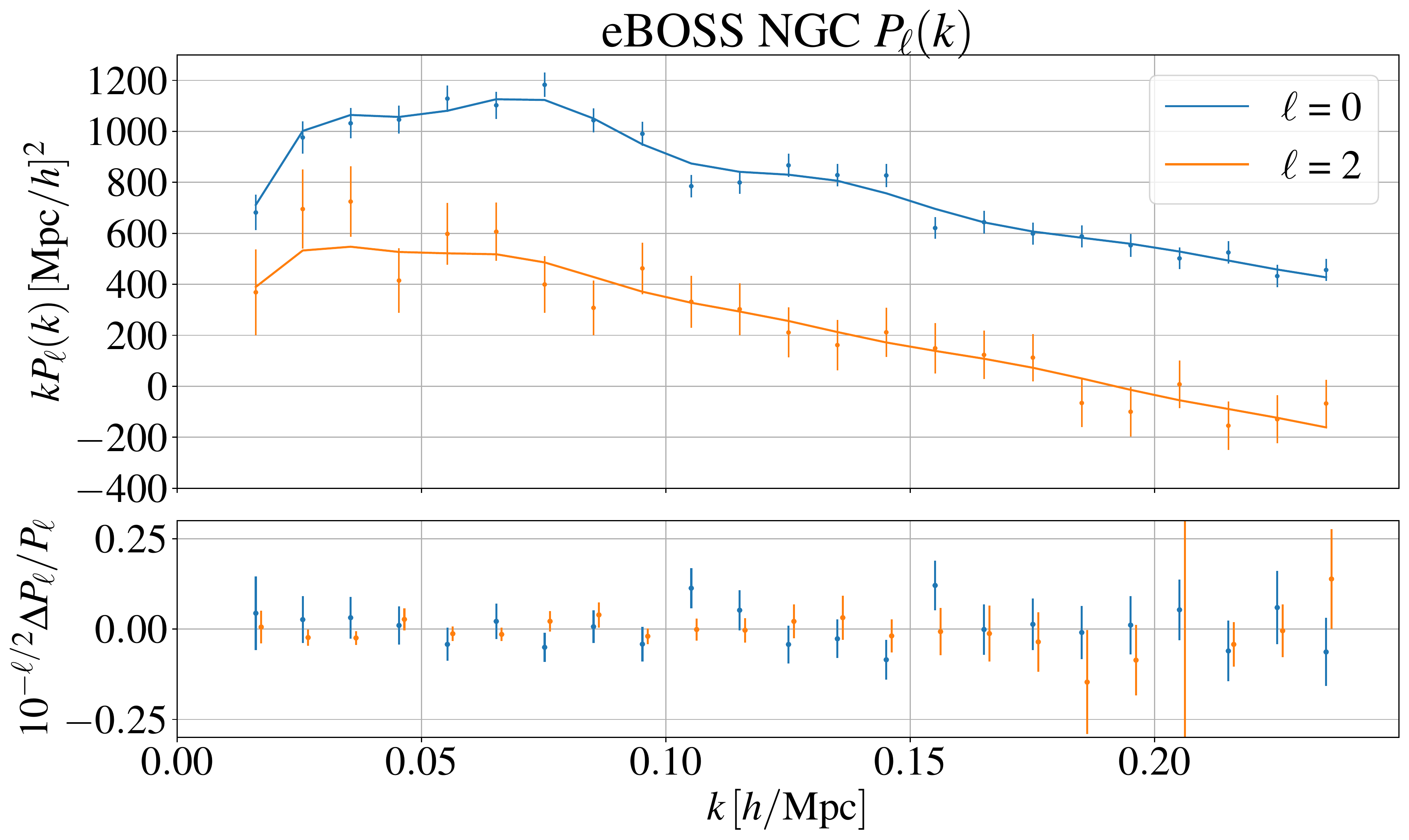}
    \includegraphics[width=0.7\columnwidth]{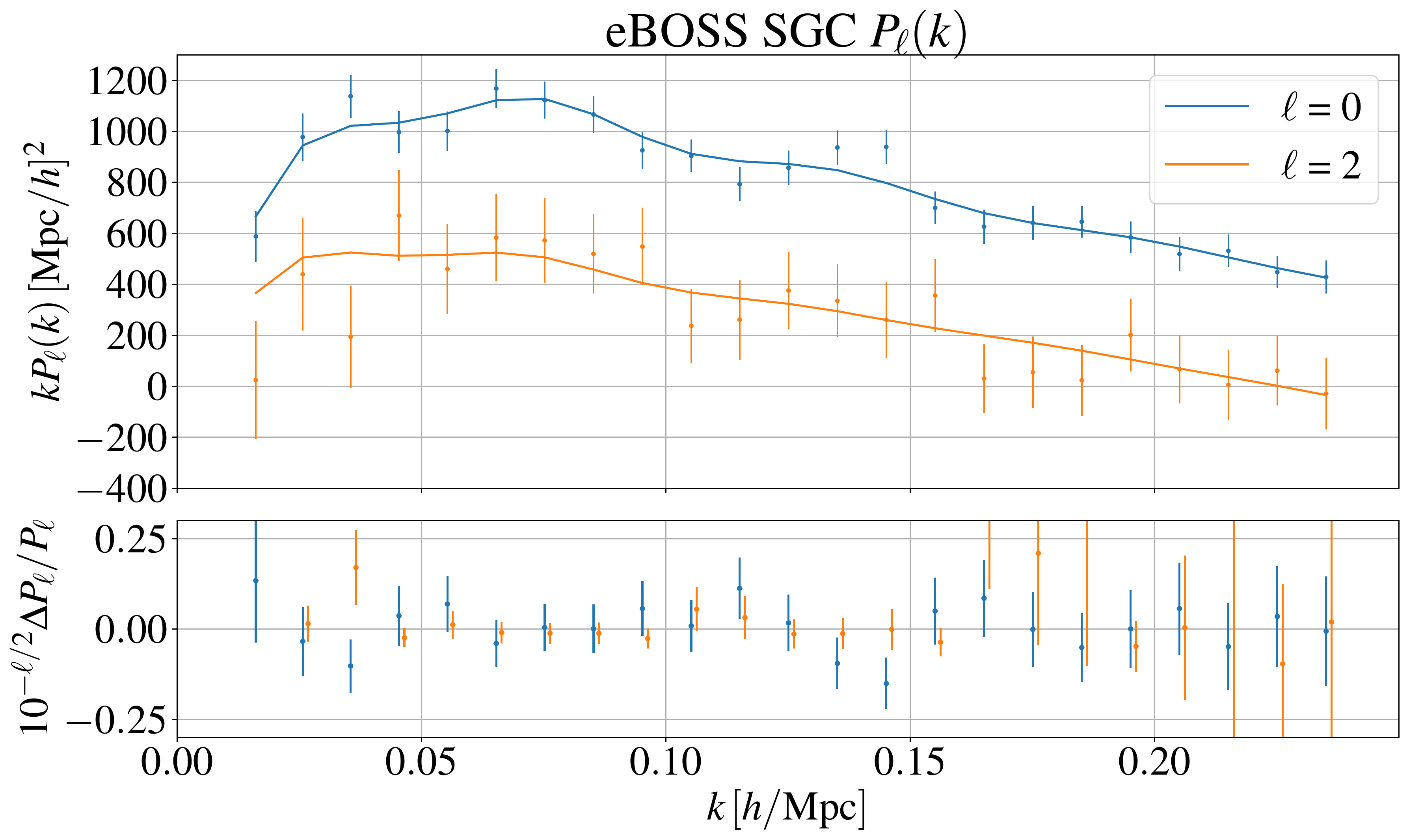}
    \includegraphics[width=0.7\columnwidth]{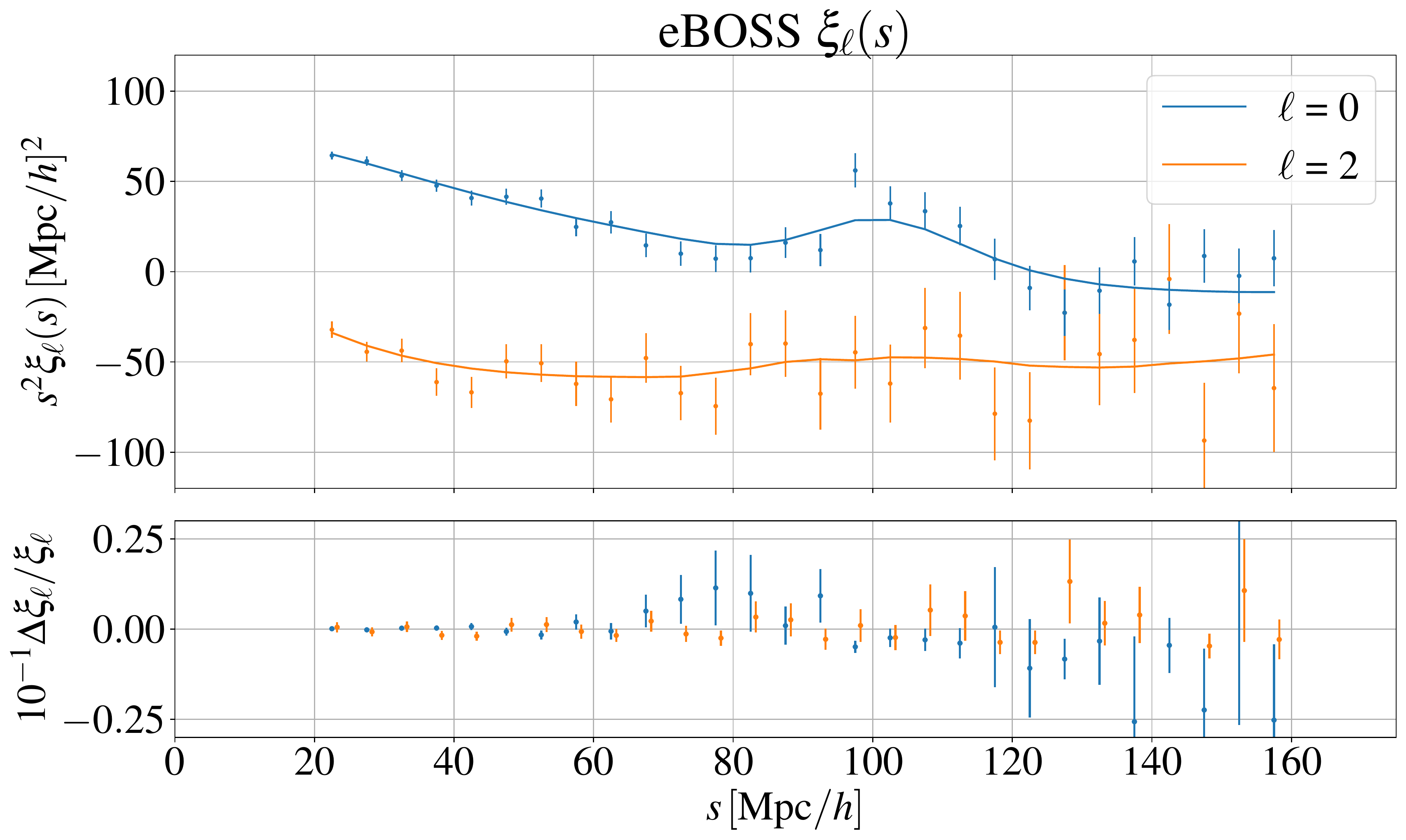}
    \caption{\footnotesize {\it Upper} - Best-fit predictions of the monopole and quadrupole of the power spectrum $P_\ell(k)$ for the NGC skycut of eBOSS QSOs against the data. We also plot the residuals with respect to the data. {\it Middle} - Same but with the eBOSS SGC skycut. {\it Lower} - Same but with the eBOSS correlation function $\xi_\ell(s)$ measured on the combination of NGC-SGC skycuts.}
    \label{fig:power_spectrum_eBOSS_FS}
\end{figure*}

\begin{table*}[ht!]
\center
\scriptsize
\begin{tabular}{|l|c|c|c|c|c|c|}
 \hline
  best-fit & \multirow{2}{*}{BOSS} & \multirow{2}{*}{eBOSS} & \multirow{2}{*}{eBOSS $\xi_\ell(s)$} & \multirow{2}{*}{eBOSS + BOSS} & BOSS & eBOSS + BOSS \\
  $\mu_{-\sigma_l}^{+\sigma_u}$ & & & & & + ext.BAO + Pan & + ext.BAO + Pan \\
 \hline

 \multirow{2}{*}{$\Omega_{\rm m}$} & $0.2981$ & $0.331$ & $0.306$ &  $0.2986$ & $0.2986$ & $0.2991$ \\
 & $0.2978^{+0.0082}_{-0.0083}$ & $0.327_{-0.039}^{+0.031}$  & $0.311^{+0.032}_{-0.037}$ & $0.2981^{+0.0077}_{-0.0079}$  &  $0.2979^{+0.0075}_{-0.0076}$ &  $0.2985^{+0.0066}_{-0.0071}$\\
 \hline
 
 \multirow{2}{*}{$h$} & $0.6839$ & $0.646$ & $0.648$ &  $0.6814$ & $0.6813$ & $0.6793$\\
 & $0.6846^{+0.0086}_{-0.0086}$ & $0.655^{+0.033}_{-0.034}$ & $0.655^{+0.026}_{-0.031}$ & $0.6827^{+0.0078}_{-0.0085}$ & $0.6812^{+0.0078}_{-0.0079}$ &  $0.6803^{+0.0072}_{-0.0078}$\\
 \hline
 
 \multirow{2}{*}{$\sigma_8$} & $0.811$ & $0.943$ & $0.922$ &  $0.840$ & $0.809$ & $0.840$\\
 & $0.763^{+0.038}_{-0.045}$ & $0.880^{+0.076}_{-0.089}$ & $0.888^{+0.084}_{-0.085}$ & $0.787^{+0.036}_{-0.039}$ & $0.762^{+0.038}_{-0.044}$ &  $0.788^{+0.037}_{-0.037}$\\
 \hline

 \multirow{2}{*}{$\omega_{\text{cdm}}$} & $0.1164$ & $0.1154$ & $0.106$ & $0.1157$ & $0.1156$ & $0.1150$\\
 & $0.1166^{+0.0047}_{-0.0047}$ & $0.1162^{+0.0077}_{-0.0079}$ & $0.110^{+0.010}_{-0.010}$ & $0.1160^{+0.0038}_{-0.0043}$ & $0.1153^{+0.0042}_{-0.0042}$ & $0.1152^{+0.0035}_{-0.0037}$\\
 \hline
 
 \multirow{2}{*}{$\text{ln}(10^{10} A_s)$} & $3.08$ & $3.42$ & $3.49$ & $3.16$ & $3.08$ & $3.17$\\
 & $2.95^{+0.12}_{-0.12}$ & $3.26^{+0.20}_{-0.21}$ & $3.36^{+0.21}_{-0.24}$ & $3.02^{+0.11}_{-0.11}$ & $3.03^{+0.12}_{-0.12}$ & $3.17^{+0.10}_{-0.10}$\\
 \hline
 
 \multirow{2}{*}{$S_8$} & $0.808$
 & $0.991$ & $0.931$ & $0.838$ & $0.807$ & $0.839$ \\
 & $0.761^{+0.040}_{-0.046}$
 & $0.918^{+0.089}_{-0.123}$ & $0.903^{+0.096}_{-0.115}$ & $0.785^{+0.037}_{-0.040}$ & $0.759^{+0.039}_{-0.045}$ & $0.786^{+0.038}_{-0.038}$ \\
\hline

$\chi^2_{\rm min}$ &  157.9  &  57.1  &  53.9 & 217.8 & 1191.1 & 1251.0	\\
\hline
$N_{\rm data}$ & 170 & 92 &  56 & 262 & 1224  & 1316 	\\
\hline
$p$-value & 0.13 & 0.94 & 0.20 & 0.47 & 0.50 & 0.64  \\
\hline
\end{tabular}
\caption{\footnotesize
Cosmological results (best-fit, posterior mean, and $68\%$ CL) of different combinations of LSS data, including BOSS and eBOSS, for our base-$\Lambda$CDM model. 
For each dataset we also report its best-fit $\chi^2_{\rm min}$, the number of data bins $N_{\rm data}$, and the associated $p$-values. In the following, ``LSS'' refers to eBOSS + BOSS + ext-BAO + Pantheon.}
\label{tab:parameter_BOSS_eBOSS}
\end{table*}

\begin{table*}[h!]
\scriptsize
\center
\begin{tabular}{|l|c|c|c|c|c|c|c|c|c|c|}
 \hline
 Parameter & \multicolumn{2}{|c|}{eBOSS - BOSS} & \multicolumn{2}{|c|}{BOSS - Planck} & \multicolumn{2}{|c|}{eBOSS - Planck} & \multicolumn{2}{|c|}{(eBOSS + BOSS) - Planck} & \multicolumn{2}{|c|}{LSS - Planck}\\
 \hline 
 & b-f & $\mu$ & b-f & $\mu$ & b-f & $\mu$ & b-f & $\mu$ & b-f & $\mu$ \\
 \hline
 $\Omega_{\rm m}$ & 0.89 & 0.76 & -1.60 & -1.58 & 0.42 & 0.30 & -1.61 & -1.62 & -1.67 & -1.68 \\
 $h$ & -1.08 & -0.84 & 1.05 & 1.08 & -0.79 & -0.53 & 0.84 & 0.93 & 0.66 & 0.72 \\
 $\sigma_8$ & 1.41 & 1.25 & -0.03 & -1.13 & 1.56 & 0.82 & 0.72 & -0.63 & 0.75 & -0.63 \\
 $\omega_{\text{cdm}}$ & -0.11 & -0.05 & -0.75 & -0.69 & -0.58 & -0.47 & -1.04 & -0.95 & -1.32 & -1.26 \\
 $\text{ln}(10^{10} A_s)$ & 1.43 & 1.30 & 0.26 & -0.76 & 1.82 & 1.05 & 1.04 & -0.19 & 1.17 & -0.09 \\
 $S_8$ & 1.55 & 1.33 & -0.55 & -1.57 & 1.43 & 0.77 & 0.11 & -1.15 & 0.13 & -1.15 \\
\hline
\end{tabular}
\caption{\footnotesize $\sigma$-deviations between the $\Lambda$CDM cosmological parameters reconstructed from eBOSS, BOSS, their combination, and {\it Planck}. 
For a given cosmological parameter, the $\sigma$-deviation metric is computed, assuming Gaussian errors, as $(\mu_{1} - \mu_{2})/\sqrt{\sigma_{1}^2 + \sigma_{2}^2}$, where $\mu_i$ are either the means ($\mu$) or the best-fits (b-f) obtained from the two experiments $i=1,2$, while $\sigma_i$ are the associated error bars read from the $68\%$-credible intervals.}
\label{tab:sigma_deviation}
\end{table*}

\subsection{Flat $\Lambda$CDM from the EFT analysis of eBOSS}

\paragraph{Goodness of fit} Before commenting over the reconstructed cosmological parameters, let us first assess the goodness of fit. 
We plot in figure~\ref{fig:power_spectrum_eBOSS_FS}, using the best-fit parameters listed in table~\ref{tab:parameter_BOSS_eBOSS}, the theoretical prediction of the monopole and quadrupole of the power spectrum, as well as the correlation function, against the data. One can see that there are not particular features in the residuals.
We list in table~\ref{tab:parameter_BOSS_eBOSS} the $\chi^2_{\rm min}$ and degrees of freedom of each fit. Assuming that all data points and parameters are uncorrelated,  we find that the $p$-values associated with the different fits are acceptable both for our analyses in Fourier and configuration space, which tell us that our model is a good description of the data, up to the scale cut chosen in section~\ref{sec:scalecut}.

\begin{figure*}[p]
    \centering
    \includegraphics[width=0.49\columnwidth]{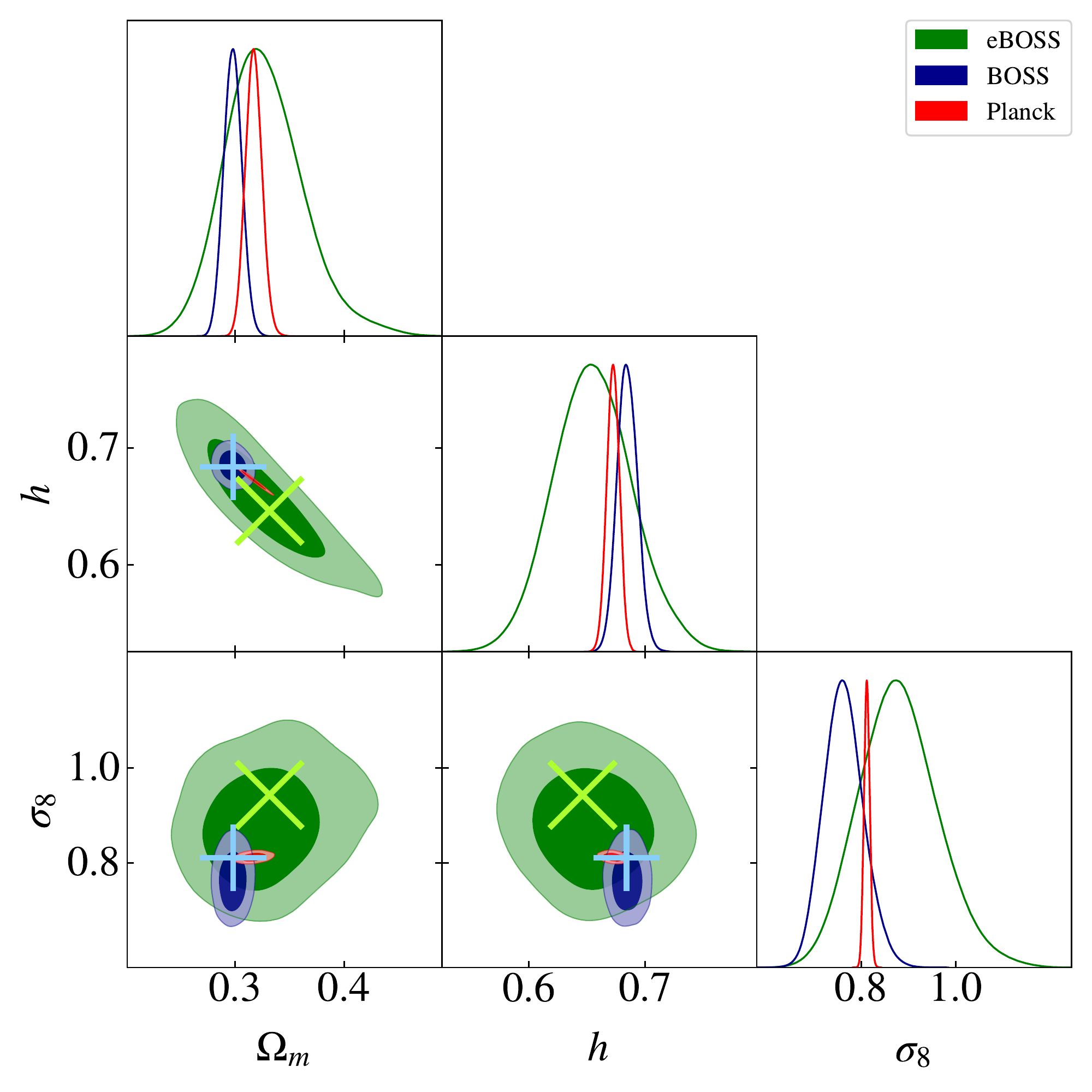}
    \includegraphics[width=0.49\columnwidth]{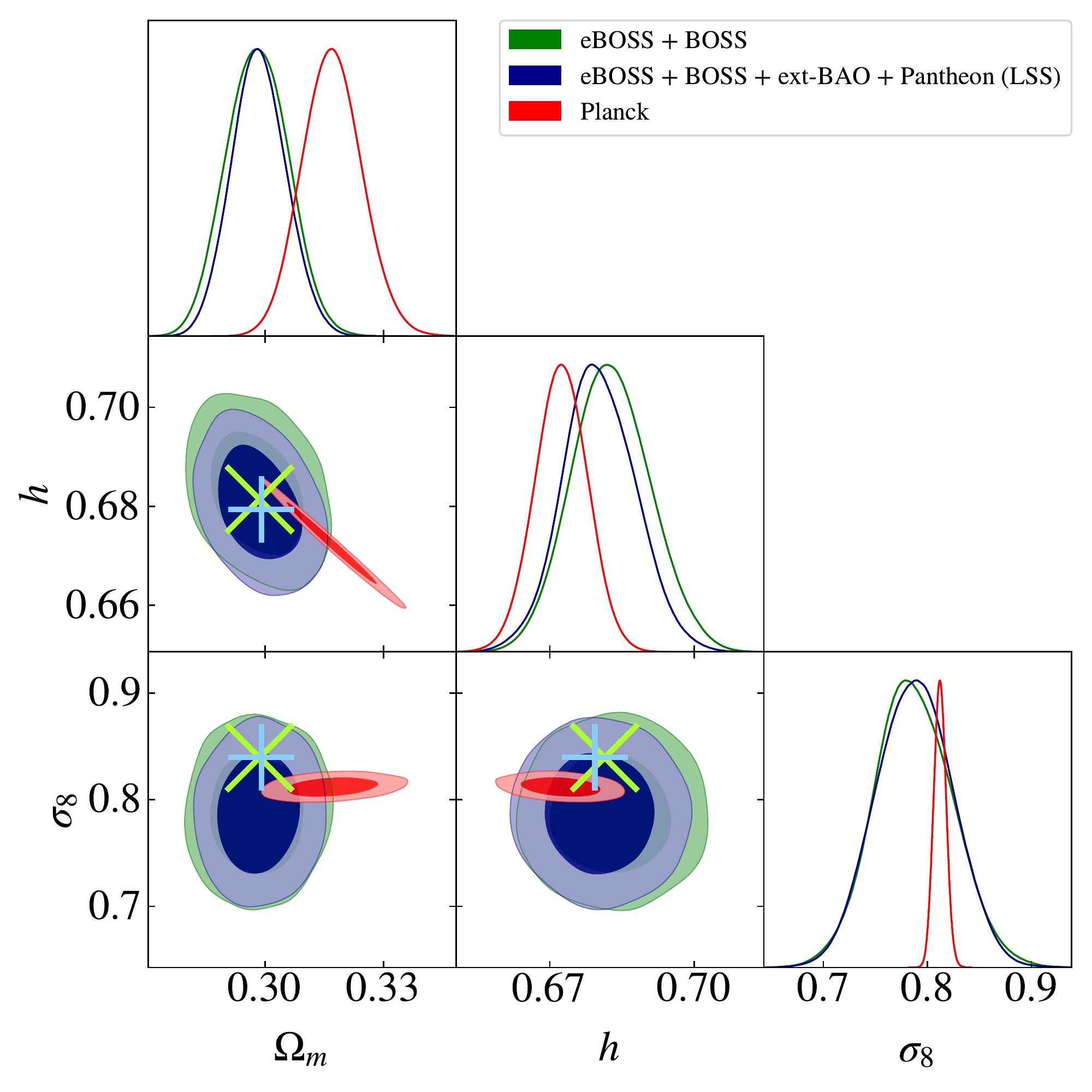}
    \includegraphics[width=0.49\columnwidth]{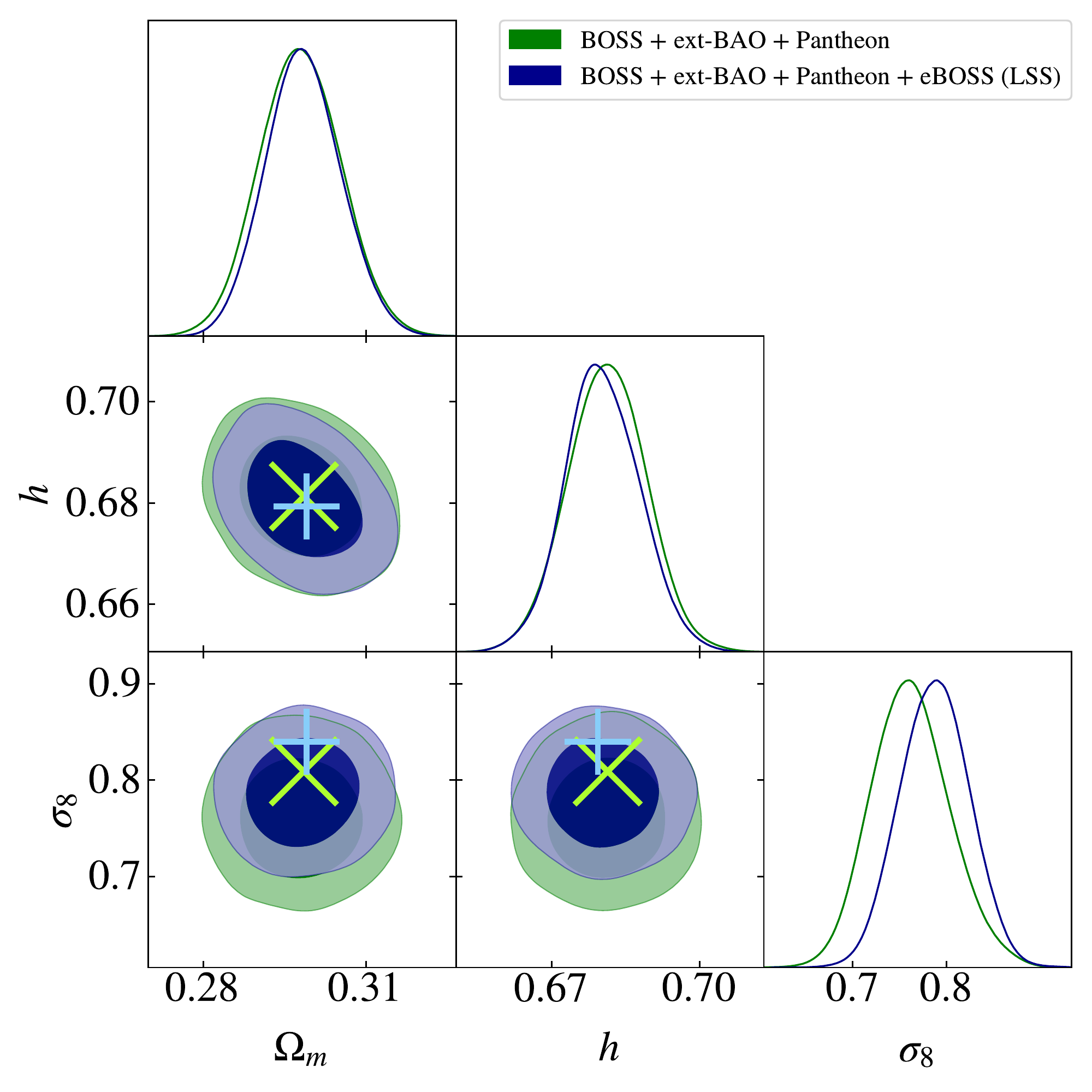}
    \includegraphics[width=0.49\columnwidth]{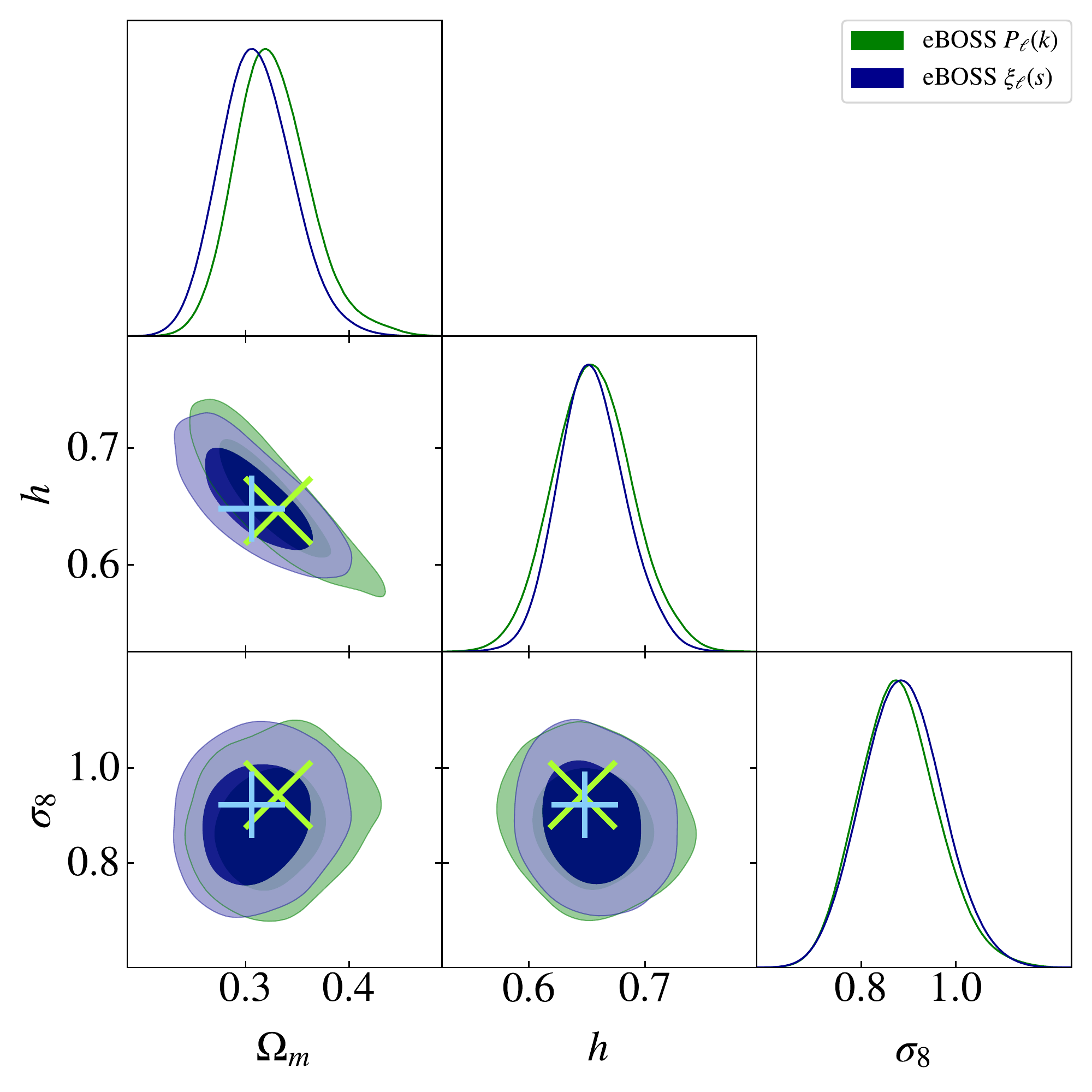}
    
        \caption{\footnotesize Triangle plots (1D and 2D posterior distributions) of the cosmological parameters reconstructed from the base-$\Lambda$CDM analyses performed in this work of the following datasets:  
        {\it Upper left} - eBOSS compared to BOSS or {\it Planck}.
        {\it Upper right} - eBOSS + BOSS and eBOSS + BOSS + ext-BAO + Pantheon, (\emph{i.e.}, the LSS dataset). {\it Planck} is shown for comparison.
        {\it Lower left} - eBOSS + BOSS + ext-BAO + Pantheon with and without eBOSS, to gauge the impact of the addition of eBOSS.
        {\it Lower right} - eBOSS power spectrum $P_\ell(k)$ or correlation function $\xi_\ell(s)$.
        The crosses represent the best-fit values. }
    \label{fig:main_results}
\end{figure*}

\paragraph{$68\%$-credible interval} 
The 1D and 2D posterior distributions from eBOSS, analysed alone or in combination with other LSS probes, are shown in figure~\ref{fig:main_results}, with the corresponding $68\%$-credible intervals and best-fit values given in table~\ref{tab:parameter_BOSS_eBOSS}. 
We also display posteriors obtained with {\it Planck} data for comparison.
As it can be read off from table~\ref{tab:parameter_BOSS_eBOSS}, from eBOSS alone, we reconstruct at $68\%$ CL within the base-$\Lambda$CDM model, $\Omega_m$, $h$, and $\sigma_8$ to $11\%$, $5\%$, and $9\%$ precision, respectively. 
The eBOSS full-shape analysis in configuration space leads to comparable error bars at $\lesssim 20\%$ (and consistency on $\Omega_m$, $h$, and $\sigma_8$ at $<0.4\sigma$, as already commented in section~\ref{sec:scalecut}).
The corresponding posteriors are shown in the lower right panel of figure~\ref{fig:main_results}. 

\subsection{Comparison with several LSS probes}

\paragraph{Consistency across LSS probes}
The EFT analysis of eBOSS QSOs provides independent measurements of $\Lambda$CDM parameters in a different redshift range than previous EFT analyses (recalling that $z_{\rm eff}\simeq 1.5$), and from yet another tracer. It is therefore interesting, as a consistency check of $\Lambda$CDM (and the assumptions behind the EFTofLSS), to compare the results with other cosmological probes. 
In table~\ref{tab:sigma_deviation}, we present the $\sigma$-deviations on the cosmological parameters reconstructed from BOSS, eBOSS, and {\it Planck}, analyzed under $\Lambda$CDM.~\footnote{While for BOSS and eBOSS, we remind that we analyze them fixing $\omega_b$ and $n_s$, for {\it Planck}, we let them vary and additionally vary $\tau_{\rm reio}$. 
See appendix~\ref{app:extensions} for results with $n_s$ free and a BBN prior on $\omega_b$. }
As already mentioned in section~\ref{sec:scalecut} and discussed throughout in Ref.~\cite{Simon:2022lde}, the EFT analysis, given the prior chosen in section~\ref{sec:inference}, can lead to potentially important prior volume projection effects (which do not affect the best-fit values). 
Therefore, we present in table~\ref{tab:sigma_deviation} the $\sigma$-deviation on the cosmological parameters between two experiments comparing both their means and their best-fits (shown in table~\ref{tab:parameter_BOSS_eBOSS} and in the left upper panel of figure~\ref{fig:main_results}). 
Given those two metrics, we find that all cosmological parameters are consistent at $\lesssim 1.6\sigma$ between eBOSS and BOSS. Note that the value of $H_0$ is $\sim 1\sigma$ lower for eBOSS than for BOSS, while $\sigma_8$, as well as $\text{ln}(10^{10} A_s)$ and $S_8$, are $\sim 1.5\sigma$ higher.

\paragraph{Combining LSS probes} 
We present constraints from combining eBOSS + BOSS data in figure~\ref{fig:main_results}. 
Posteriors are also given in table~\ref{tab:parameter_BOSS_eBOSS}. Combining eBOSS with BOSS, we reconstruct $\Omega_m$, $h$, and $\sigma_8$ to $3\%$, $1\%$, and $5\%$ precision at $68\%$ CL.
This represents an improvement of about $10\%$ over BOSS alone.
If the improvement in the constraints within $\Lambda$CDM is somewhat marginal, in the next section we show that the addition of eBOSS can play a significant role in extended models, in particular in constraining the total neutrino mass. For better comparison with the official eBOSS analysis~\cite{eBOSS:2020yzd}, we also present in figure~\ref{fig:main_results} results obtained when combining with a compilation of independent BAO data dubbed ``ext-BAO'' and the Pantheon SN1a sample.
These additional data, on top of eBOSS + BOSS, further improves the constraints on $\Omega_m$ and $h$ by about $10\%$ (see upper right panel of figure~\ref{fig:main_results}). 

\paragraph{Comparison with other works} Our joint constraints on flat $\Lambda$CDM from  eBOSS + BOSS using the EFTofLSS can be compared with constraints from previous full-shape analysis, \emph{e.g.}, from Refs. \cite{Semenaite:2021aen,Neveux:2022tuk,Brieden:2022lsd}, 
with the caveat that there are some differences in the modeling, the data combination considered, the scale cuts, and the priors on the cosmological parameters $\omega_b$ and $n_s$. 
Concretely, we find that when considering similar data we are consistent at $\lesssim 0.4\sigma$ on all cosmological parameters with Ref.~\cite{Semenaite:2021aen} when freeing $\omega_b$ and $n_s$ (see appendix~\ref{app:extensions}).~\footnote{In Ref.~\cite{Semenaite:2021aen}, our constraints can be compared with the results for ``BOSS + eBOSS'' with ``Wide priors'' from their table 2, as they consider the same redshift bins of BOSS LRG LOWZ + CMASS and eBOSS QSO that we analyze. However, their analysis is carried in configuration-space wedges restricted to scales $20 < s / [{\rm Mpc}/h] < 160$ instead. 
Their model uses a different parameterization for the galaxy biases, no counterterms, and a different treatment for the BAO smearing than our IR-resummation. Albeit small differences in the treatment of the BAO smearing, redshift-space distortions, and nuisance parametrization, we believe that the modeling in~\cite{Semenaite:2021aen} is effectively not so different than eq.~\eqref{eq:Xi_02}, for the following reason. Although they do not include counterterms, in configuration space, the stochastic terms are absent as they are just Dirac-$\delta$ distributions at vanishing seperation, as we explained in section~\ref{sec:model}. Plus, the remaining counterterms are very steep and thus potentially negligible at the scales analyzed. This is not the case in Fourier space: there, the counterterms are substantial contributions at the scales analyzed. } 
With the results of Ref.~\cite{Brieden:2022lsd}, we find consistency at $0.3\sigma, \, 0.1\sigma, \, 1.9\sigma$ on $\Omega_m, \, h, \, \sigma_8$, with fixed $\omega_b$ and $n_s$.~\footnote{Our results can be compared with the ``BBN + ShapeFit'' analysis from table 8 in Ref.~\cite{Brieden:2022lsd}. 
They consider the two redshift bins BOSS LRG $0.2<z<0.6$ and $0.4<z<0.6$, eBOSS LRG $0.6<z<1$, but the same eBOSS QSO redshift bins than us, with scale cuts $0.02 < k / [h/{\rm Mpc}] < 0.15$ for the LRG bins and $0.02 < k / [h/{\rm Mpc}] < 0.30$ for the QSO bins. 
They use a different model to describe the perturbation theory contributions and no counterterms, as well as a different approach that extends the template-based BAO/$f\sigma_8$ analysis with a new ``shape parameter'',  from which cosmological constraints can then be derived. }
Finally, we find consistency at $1.1\sigma, \, 1.7\sigma, \, 2\sigma$ on $\Omega_m, \, h$, $\sigma_8$ with the results of Ref.~\cite{Neveux:2022tuk}, with fixed $\omega_b$ and $n_s$.~\footnote{Our results can be compared with the case ``3 surveys'' in Ref.~\cite{Neveux:2022tuk} with ``$\omega_b$ \& $n_s$'' prior from their table 3.
They consider as redshift bins BOSS LRG low-$z$ ($0.2<z<0.5$), eBOSS LRG ($0.6 < z < 1$) instead of CMASS, but the same eBOSS QSO sample, with scale cuts respectively $0.02 < k / [h/{\rm Mpc}] < 0.15 $, $0.02 < k / [h/{\rm Mpc}] < 0.15$, and $0.02 < k / [h/{\rm Mpc}] < 0.30$. 
They use a different model on the perturbation theory contributions and no counterterms. }
This tells us that, despite the various choices of those analyses, our results are in broad agreement with these results. 
It would be interesting to understand better how the small differences we find arise. We leave this to future work.

\begin{figure*}[ht]
\begin{minipage}{0.3\linewidth}
\centering
\scriptsize
\begin{tabular}{|l|c|}
\hline
best-fit & LSS + Planck \\
$\mu_{-\sigma_l}^{+\sigma_u}$ &  \\
\hline
\multirow{2}{*}{$\Omega_{\rm m}$} & $0.3081$ \\
&  $0.3081^{+0.0052}_{-0.0052}$ \\
\hline

\multirow{2}{*}{$h$} & $0.6787$ \\
& $0.6789^{+0.0039}_{-0.0040}$ \\
\hline

\multirow{2}{*}{$\sigma_8$} & $0.8096$\\
& $0.8095^{+0.0057}_{-0.0057}$\\
\hline

\multirow{2}{*}{$\omega_{\text{cdm}}$} & $0.1188$\\
& $0.1189^{+0.0009}_{-0.0009}$ \\
\hline

\multirow{2}{*}{$\text{ln}(10^{10} A_s)$} & $3.052$ \\
& $3.051^{+0.013}_{-0.015}$ \\
\hline

\multirow{2}{*}{$S_8$} & $0.820$\\
& $0.820^{+0.010}_{-0.010}$ \\
\hline

\multirow{2}{*}{$n_s$} & $0.9665$ \\
& $0.9674^{+0.0036}_{-0.0038}$ \\
\hline

\multirow{2}{*}{$\tau_{\rm reio}$} & $0.0592$ \\
& $0.0585^{+0.0064}_{-0.0075}$ \\
\hline

$\chi^2_{\rm min}$ &  4026.4\\
\hline
\end{tabular} 
\end{minipage}\hfill
\begin{minipage}{0.7\linewidth}
\includegraphics[width=1.\textwidth]{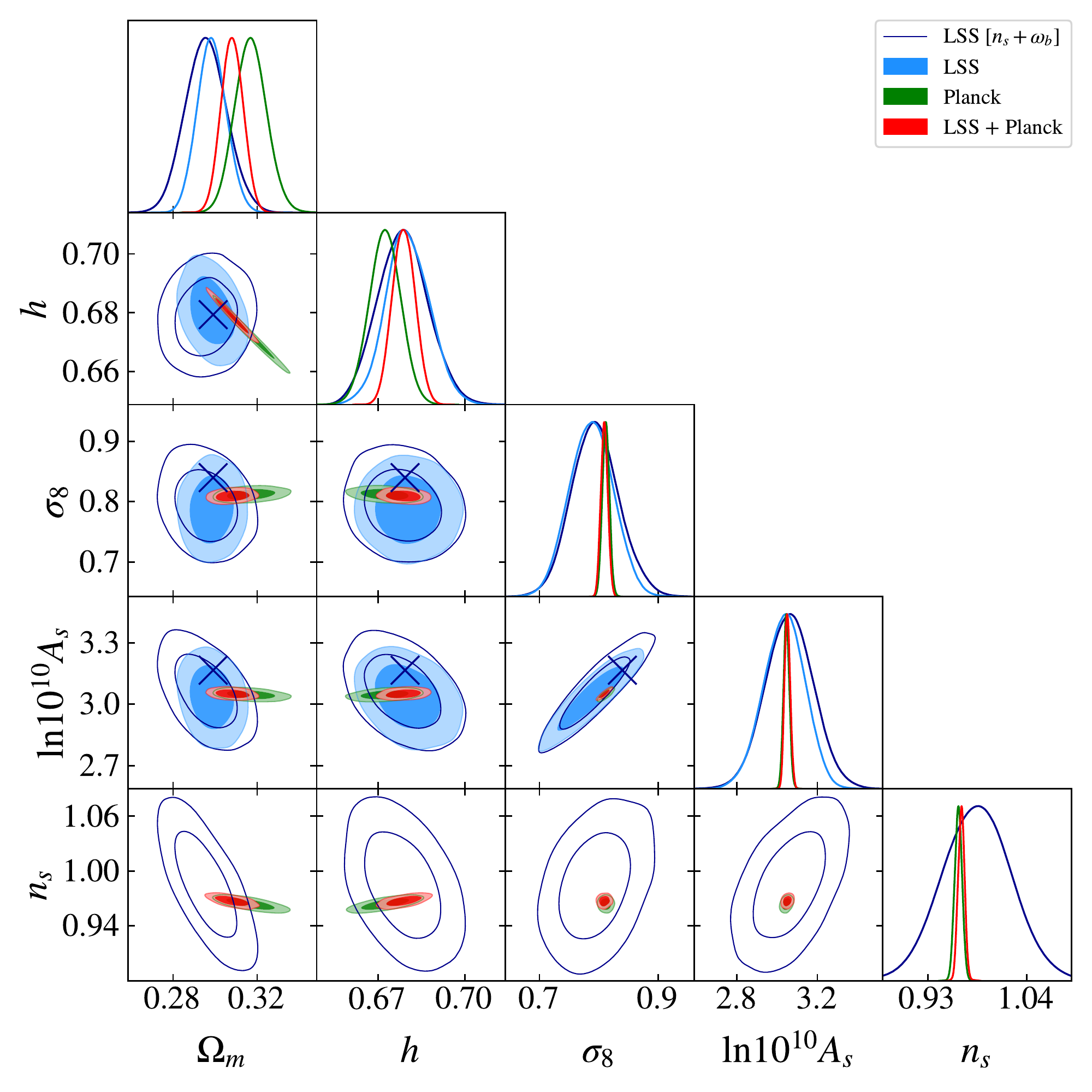}
\end{minipage}
\caption{\footnotesize 
{\it Left} - Cosmological results (best-fit, posterior mean, and $68\%$ CL) of eBOSS + BOSS + ext-BAO + Pantheon in combination with {\it Planck}. 
We also report the best-fit $\chi^2_{\rm min}$ of this analysis. 
{\it Right} -  Triangles plots (1D and 2D posterior distributions) of the cosmological parameters reconstructed from eBOSS + BOSS + ext-BAO + Pantheon (referred as LSS), {\it Planck}, or their combination, within the base-$\Lambda$CDM model. 
When {\it Planck} is analyzed, $n_s$ and $\omega_b$ are set free. 
The empty contours correspond to the LSS analysis of appendix~\ref{app:extensions} where we let $n_s$ free and  $\omega_b$ vary within a BBN prior. 
The blue crosses represent the LSS best-fit values for our base-$\Lambda$CDM model. 
}
\label{fig:LSS_Planck}
\end{figure*}

\subsection{Comparison with Planck}

\paragraph{Consistency with Planck} The combination of eBOSS + BOSS allows to determine $\Omega_m$ and $h$ at a precision similar to {\it Planck}, with error bars only larger by  $5\%$ and $50\%$ respectively. Once ext-BAO and Pantheon are included, we obtain lower error bar of $5\%$ and higher error bar of $40\%$ for $\Omega_m$ and $h$ respectively. Indeed, it is known that there is a rather large geometric degeneracy between $\Omega_m$ and $h$ in the CMB, as displayed in figure~\ref{fig:main_results}, while LSS data benefit from measuring the BAO (and additional shape features) over volumes ({\it i.e.}, in two directions, parallel and perpendicular to the line-of-sight), and covering a large range of redshifts.  In contrast, the amplitude parameter $\sigma_8$ is constrained roughly $\sim 6$ times better by {\it Planck}.
Given the two metrics used in table~\ref{tab:sigma_deviation}, we find that all cosmological parameters are consistent at $\lesssim 1.8\sigma$ between {\it Planck} and the various combinations of LSS data.
More specifically, as it can be seen from table~\ref{tab:sigma_deviation} and in the upper right panel of figure~\ref{fig:main_results}, we find that the combination of eBOSS + BOSS leads to cosmological parameters consistent at $\lesssim 1.0\sigma$ with the ones inferred from {\it Planck}, except for $\Omega_m$ where we find a $\sim 1.6$ $\sigma$-deviation. 
These conclusions are unchanged with the inclusion of ext-BAO and Pantheon.  
The consistency between LSS and CMB experiments represent a nontrivial check of the $\Lambda$CDM model, given that those are different experiments in the target objects they probe, their technical design (and associated systematic errors), and the redshift range they cover.
Further considerations on consistency of beyond-$\Lambda$CDM models are discussed in section~\ref{sec:ext_LCDM}.

\paragraph{$S_8$ and $H_0$ tension} From the analyses presented here, we find no tension on $H_0$ or $S_8$ between LSS and {\it Planck}.  Nevertheless, let us remark that, since the error bars on the cosmological parameters inferred from the LSS data are larger than the ones obtained by {\it Planck}, our results are in reasonable agreement with the lensing measurements of $S_8$. While {\it Planck} is known to be in $3\sigma$ tension with KIDS~\cite{KiDS:2020suj}, we find here agreement at about $\sim 1.8\sigma$, considering either the best-fit value of eBOSS + BOSS or LSS. 
As for $H_0$, we find that the inferred value is, depending on the addition of ext-BAO + Pantheon to the EFT analyses of BOSS + eBOSS, in $\sim 3.6-4.0\sigma$ tension with the one obtained by SH0ES~\cite{Riess:2021jrx}. This level of tension is comparable to that recently determined in Ref.~\cite{Schoneberg:2022ggi} with updated BBN predictions and the recently-developed ShapeFit analysis \cite{Brieden:2022lsd}. Yet, we stress that this CMB-independent determination of $H_0$ does not marginalize over the sound horizon information, as advertised in Refs.~\cite{Philcox:2020xbv,Farren:2021grl,Philcox:2022sgj,Smith:2022iax}, and therefore should not be interpreted as a constraint against models affecting the pre-recombination era in order to resolve the Hubble tension.

\paragraph{Combination with Planck}
In figure~\ref{fig:LSS_Planck}, we show the results from the combination of eBOSS + BOSS + ext-BAO + Pantheon with {\it Planck}. 
Compared to {\it Planck} alone, the constraints on $\Omega_m$ and $h$ are improved by $\sim 30\%$, as we can clearly see that LSS data break the partial degeneracy present in the CMB in the $\Omega_m - h$ plane. 
This is further accompanied by an improvement in the constraints on $\omega_{cdm}$, $S_8$, and $n_s$  of about $28\%$, $24\%$, and $13\%$, respectively. On the other hand, the amplitude parameter $\sigma_8$ and $A_s$ are not significantly impacted, as they are very tightly constrained by {\it Planck} alone. 
Note that the best-fit values are not far from the mean with respect to the error bars for all cosmological parameters, as expected from the relatively large data volume of {\it Planck} which lead to posteriors that are much more Gaussian (see Refs.~\cite{Simon:2022lde,Planck:2013nga} for related discussions).

\section{Extensions to the flat $\Lambda$CDM model}
\label{sec:ext_LCDM}

\begin{table*}[p]
\center
\tiny 
\scalebox{1.1}{\begin{tabular}{|l|c|c|c|c|}
\hline
\multicolumn{5}{|c|}{BOSS + ext-BAO + Pantheon} \\
\hline
best-fit  & \multirow{2}{*}{$\Omega_k\Lambda$CDM} & \multirow{2}{*}{$w_{0}$CDM} & \multirow{2}{*}{$\nu \Lambda$CDM} & \multirow{2}{*}{$N_{\rm eff} \Lambda$CDM} \\
$\mu_{-\sigma_l}^{+\sigma_u}$& & & & \\
\hline
\multirow{2}{*}{$\Omega_{\rm m}$} & $0.2962$ & $0.2987$ & $0.2982$ & $0.298$  \\
& $0.2955_{-0.0074}^{+0.0079}$ & $0.2967_{-0.0085}^{+0.0080}$ &$0.3049_{-0.0110}^{+0.0094}$ & $0.291_{-0.010}^{+0.010}$  \\
\hline

\multirow{2}{*}{$h$} & $0.6853$ & $0.681$ & $0.6811$ & $0.685$ \\
& $0.6886_{-0.0110}^{+0.0099}$ & $0.684_{-0.011}^{+0.010}$ & $0.6826_{-0.0082}^{+0.0080}$ &  $0.714_{-0.044}^{+0.026}$ \\
\hline

\multirow{2}{*}{$\sigma_8$} & $0.801$ & $0.810$ & $0.810$ & $0.809$ \\
& $0.749_{-0.043}^{+0.044}$ & $0.762_{-0.044}^{+0.042}$ & $0.763_{-0.044}^{+0.039}$ & $0.758_{-0.046}^{+0.041}$ \\
\hline

\multirow{2}{*}{$\omega_{\text{cdm}}$} & $0.1161$ & $0.1155$ & $0.1154$ & $0.1168$ \\
& $0.1172_{-0.0043}^{+0.0046}$ & $0.1158_{-0.0047}^{+0.0046}$ & $0.1179_{-0.0055}^{+0.0047}$ & $0.1256_{-0.0150}^{+0.0078}$ \\
\hline

\multirow{2}{*}{$\text{ln}(10^{10} A_s)$} & $3.01$ & $3.09$ & $ 3.09$ & $3.08$ \\
& $2.84_{-0.17}^{+0.16}$ & $2.95_{-0.13}^{+0.14}$ & $3.00_{-0.13}^{+0.12}$ &  $2.95_{-0.12}^{+0.12}$\\
\hline

\multirow{2}{*}{$S_8$} & $0.796$ & $0.808$ & $0.808$ & $0.806$ \\
& $0.744^{+0.045}_{-0.044}$ & $0.758^{+0.044}_{-0.044}$ & $0.769^{+0.040}_{-0.045}$ & $0.748^{+0.044}_{-0.048}$ \\
\hline

\multirow{2}{*}{$\Omega_k$} &  $-0.023$ & \multirow{2}{*}{--} & \multirow{2}{*}{--} & \multirow{2}{*}{--} \\
& $-0.032^{+0.028}_{-0.031}$ &&&  \\
\hline

\multirow{2}{*}{$w_{0}$} & \multirow{2}{*}{--} & $-0.998$ & \multirow{2}{*}{--} & \multirow{2}{*}{--} \\
&& $-1.015_{-0.042}^{+0.042}$ &&  \\
\hline

\multirow{2}{*}{$\sum m_{\nu}$ [$e$V]} & \multirow{2}{*}{--} & \multirow{2}{*}{--} & $0.052$  & \multirow{2}{*}{--} \\
&&& $<0.429$ &  \\
\hline

\multirow{2}{*}{$\Delta N_{\rm eff}$} & \multirow{2}{*}{--} & \multirow{2}{*}{--} & \multirow{2}{*}{--} & $0.09$ \\
&&&& $0.82^{+0.62}_{-1.08}$ \\
\hline
\hline
$\chi^2_{\rm min}$ & 1190.8 & 1191.1 & 1191.1 & 1191.1 \\ 
\hline
$\Delta\chi^2_{\rm min}$ & -0.3 & 0 & 0 & 0 \\
\hline
\end{tabular}}

\medskip

\scalebox{1.1}{\begin{tabular}{|l|c|c|c|c|}
\hline
\multicolumn{5}{|c|}{eBOSS + BOSS + ext-BAO + Pantheon (LSS)} \\
\hline
best-fit  & \multirow{2}{*}{$\Omega_k\Lambda$CDM} & \multirow{2}{*}{$w_{0}$CDM} & \multirow{2}{*}{$\nu \Lambda$CDM} & \multirow{2}{*}{$N_{\rm eff} \Lambda$CDM} \\
$\mu_{-\sigma_l}^{+\sigma_u}$ & & & & \\
\hline

\multirow{2}{*}{$\Omega_{\rm m}$} & $0.2952$ & $0.2987$ & $0.2962$ & $0.3029$  \\
& $0.2945_{-0.0081}^{+0.0072}$ & $0.3042_{-0.0096}^{+0.0084}$ & $0.3017_{-0.0097}^{+0.0076}$ & $0.2950_{-0.0093}^{+0.0099}$  \\
\hline

\multirow{2}{*}{$h$} & $0.6858$ & $0.680$ & $0.6789$ & $0.664$ \\
& $0.6882_{-0.0094}^{+0.0098}$ & $0.683_{-0.011}^{+0.011}$ & $0.6810_{-0.0072}^{+0.0078}$ & $0.696_{-0.039}^{+0.017}$\\
\hline

\multirow{2}{*}{$\sigma_8$} & $0.824$ & $0.835$ & $0.838$ & $0.843$  \\
& $0.775_{-0.050}^{+0.032}$ & $0.744_{-0.041}^{+0.040}$ & $0.787_{-0.040}^{+0.035}$ & $0.787_{-0.043}^{+0.033}$ \\
\hline

\multirow{2}{*}{$\omega_{\text{cdm}}$} & $0.1158$ & $0.1150$ & $0.1142$ & $0.1107$ \\
& $0.1165_{-0.0031}^{+0.0041}$ & $0.1189_{-0.0038}^{+0.0043}$ & $0.1164_{-0.0044}^{+0.0040}$ & $0.1199_{-0.0120}^{+0.0061}$ \\
\hline

\multirow{2}{*}{$\text{ln}(10^{10} A_s)$} & $3.04$ & $3.15$ & $3.14$ & $3.18$ \\
& $2.90_{-0.15}^{+0.14}$ & $2.86_{-0.13}^{+0.13}$ & $3.04_{-0.11}^{+0.10}$  & $3.03_{-0.11}^{+0.11}$ \\
\hline

\multirow{2}{*}{$S_8$} & $0.817$ & $0.833$ & $0.833$ &  $0.847$ \\
& $0.768^{+0.040}_{-0.047}$ & $0.749^{+0.041}_{-0.043}$ & $0.789^{+0.036}_{-0.041}$ & $0.780^{+0.039}_{-0.039}$ \\
\hline

\multirow{2}{*}{$\Omega_k$} & $-0.034$ & \multirow{2}{*}{--} & \multirow{2}{*}{--} & \multirow{2}{*}{--} \\
& $-0.039^{+0.028}_{-0.029}$ &&&  \\
\hline

\multirow{2}{*}{$w_{0}$} & \multirow{2}{*}{--} & $-1.002$ & \multirow{2}{*}{--} & \multirow{2}{*}{--} \\
&& $-1.038_{-0.041}^{+0.041}$ &&  \\
\hline

\multirow{2}{*}{$\sum m_{\nu}$ [$e$V]} & \multirow{2}{*}{--} & \multirow{2}{*}{--} & $0.002$ & \multirow{2}{*}{--} \\
&&& $<0.274$ &  \\
\hline

\multirow{2}{*}{$\Delta N_{\rm eff}$} & \multirow{2}{*}{--} & \multirow{2}{*}{--} & \multirow{2}{*}{--} & $-0.37$ \\
&&&& $0.40^{+0.44}_{-0.91}$ \\
\hline
\hline
$\chi^2_{\rm min}$ & 1250.0 & 1250.9 & 1250.6 & 1250.8 \\ 
\hline
$\Delta\chi^2_{\rm min}$ & -1.0 & -0.1 & -0.4 & -0.2 \\
\hline
\end{tabular}}

\caption{\footnotesize {\it Upper} - Cosmological results (best-fit, posterior mean, and $68\%$ CL) from BOSS + ext-BAO + Pantheon for several model extensions to $\Lambda$CDM. Note that we quote the $95\%$ CL bound for $\sum m_{\nu}$. For each dataset we also report its best-fit $\chi^2$, and the $\Delta \chi^2$ with respect to the analogous $\Lambda$CDM best-fit model. {\it Lower} - Same, but this time with the addition of eBOSS data.}
\label{tab:parameter_extansion}
\end{table*}

\begin{table*}[ht]
\center
\tiny
\scalebox{1.1}{\begin{tabular}{|l|c|c|c|c|}
\hline
\multicolumn{5}{|c|}{LSS + Planck} \\
\hline
best-fit  & \multirow{2}{*}{$\Omega_k\Lambda$CDM} & \multirow{2}{*}{$w_{0}$CDM} & \multirow{2}{*}{$\nu \Lambda$CDM} & \multirow{2}{*}{$N_{\rm eff}\Lambda$CDM} \\
$\mu_{-\sigma_l}^{+\sigma_u}$ & & & & \\
\hline

\multirow{2}{*}{$\Omega_{\rm m}$} & 0.3069 & 0.3015 &  0.3039 & 0.3080  \\
& $0.3065_{-0.0054}^{+0.0051}$ &  $0.3013_{-0.0073}^{+0.0071}$ & $0.3058_{-0.0059}^{+0.0055}$ & $0.3090_{-0.0054}^{+0.0063}$  \\
\hline

\multirow{2}{*}{$h$} & 0.6809 & 0.6877 & 0.6825 & 0.680 \\
& $0.6813_{-0.0055}^{+0.0059}$ & $0.6878_{-0.0081}^{+0.0076}$ & $0.6809_{-0.0044}^{+0.0045}$ & $0.675_{-0.011}^{+0.010}$\\
\hline

\multirow{2}{*}{$\sigma_8$}& 0.8100 & 0.821 & 0.8203 & 0.811  \\
& $0.8109_{-0.0069}^{+0.0068}$ & $0.821_{-0.010}^{+0.011}$ & $0.8144_{-0.0071}^{+0.0098}$ & $0.806_{-0.010}^{+0.009}$ \\
\hline

\multirow{2}{*}{$\omega_{\text{cdm}}$} & 0.1192 & 0.1195 & 0.1192 & 0.1191 \\
& $0.1192_{-0.0012}^{+0.0013}$ & $0.1194_{-0.0010}^{+0.0009}$ & $0.1189_{-0.0009}^{+0.0009}$ & $0.1177_{-0.0026}^{+0.0027}$ \\
\hline

\multirow{2}{*}{$\text{ln}(10^{10} A_s)$} & 3.050 & 3.045 & 3.043 & 3.052 \\
& $3.051_{-0.015}^{+0.014}$ & $3.046_{-0.015}^{+0.013}$ &  $3.048_{-0.015}^{+0.014}$ & $3.047_{-0.017}^{+0.014}$ \\
\hline

\multirow{2}{*}{$S_8$} & 0.819 & 0.823 & 0.826 & 0.821 \\
& $0.820^{+0.010}_{-0.010}$ & $0.822^{+0.010}_{-0.010}$
 & $0.822^{+0.010}_{-0.010}$ & $0.818^{+0.010}_{-0.010}$ \\
\hline

\multirow{2}{*}{$n_s$} & 0.9650 & 0.9646 & 0.9658 &  0.9665\\
& $0.9665_{-0.0043}^{+0.0042}$ & $0.9659_{-0.0039}^{+0.0038}$ & $0.9673_{-0.0037}^{+0.0037}$ & $0.9650_{-0.0070}^{+0.0062}$ \\
\hline

\multirow{2}{*}{$\tau_{\rm reio}$} & 0.0581 & 0.0555 & 0.0548 & 0.0593 \\
& $0.0582_{-0.0081}^{+0.0068}$ & $0.0555_{-0.0074}^{+0.0068}$ & $0.05716_{-0.0076}^{+0.0070}$ & $0.0581_{-0.0074}^{+0.0073}$ \\
\hline

\multirow{2}{*}{$\Omega_k$} & $0.0007$ & \multirow{2}{*}{--} & \multirow{2}{*}{--} & \multirow{2}{*}{--} \\
& $0.0008^{+0.0018}_{-0.0017}$ &&&  \\
\hline

\multirow{2}{*}{$w_{0}$} & \multirow{2}{*}{--} & $-1.040$ & \multirow{2}{*}{--} & \multirow{2}{*}{--} \\
&& $-1.039^{+0.029}_{-0.029}$ &&  \\
\hline

\multirow{2}{*}{$\sum m_{\nu}$ [$e$V]} & \multirow{2}{*}{--} & \multirow{2}{*}{--} & $9 \times 10^{-5}$ & \multirow{2}{*}{--} \\
&&& $<0.093$ &  \\
\hline

\multirow{2}{*}{$\Delta N_{\rm eff}$} & \multirow{2}{*}{--} & \multirow{2}{*}{--} & \multirow{2}{*}{--} & $0.02$ \\
&&&& $-0.07^{+0.15}_{-0.16}$  \\
\hline
\hline
$\chi^2_{\rm min}$ & 4025.9 & 4025.0 &  4023.3 & 4026.1 \\ 
\hline
$\Delta\chi^2_{\rm min}$ & -0.5 & -1.4 & -3.1 & -0.3 \\
\hline
\end{tabular}}

\caption{\footnotesize Cosmological results (best-fit, posterior mean, and $68\%$ CL) from LSS + {\it Planck} for several model extensions to $\Lambda$CDM. Note that we quote $95\%$ CL bound for $\sum m_{\nu}$. For each dataset we also report its best-fit $\chi^2$, and the $\Delta \chi^2$ with respect to the analogous $\Lambda$CDM best-fit model.}
\label{tab:parameter_extansion_Planck}
\end{table*}

\begin{figure*}[ht]
    \centering
    \includegraphics[trim=7cm 0cm 7cm 0cm, width=1\columnwidth]{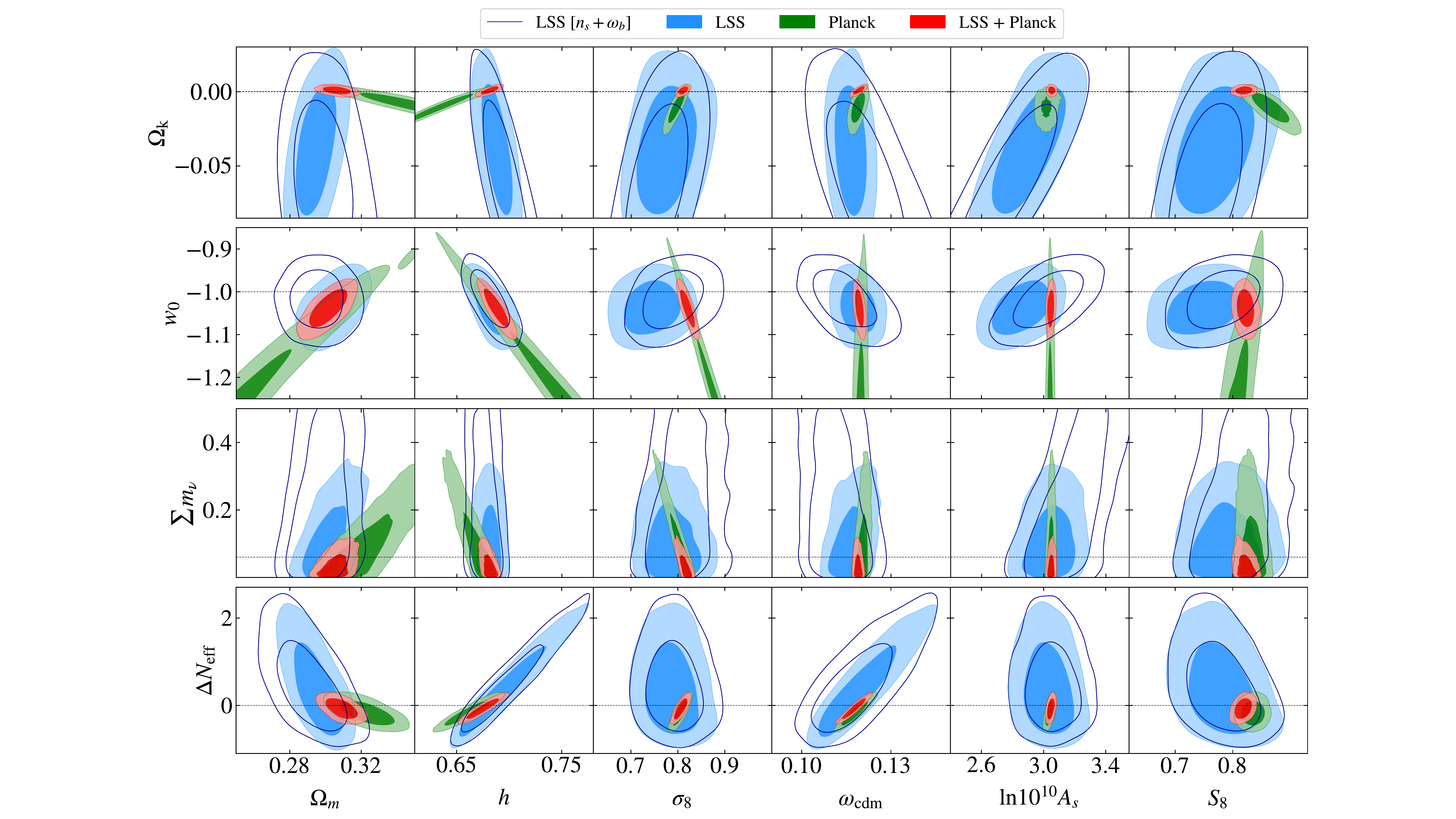}
        \caption{\footnotesize 2D posterior distributions of cosmological parameters in several model extensions to $\Lambda$CDM reconstructed from analyzing LSS data (\emph{i.e.}, eBOSS + BOSS + ext-BAO + Pantheon), with and without freeing $n_s$ and $\omega_b$, compared with {\it Planck} and their combination. }
    \label{fig:extension}
\end{figure*}

In this section, we present the results of analyses of several extensions to the flat $\Lambda$CDM model, namely the curvature density fraction $\Omega_k$, the equation of state of dark energy $w_{0}$, the neutrino mass $\sum m_{\nu}$, and the number of relativistic degrees of freedom $N_{\rm eff}$. For each model, we list in table~\ref{tab:parameter_extansion} the results of LSS analyses with and without eBOSS to highlight the role of the EFT likelihood of eBOSS QSO data in constraining these extensions.
In appendix~\ref{app:extensions}, we provide the results of analyses including the variation of $n_s$ (within a uninformative flat prior) and $\omega_b$ (within a BBN prior), and show the differences with the baseline analyses that fix these parameters. 
We also combine the LSS datasets with {\it Planck} and provide reconstructed parameters in table~\ref{tab:parameter_extansion_Planck}.
For comparison (and although not explicitly listed), we perform the same series of analyses using the BAO/$f\sigma_8$ information for the eBOSS and BOSS data.
Finally, in figure~\ref{fig:extension} we plot, for each model, the 2D posterior distributions obtained from analyzing the combination of the LSS datasets (with and without the $n_s$ and $\omega_b$ variations) compared to {\it Planck}, and their combination.

\subsection{$\Omega_k\Lambda {\rm CDM}$}

In this section, we consider the $\Lambda$CDM model with the addition of the curvature density fraction, $\Omega_k$, by imposing a large flat prior on this parameter. We derive $\Omega_k = -0.032^{+0.028}_{-0.031}$ at $68\%$ CL (with a best-fit value at $-0.023$) from the analysis without eBOSS, while we derive $\Omega_k = -0.039^{+0.028}_{-0.029}$ at $68\%$ CL (with a best-fit value at $-0.034$) for the analysis with eBOSS.
These analyses allow us to highlight several important points for the LSS analysis:
\begin{itemize}
    \item With the LSS data only, we find $\Omega_k$ compatible with zero curvature at $1.2\sigma$ (considering the best-fit). When we vary $n_s$ and $\omega_b$ (see appendix~\ref{app:extensions}), this compatibility is increased to $0.6\sigma$.
    \item The addition of eBOSS data does not significantly reduce the $68\%$ constraints on $\Omega_k$. However it improves the constraint at $95\%$ CL from $-0.032^{+0.062}_{-0.057}$ to $-0.039^{+0.054}_{-0.052}$, which corresponds to an improvement of $\sim 10\%$.
    \item The EFT analysis significantly improves the constraints on $\Omega_k$ (by $\sim 50\%$) compared to the conventional BAO/$f\sigma_8$ analysis ($-0.037^{+0.067}_{-0.053}$ at 68\% C.L., with a best-fit value at $-0.006$).
    \item
    Note that, as visible on figure~\ref{fig:extension}, we find no tension between LSS and {\it Planck} when curvature is allowed to vary as long as CMB lensing is included in the fit to {\it Planck}: at $95\%$ CL, {\it Planck} finds $\Omega_k=-0.011^{+ 0.013}_{-0.012}$ while our combination of LSS data leads to $\Omega_k = -0.039 \pm 0.053$.
\end{itemize}

When combining LSS and {\it Planck}, we reconstruct $\Omega_k = 0.0008^{+0.0018}_{-0.0017}$ at $68\%$ CL and $\Omega_k = 0.0008 \pm 0.0034$ at $95\%$ CL (with a best-fit value at $-0.0007$), which allows us to highlight that:
\begin{itemize}
    \item The combination of LSS and {\it Planck} leads to a strong constraint on the $\Omega_k$ parameter, thanks to the redshift leverage between LSS probes and the last-scattering surface, allowing the breaking of  degeneracies in the $\Omega_k$-$\Omega_m$ and $\Omega_k$-$H_0$ planes.
    \item The combination with {\it Planck} data excludes the (slightly favored) negative values of $\Omega_k$ reconstructed from LSS data alone, and we find that cosmological data are in very good agreement with $\Omega_k=0$.
    \item Our constraints are better at $10\%$ than the {\it Planck} + BAO constraints of Ref.~\cite{Planck:2018vyg} ($0.0007 \pm 0.0037$ at $95\%$ CL), and are similar within $5\%$ to the constraints obtained from the ShapeFit method ($0.0015 \pm 0.0016$ at $68\%$ CL) \cite{Brieden:2022lsd}\footnote{We note that this comparison should be taken with caution, as we do not use the same exact dataset as in Ref.~\cite{Brieden:2022lsd}. In particular, authors therein consider the eBOSS LRG sample, which is not the case in our analysis, while we consider the Pantheon data, which is not the case in their analysis.} or from the equivalent combined analysis including instead standard BAO/$f\sigma_8$ method ($0.0013 \pm 0.0017$ at $68\%$ CL) derived in our work.
    See also Ref.~\cite{Bel:2022iuf} for a work presenting CMB-independent constraints on spatial curvature from beyond the standard BAO/$f\sigma_8$ that also support a flat Universe.
\end{itemize}

\subsection{$w_{0}{\rm CDM}$}

\begin{figure*}
    \centering
    \includegraphics[trim=0.2cm 0cm 0cm 0.2cm, width=1\columnwidth]{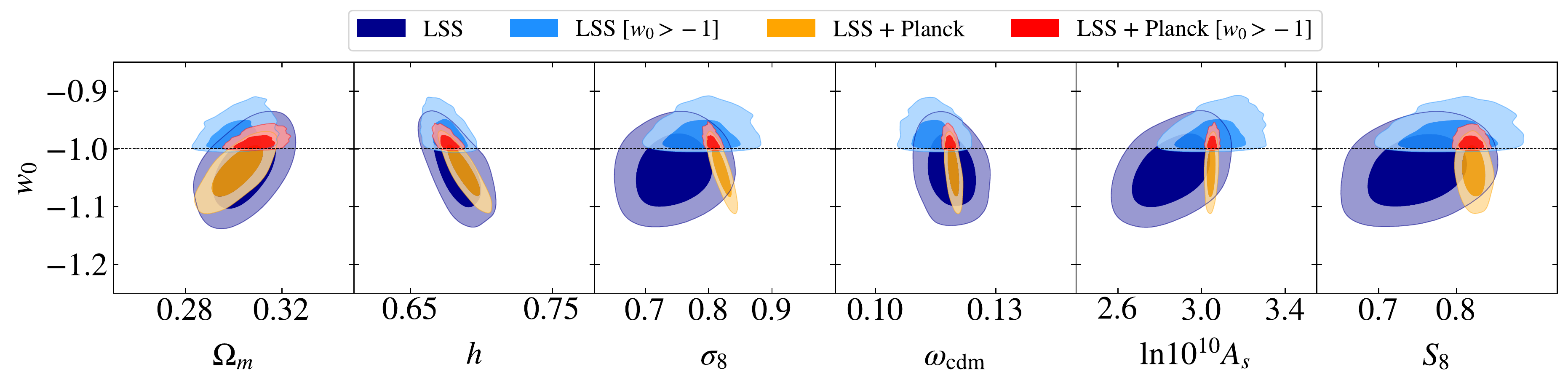}
        \caption{\footnotesize 2D posterior distributions of LSS (\emph{i.e.}, eBOSS + BOSS + ext-BAO + Pantheon), and the combination between LSS and {\it Planck} for the $w_0$CDM model. For both analyses, we show the results with the uninformative prior on $w_0$ and with the $w_0\geq-1$ prior. }
    \label{fig:w0_prior}
\end{figure*}

We now turn to the dark energy equation of state $w_0$ in the context of smooth quintessence, \emph{i.e.}, a quintessence field with no perturbations, and carry out two analyses: one in which we impose a large flat prior on $w_0$, and another in which we restrict $w_0 \geq -1$ to stay in the physical region. With the former prior, we obtain $w_0=-1.015 \pm 0.042$ at $68\%$ CL (with a best-fit value at $-0.998$) for the analysis without eBOSS, and $w_0=-1.038 \pm 0.041$ at $68\%$ CL (with a best-fit value at $-1.002$) for the analysis with eBOSS. We can conclude that:
\begin{itemize}
    \item With the LSS data, we find no evidence for a universe with $w_0 \ne -1$. This conclusion does not change when considering the variation of $n_s$ and $\omega_b$.
    \item The inclusion of eBOSS data does not improve the constraints at $68\%$ CL and at $95\%$ CL. This is expected as the effective redshift of the eBOSS data, namely $z=1.52$ (which is significantly higher than the effective redshift of the BOSS data, $z=0.32-0.57$), is well above the beginning of the dark energy dominated universe, $z_{\Lambda}\sim 0.3$. 
    \item  The EFT analysis improves the constraints on $w_0$ by $\sim 20\%$ compared to the BAO/$f\sigma_8$ analysis ($-0.961^{+0.054}_{-0.042}$ at 68\% CL,  with a best-fit value at $-0.931$). 
    This improvement can be understood as follow. 
    Roughly, for each mode $k$ in our analysis, the amplitudes of the monopole and the quadrupole are given by $b_1^2 D^2 A_s \mathcal{T}(k)$ and $4/3 \cdot b_1 f D^2 A_s \mathcal{T}(k)$, respectively, where $D$ is the growth factor and $\mathcal{T}(k)$ describes a $k$-dependence fixed by the transfer function of the linear power spectrum, roughly scaling as $\mathcal{T}(k) \propto (1+\log(k/k_{\rm eq}))^2 (k/k_*)^{n_s-4}$, where $k_*$ is the pivot scale. The transfer function captures modes that have re-entered the horizon during radiation domination, $k> k_{\rm eq}$, where $k_{\rm eq} \sim 0.01 \hinvMpc$ is the radiation-matter equality scale, and are log-enhanced (see \textit{e.g.}, Ref.~\cite{Smith:2022iax}). 
    The ratio of the monopole and quadrupole allows to obtain  $f \sigma_8 \propto f D^2 A_s$, but does not contain information on $\mathcal{T}(k)$. 
    The EFT analysis allows us to make use of the $k$-dependence of the full-shape that contains information beyond the (scale-independent) BAO/$f\sigma_8$ analysis, in particular about $k_{\rm eq} \propto \Omega_m h^2$, in turn providing an extra handle on $\Omega_m h^2$ with no degeneracy with $w_0$. This allows to break the  $w_0-\Omega_m$ degeneracy that remains with BAO/$f\sigma_8$, and therefore to improve constraints on $w_0$  \cite{DAmico:2020kxu,DAmico:2020tty}.
    \item The constraints we derived in this work are much better than {\it Planck} ($w_0=-1.57_{-0.40}^{+0.50}$ at $95\%$~CL for the high-$\ell$ TT, TE, EE + lowE + lensing analysis \cite{Planck:2018vyg}), which is expected as (the primary) {\it Planck} CMB data is largely unable to constrain the behaviour of the Universe at low redshift.
    
\end{itemize}

When combining the LSS data with {\it Planck}, we obtain $w_0=-1.039 \pm 0.029$ at $68\%$ CL and  $w_0=-1.039^{+0.055}_{-0.059}$ at $95\%$ CL (with a best-fit value at $-1.040$), which allows us to highlight that:
\begin{itemize}
    \item The $w_0$ best-fit value is $\sim 1.4\sigma$ below $-1$, with $\Delta \chi^2 = -1.4$ with respect to the $\Lambda$CDM model.
    \item The addition of LSS data select values of $w_0$ close to $-1$, located in the 2-$\sigma$ region reconstructed from {\it Planck} data.
    \item Our constraints are better at $43\%$ than the {\it Planck} + BAO constraints of Ref.~\cite{Planck:2018vyg}  ($-1.04 \pm 0.10$ at $95\%$ CL) and better than the constraints reported using ShapeFit ($-1.093_{-0.044}^{+0.048}$ at $68\%$ CL)~\cite{Brieden:2022lsd} at $37\%$. However, these constraints are similar at $<1\%$ to the equivalent combined analysis including instead the standard BAO/$f\sigma_8$ ($-1.043 \pm 0.030$ at $68\%$ CL), indicating the improvement mostly comes from our inclusion of Pantheon data, when combining with {\it Planck}.
\end{itemize}

In figure~\ref{fig:w0_prior}, we show results of the analyses with the $w_0 \geq -1$ prior, to restrict the parameter space in the physical region of smooth quintessence (see \emph{e.g.}, discussions in Ref.~\cite{DAmico:2020tty}). 
We obtain $w_0 < -0.932$ at $95\%$ CL for the LSS analysis, while the constraint improves to $w_0 < -0.965$ at $95\%$ CL for the LSS + {\it Planck} analysis. 
In addition, one can see that this new prior shifts the 2D posteriors inferred from the LSS data in a non-negligible way, while it remains globally stable for the LSS + {\it Planck} analysis. Note that, for these analyses, $\Delta \chi^2 = 0$ with respect to $\Lambda$CDM, since we obtain best-fit values of $w_0 = -1$.

\subsection{$\nu \Lambda {\rm CDM}$}

We now turn to variation of the total neutrino mass $\sum m_{\nu}$. Following the convention used by {\it Planck}, we consider one massive neutrino and two massless ones. 
We derive $\sum m_{\nu}<0.429e$V at $95\%$ (with a best-fit value at $0.052e$V) from the analysis without eBOSS, while we derive $\sum m_{\nu}<0.274e$V (with a best-fit value at $0.002e$V) from the analysis that includes eBOSS. We can conclude that:
\begin{itemize}
    \item A non-zero value of the total neutrino mass is not favored.
    \item The eBOSS data improve the upper limit on the sum of neutrino masses by $\sim 40\%$ with respect to the analysis of LSS data without eBOSS. 
    This significant improvement is due to the fact that the neutrino energy density is higher at the epoch probed by eBOSS QSO than at the epoch probed by BOSS.
    \item The EFT analysis significantly improves the constraints on $\sum m_{\nu}$ (by a factor of $\sim 18$) over the BAO/$f\sigma_8$ constraint ($\sum m_{\nu}<4.84e$V at 95\% CL,  with a best-fit value at $0.74e$V). This is expected as BAO/$f\sigma_8$ gains constraining mostly from the geometrical impact of massive neutrinos (as the determination of $f\sigma_8$ assumes scale-independence), which is largely degenerate with the CDM density. The inclusion of the full-shape of the power spectrum breaks that degeneracy. 
    The reason is the following: 
    the EFT full-shape analysis allows to exploit the scale-dependence of the power spectrum that contains both information on the characteristic power suppression induced by neutrinos at the perturbation level, as well as on $k_{\rm eq}\propto \Omega_m h^2$ (through the log-enhancement at small-scales), therefore breaking the degeneracy with neutrino masses at the background level.
    \item The LSS constraint derived in this work is only $\sim 10\%$ weaker than the {\it Planck} constraint ($\sum m_{\nu}< 0.241e$V) obtained from high-$\ell$ TT, TE, EE + lowE + lensing \cite{Planck:2018vyg}.

\end{itemize}
We note here that unlike the other model extension, co-varying $n_s$ and $\omega_b$ has a significant impact on the reconstructed parameters, as seen in appendix~\ref{app:extensions}. 
In particular, we find $\sum m_{\nu}< 0.777e$V at 95\% CL from the full combination of LSS datasets, which is 2.8 times weaker than what we found when fixing those parameters. 
We show in appendix~\ref{app:extensions} that (perhaps unsurprisingly) this relaxation is due to the marginalization over $n_s$, which is strongly degenerate with the power suppression induced by a non-zero neutrino mass.\\

When combining LSS with {\it Planck}, we derive the strong constraint $\sum m_{\nu}< 0.093e$V at $95\%$ CL. We conclude that:
\begin{itemize}
    \item This analysis disfavors the inverse hierarchy at $\sim 2.2\sigma$, as the minimal sum of neutrino masses allowed by oscillation experiments is $\sim 0.1e$V~\cite{Esteban:2018azc}.
    \item The best-fit value of $\sum m_{\nu}$ is very close to $ 0e$V, with $\Delta\chi^2_{\rm min} = - 3.1$ compared to the base-$\Lambda$CDM analysis which sets the mass of a neutrino at $0.06e$V. 
    Note that the contribution to the $\Delta\chi^2_{\rm min}$ is about $-2$ from {\it Planck} and $-1$ from eBOSS + BOSS data. Taken at face value, our analysis therefore seems to slightly favor a universe {\em without} massive neutrinos compared to the base-$\Lambda$CDM model, in agreement with what was found for high-$\ell$ TT, TE, EE + lowE + BAO analysis in Ref.~\cite{Planck:2018vyg} (where the best-fit is $0.0009e$V), and may be connected to the lensing amplitude anomaly, see \emph{e.g.}, discussions in Ref.~\cite{Planck:2013nga}.
    \item This constraint is better than the {\it Planck} high-$\ell$ TT, TE, EE + lowE + BAO constraint~\cite{Planck:2018vyg}: $\sum m_{\nu}< 0.120e$V. However, our constraint is slightly weaker than the ShapeFit constraint~\cite{Brieden:2022lsd}: $\sum m_{\nu}< 0.082e$V. This may be due to their inclusion of eBOSS LRG data in addition to QSO. Nevertheless, these constraints are competitive with recent constraints from the Ly-$\alpha$ forest power spectrum~\cite{Palanque-Delabrouille:2019iyz}. 
    Note that the bound on the total neutrino mass derived in this work is somewhat relaxed with respect the one derived from the equivalent combined analysis including instead standard BAO/$f\sigma_8$: $\sum m_{\nu} < 0.080e$V (derived in this work). 
    Such relaxation was already observed when combining {\it Planck} with the EFT likelihood of BOSS~\cite{Ivanov:2019hqk} (see also Ref.~\cite{DAmico:2020kxu}). 
    Importantly, this does not imply that the EFT likelihood has less statistical power than the BAO/$f\sigma8$ likelihood, but rather results from the differences in the modeling of the power spectrum in the mildly nonlinear regime in the presence of massive neutrinos. 
\end{itemize}

We perform two complementary analyses where we consider the sum of the neutrino masses either under the assumption of normal ($\sum m_{\rm  \nu, \, NH}$) or inverted ($\sum m_{\rm  \nu, \, IH}$) hierarchy. In summary, we find that:

\begin{itemize}
    \item For the normal hierarchy (NH), $\sum m_{\rm  \nu, \, NH}< 0.469e$V at 95 \% CL for the analysis without eBOSS, and $\sum m_{\rm  \nu, \, NH}< 0.308e$V with eBOSS (with a best-fit value compatible with the prior lower bound). 
    This represents a $\sim 50\%$ stronger constraint thanks to eBOSS. 
    When varying $n_s$ and $\omega_b$, we get $\sum m_{\rm  \nu, \, NH}< 0.633e$V, which is $\sim 2$ times larger than the analysis where these two parameters are kept fix. 
    Yet, the combination of LSS + {\it Planck} leads to $\sum m_{\rm  \nu, \, NH}< 0.134e$V (with a best-fit value compatible with the prior lower bound, \emph{i.e.}, $0.06e$V).
    \item Similarly, for the inverse hierarchy (IH) we reconstruct $\sum m_{\rm  \nu, \, IH}< 0.337e$V for the LSS analysis and $\sum m_{\rm  \nu, \, IH}< 0.177e$V in combination with {\it Planck} (with a best-fit value compatible with the prior lower bound for both analyses, \emph{i.e.}, $0.1e$V).
    \item We obtain, for LSS + {\it Planck},  $\Delta\chi^2_{\rm min} = + 3.7$ between the analysis with the NH and the one assuming two massless neutrinos. This is because the LSS + {\it Planck} analysis has a preference for zero mass (see above).
    \item Similarly, we obtain, for LSS + {\it Planck}, $\Delta\chi^2_{\rm min} = + 7.1$ between the analysis with the IH and the one with two massless neutrinos. Interestingly, we find $\Delta\chi^2_{\rm min} = + 3.4$ between the IH and NH analyses, implying that the IH is disfavoured by LSS + {\it Planck} compared to the NH, as already mentioned above (see Refs.~\cite{Jimenez:2022dkn,Gariazzo:2022ahe} for detailed discussions about preference against the IH from a compilation of data).
\end{itemize}

\subsection{$N_{\rm eff} \Lambda {\rm CDM}$}

Finally, we co-vary the effective number of relativistic species, $N_{\rm eff}$. Here we consider $\Delta N_{\rm eff} = N_{\rm eff} - 3.044$, where $3.044$ is the standard model prediction \cite{Bennett:2020zkv}. We derive $\Delta N_{\rm eff} = 0.82^{+0.62}_{-1.08}$ at $68\%$ CL (with a best-fit value at $0.09$) from the analysis without eBOSS, while we derive $\Delta N_{\rm eff} = 0.40^{+0.44}_{-0.91}$ at $68\%$ CL (with a best-fit value at $-0.37$) for the analysis with eBOSS. These analyses allow us to conclude that:
\begin{itemize}
    \item The inclusion of eBOSS data changes the best-fit value of $\Delta N_{\rm eff}$, favoring a negative value, while the analysis without eBOSS favored a positive one. Nevertheless, constraints with and without eBOSS are in statistical agreement, with a significantly stronger constraints  on $\Delta N_{\rm eff}$ (by $\sim 30\%$) once eBOSS is included. 
    The clear improvement in this constraint is due to the fact that $\Delta N_{\rm eff}$ leaves an impact on the matter power spectrum by affecting the BAO and the amplitude on scales $k>k_{\rm eq}$, where $k_{\rm eq}$ is the equality scale: the presence of additional energy density affects the expansion rate of the Universe, reducing the sound horizon and delaying the onset of matter domination (see \emph{e.g.}, Ref.~\cite{Lesgourgues:2013sjj} for a review). The addition of a data point at $z\sim1.5$ thus improves the constraints.
    \item Interestingly, the conventional BAO/$f\sigma_8$  analysis is unable to constrain this parameter since it is mostly sensitive to the background radiation density that is negligible at low-redshift. The EFT therefore provides a new way of probing the contribution of extra relativistic species using BOSS and eBOSS data.
    \item Current LSS constraint is weaker than the {\it Planck} constraint ($\Delta N_{\rm eff} = -0.15^{+0.36}_{-0.38}$ at $95\%$~CL from high-$\ell$ TT, TE, EE + lowE + lensing \cite{Planck:2018vyg}), but it would be interesting to test whether higher accuracy LSS data can provide an independent constraint competitive with that obtained from the next generation CMB data.
\end{itemize}

When we combine the LSS analysis with {\it Planck} data, we reconstruct $\Delta N_{\rm eff} = -0.07^{+0.15}_{-0.16}$ at $68\%$ CL and $\Delta N_{\rm eff} = -0.07^{+0.30}_{-0.29}$ at $95\%$ CL (with a best-fit value at $0.02$). We conclude that:
\begin{itemize}
    \item This is compatible with the standard model value and this represents a significant improvement over the results from {\it Planck} alone.
    \item Our constraints are better by $12\%$ than the {\it Planck} + BAO constraints of Ref.~\cite{Planck:2018vyg} ($-0.06_{-0.33}^{+0.34}$ at $95\%$ CL) and by $20\%$ than the ShapeFit constraints ($0.07 \pm 0.38$ at $95\%$ CL) \cite{Brieden:2022lsd}. In addition, we note that our full-shape constraints are also slightly better than those obtained with the equivalent combined analysis including instead the standard BAO/$f\sigma_8$ ($0.05^{+0.33}_{-0.31}$ at 95\% CL.).
\end{itemize}

\section{Conclusion}\label{sec:conclusion}

In this paper we have performed the first EFT analysis of the eBOSS QSO full-shape data. We have combined this analysis with other LSS data in order to obtain independent constraints from {\it Planck}. As results are in good agreement with {\it Planck}, we have combined LSS and CMB probes in order to break the degeneracies present in the CMB, especially within models beyond $\Lambda$CDM. We summarise our main results here.

\paragraph{Determining the scale cut} In order to adequately study the eBOSS QSOs data, we have determined the maximum scale $k_{\rm max}$ at which the EFT full-shape analysis is valid. This scale is chosen such that the theoretical error is smaller than the data error bars and does not cause a significant shift in the cosmological results.   By fitting the full-shape of the mean over all EZmock realizations with the EFTofLSS at one-loop, as well as with the addition of the dominant next-to-next-leading order terms, we determine that for a scale cut $k_{\rm max} =\ 0.24\hinvMpc$, the shift in all cosmological and EFT parameters is below a reasonable threshold (of $< 1/3\sigma$). In addition, cosmological results obtained using the correlation function of eBOSS QSO are in overall good agreement.
It is interesting to note that the scale cut for eBOSS QSO full-shape is restricted by a rather large ``dispersion'' scale $\kr \sim 0.25 \hinvMpc$, entering the renormalization of the products of velocity operators appearing in the redshift-space expansion of the density field. 
This might provide indication that quasars selected by the eBOSS are preferentially populating satellite galaxies rather than central ones, as also argued from the perspective of halo occupation~\cite{2021MNRAS.504..857A}.

\paragraph{eBOSS QSOs flat $\Lambda$CDM cosmological results} The EFT analysis of eBOSS QSOs provides independent measurements of the $\Lambda$CDM parameters in a different redshift range ($z\sim 1.5$) than previous EFT analyses, and from yet another tracer.
Interestingly, we find good consistency between the EFT analysis of eBOSS QSO and the BOSS LRG data.  
For the $H_0$ parameter in particular, we find consistency at $\sim 1.0\sigma$ between these two datasets.
Additionally, we find that eBOSS favors a higher value for $\sigma_8$ (and $S_8$) at $\sim 1.5\sigma$ than the ones reconstructed using BOSS or {\it Planck}. 
Therefore, in general, the combination of eBOSS to the other cosmological probes tend to lift the value of the clustering amplitude. 
The addition of the EFT likelihood of eBOSS QSOs on top of the EFT likelihood of BOSS LRG (as well as in combination with ext-BAO + Pantheon) improves the $68\%$ CL error bars by about $10\%$.

\paragraph{Consistency with Planck} Interestingly, we found that all cosmological parameters are consistent at $\lesssim 1.0\sigma$ between eBOSS + BOSS and {\it Planck} data, except $\Omega_m$ that is consistent at $\sim1.6\sigma$. 
This consistency is a non-trivial check of the $\Lambda$CDM model and the many associated assumptions, as we considered very different data both in terms of redshift and in terms of the objects being probed. This may hint that the $S_8$ tension, unless causes by a systematic error, is restricted to scales smaller than $k\sim 0.2\hinvMpc$, or originate only at very late times (see \emph{e.g.}, Refs.~\cite{FrancoAbellan:2020xnr,FrancoAbellan:2021sxk,Simon:2022ftd,Poulin:2022sgp}).

\paragraph{Extensions of the flat $\Lambda$CDM model} In addition to further testing the $\Lambda$CDM model, we have assessed that eBOSS data help improving constraints on extended cosmological models in which the late-time background dynamics departs from flat $\Lambda$CDM.
Using a combination of LSS datasets, \emph{i.e.}, eBOSS + BOSS + ext-BAO + Pantheon, we obtain competitive constraints on the curvature density fraction $\Omega_k = -0.039 \,  \pm 0.029$, the dark energy equation of state $w_0 = -1.038 \,\pm 0.041$, the effective number of relativistic species $N_{\rm eff} = 3.44^{+0.44}_{-0.91}$ at $68\%$ CL, and the sum of neutrino masses $\sum m_\nu< 0.274 e$V at $95\%$ CL.
These constraints represent a significant improvement over the standard BAO/$f\sigma_8$ method. First, $\Omega_k$ and $w_0$ are better constrained by 50\% and 20\% respectively (at 68\% CL). Second, the application of the EFT to BOSS and eBOSS allows to tremendously improve constraints on the $\sum m_\nu$ (by a factor $\sim 18$) as the EFT allows to gain sensitivity to the power suppression of neutrinos, while conventional analysis are mostly sensitive to their effect on the angular diameter distance. In the same vein,  the EFT likelihoods of eBOSS and BOSS allows for a novel independent determination of the effective number of relativistic species, while $N_{\rm eff}$ is basically unconstrained from the standard BAO/$f\sigma_8$ analysis that is mostly sensitive to the (almost negligible) background contribution of radiation at late-times.

Including {\it Planck} data, contraints significantly improve thanks to the large lever arm in redshift between LSS and CMB measurements. 
In particular, we obtain $\Omega_k = 0.0008 \pm 0.0018$ and $w_0=-1.039 \pm 0.029$ at $68\%$ CL. 
In addition, we obtain the stringent constraint $\sum m_\nu < 0.093e$V, competitive with recent Lyman-$\alpha$ forest power spectrum bound \cite{Palanque-Delabrouille:2019iyz}, and $ N_{\rm eff} = 2.97 \pm 0.16$ at $68\%$ CL ($\Delta N_{\rm eff} < 0.23 $ at 2$\sigma$).
Note that, unlike the limits obtained using LSS data only, similar results are obtained when considering conventional BAO/$f\sigma_8$ data, due to the fact that {\it Planck} constraining power still largely dominates over that of BOSS and eBOSS. 
However, we expect that for significantly larger volume of data with, \emph{e.g.}, DESI~\cite{Aghamousa:2016zmz} or Euclid~\cite{Amendola:2016saw}, the EFT analysis will allow to improve over conventional analysis even when combining with {\it Planck}.

\

Our work demonstrates that eBOSS QSO data can help breaking model degeneracies in certain extensions to $\Lambda$CDM, as they sit at an intermediate redshift between BOSS and {\it Planck}. Furthermore, we have shown the importance of going beyond conventional BAO/$f\sigma_8$ analysis with the EFTofLSS in order to constrain simple extensions to $\Lambda$CDM without the inclusion of {\it Planck} data.
We leave for future work the application of the EFT full-shape analysis of eBOSS QSO data to more sophisticated beyond-$\Lambda$CDM physics, such as, \emph{e.g.}, decaying dark matter~\cite{Simon:2022ftd} and early dark energy~\cite{Simon:2022lde}. 
Our analysis can also be extended to the eBOSS LRG~\cite{Gil-Marin:2020bct,Bautista:2020ahg} and eBOSS ELG data~\cite{deMattia:2020fkb} (see  Ref.~\cite{Ivanov:2021zmi} for an EFT analysis of eBOSS ELG), which can provide new consistency tests of the $\Lambda$CDM model, and refine our ways to look for deviations from $\Lambda$CDM. \\

Note that after submitting this paper, Ref.~\cite{Chudaykin:2022nru} carried out an EFT full-shape analysis of the eBOSS QSO data using the \code{CLASS-PT} code~\cite{Chudaykin:2020aoj}. Our results are broadly consistent.

\section*{Acknowledgements}

We thank Guillermo F. Abell\'an, Rafaela Gsponer, and Tristan Smith for useful discussions and/or comments on the draft. 
This work has been partly supported by the CNRS-IN2P3 grant Dark21. 
The authors acknowledge the use of computational resources from the Excellence Initiative of Aix-Marseille University (A*MIDEX) of the “Investissements d’Avenir” programme. These results have also been made possible thanks to LUPM's cloud computing infrastructure founded by Ocevu labex, and France-Grilles.
This project has received support from the European Union’s Horizon 2020 research and innovation program under the Marie Skodowska-Curie grant agreement No 860881-HIDDeN.


\appendix
\section{What happens if we vary $n_s$ and $\omega_b$ in the LSS analyses?} \label{app:extensions}
\begin{figure*}[ht]
\begin{minipage}{0.30\linewidth}
\centering
\scriptsize
\begin{tabular}{|l|c|}
\hline
best-fit & [$n_s$ + $\omega_b$] \\
$\mu_{-\sigma_l}^{+\sigma_u}$ & (w/ eBOSS + BOSS)  \\
\hline
\multirow{2}{*}{$\Omega_{\rm m}$} & $0.288$ \\
& $0.295^{+0.011}_{-0.013}$ \\
\hline

\multirow{2}{*}{$h$} & $0.677$ \\
& $0.682^{+0.010}_{-0.010}$ \\
\hline

\multirow{2}{*}{$\sigma_8$} & $0.858$\\
& $0.797^{+0.041}_{-0.042}$\\
\hline

\multirow{2}{*}{$\omega_{\text{cdm}}$} & $0.1091$\\
& $0.1142^{+0.0063}_{-0.0081}$\\
\hline

\multirow{2}{*}{$\text{ln}(10^{10} A_s)$} & $3.24$ \\
& $3.06^{+0.13}_{-0.13}$ \\
\hline

\multirow{2}{*}{$S_8$} & $0.842$\\
& $0.790^{+0.039}_{-0.041}$\\
\hline

\multirow{2}{*}{$n_s$} & $1.015$ \\
& $0.981^{+0.048}_{-0.044}$ \\
\hline
$\chi^2_{\rm min}$ &  216.7 \\
\hline
$N_{\rm data}$ & 262\\
\hline
$p$-value & 0.47\\
\hline
\end{tabular} 
\end{minipage}\hfill
\begin{minipage}{0.60\linewidth}
\includegraphics[width=1.\textwidth]{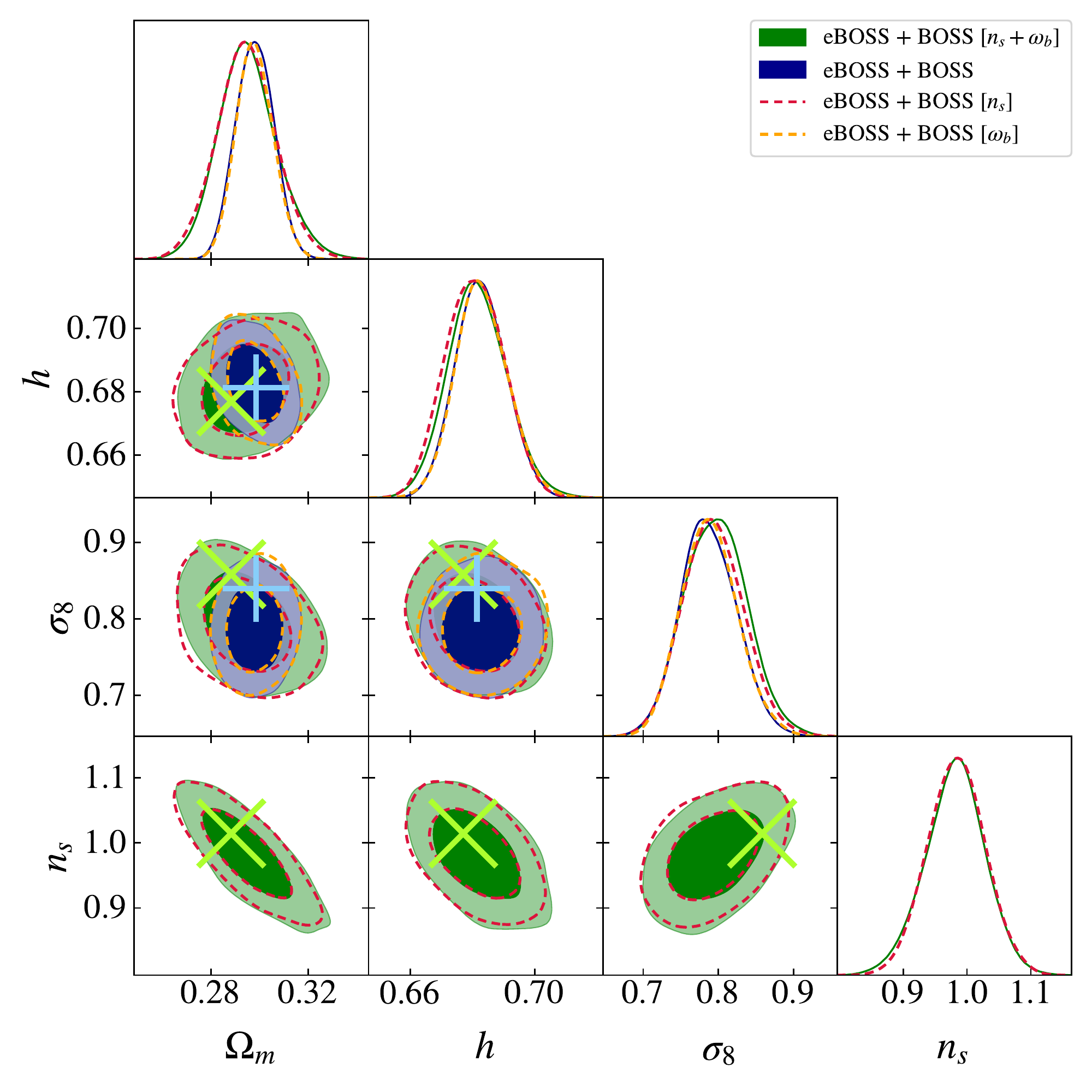}
\end{minipage}
\caption{\footnotesize {\it Left} -  Cosmological results (best-fit, posterior mean, and $68\%$ CL) of eBOSS + BOSS for the $\Lambda$CDM model. Here we let $n_s$ vary freely and restrict $\omega_b$ to a BBN prior, calling such analysis ``[$n_s$ + $\omega_b$]'', in contrast to the corresponding analyses in the main text where those are fixed.  
We also report the best-fit $\chi^2$, the number of data points $N_{\rm data}$ and the associated $p$-values.
{\it Right} -  Triangles plots (1D and 2D posterior distributions) of the cosmological parameters reconstructed from the eBOSS + BOSS base-$\Lambda$CDM analysis and the eBOSS + BOSS [$n_s$ + $\omega_b$] analysis. We also show, in dotted lines, the eBOSS + BOSS results with only the $n_s$ variation, denoted ``eBOSS + BOSS $[n_s]$'', or only the $\omega_b$ variation, denoted ``eBOSS + BOSS $[\omega_b]$''.}
\label{fig:LCDM_ns_omega_b}
\end{figure*}

In our baseline LSS analyses (\emph{i.e.}, without combination with {\it Planck} data), we set $\omega_b = 0.02233$ (the central value of big-bang nucleosynthesis (BBN) experiments~\cite{Mossa:2020gjc}) and $n_s = 0.965$ (the central value from {\it Planck} in the $\Lambda$CDM model~\cite{Planck:2018vyg}). In the following, we impose a uninformative large flat prior on $n_s$, while we impose the BBN Gaussian prior on $\omega_b$, motivated from~\cite{Mossa:2020gjc}, namely $\omega_b = 0.02233 \pm 0.00036$, to explore the impact of the variation of $n_s$ and $\omega_b$ in the LSS analyses presented in this paper.
We carried out the eBOSS + BOSS analysis either by varying one of these two parameters at a time to isolate their effects, or varying both simultaneously.
In figure~\ref{fig:LCDM_ns_omega_b}, we present these results in comparison with our base-$\Lambda$CDM analysis of eBOSS + BOSS. 

In this figure, one can easily gauge that the variation of $\omega_b$ within the BBN prior has a negligible impact on the cosmological results: we have a relative shift of $\lesssim 0.04\sigma$ between the mean values and $\lesssim 0.07\sigma$ between the best-fit values, while the $68\%$ CL constraints remain the same for all parameters. However, the variation of $n_s$ within a  uninformative large flat prior leads to a relative shift $\lesssim 0.4\sigma$ between the mean values, while the best-fit values are shifted by -1.06$\sigma$, -0.50$\sigma$, 0.52$\sigma$, -1.15$\sigma$, 0.76$\sigma$, 0.13$\sigma$ for $\Omega_m$, $h$, $\sigma_8$, $\omega_{cdm}$, $\ln10^{10}A_s$ and $S_8$ respectively. In addition, the $68\%$ CL constraints of the $n_s$-free analysis are degraded $\lesssim 10\%$ on all cosmological parameters, except for $\Omega_m$ and $\omega_{cdm}$ where the degradation reaches $57\%$ and $79\%$ respectively. Not surprisingly, we find that the analysis that combines the variation of these two parameters gives results very similar to the $n_s$-free analysis.

It is common to set the value of $n_s$ in LSS analyses (see Refs.~\cite{Ivanov:2019pdj,Chen:2021wdi,Brieden:2022lsd} for examples) to that of {\it Planck}. Several tests have been carried out, often imposing a Gaussian prior inspired by the {\it Planck} preferred value in order to evaluate the impact of the variation of $n_s$ around this value. This is for instance the case in the ShapeFit analysis \cite{Brieden:2022lsd}, where a Gaussian prior $n_s=0.965\pm0.02$ is imposed. In this analysis, which also combines eBOSS and BOSS data, it was reported that the variation of $n_s$ within this prior does not impact $h$ and $\sigma_8$ posteriors, while it slightly degrades the constraints on $\Omega_m$. This is somewhat consistent with what we observe here. However, it is essential to note that this degradation is exacerbated in our $n_s$-free analysis since we impose a large uninformative flat prior.

\begin{table*}[ht]
\center
\tiny 
\scalebox{1.1}{\begin{tabular}{|l|c|c|c|c|c|}
\hline
\multicolumn{6}{|c|}{eBOSS + BOSS + ext-BAO + Pantheon (LSS) [$n_s + \omega_b$]} \\
\hline
best-fit  & \multirow{2}{*}{$\Lambda$CDM}  & \multirow{2}{*}{$\Omega_k\Lambda$CDM} & \multirow{2}{*}{$w_{0}$CDM} & \multirow{2}{*}{$\nu \Lambda$CDM} & \multirow{2}{*}{$N_{\rm eff} \Lambda$CDM} \\
$\mu_{-\sigma_l}^{+\sigma_u}$ & & & & & \\
\hline

\multirow{2}{*}{$\Omega_{\rm m}$} & 0.2905 & 0.291 & 0.2903 & 0.290 & 0.292  \\
& $0.2957^{+0.0097}_{-0.0098}$ & $0.299_{-0.012}^{+0.010}$ & $0.2957_{-0.0110}^{+0.0096}$ & $0.299_{-0.011}^{+0.010}$ & $0.291_{-0.012}^{+0.014}$   \\
\hline

\multirow{2}{*}{$h$} & 0.6749 & 0.681 & 0.670 & 0.6753 & 0.671 \\
& $0.6786^{+0.0085}_{-0.0088}$ & $0.697_{-0.015}^{+0.013}$ & $0.682_{-0.013}^{+0.012}$ & $0.6765_{-0.0094}^{+0.0085}$ & $0.697_{-0.038}^{+0.017}$\\
\hline

\multirow{2}{*}{$\sigma_8$} & 0.858 & 0.840 & 0.866 & 0.856 & 0.859 \\
& $0.794^{+0.037}_{-0.041}$ & $0.755_{-0.047}^{+0.043}$ & $0.789_{-0.045}^{+0.042}$ & $0.811_{-0.046}^{+0.042}$ & $0.793_{-0.040}^{+0.037}$ \\
\hline

\multirow{2}{*}{$\omega_{\text{cdm}}$} & 0.1093 & 0.1120 & 0.1095 & 0.1096 & 0.108 \\
& $0.1132^{+0.0056}_{-0.0062}$ & $0.1222_{-0.0096}^{+0.0076}$ & $0.1146_{-0.0079}^{+0.0060}$ & $0.1115_{-0.0070}^{+0.0055}$ & $0.119_{-0.012}^{+0.007}$\\
\hline

\multirow{2}{*}{$\text{ln}(10^{10} A_s)$} & 3.24 & 3.14 & 3.29 & 3.22 & 3.24 \\
& $3.06^{+0.12}_{-0.12}$ & $2.76_{-0.21}^{+0.22}$ & $3.03_{-0.15}^{+0.15}$ & $3.20_{-0.21}^{+0.15}$ & $3.05_{-0.12}^{+0.12}$ \\
\hline

\multirow{2}{*}{$S_8$} & 0.844 & 0.828 & 0.852 & 0.842 & 0.847 \\
& $0.788^{+0.039}_{-0.039}$ & $0.754^{+0.040}_{-0.045}$ & $0.783^{+0.040}_{-0.044}$ & $0.809^{+0.041}_{-0.047}$ & $0.780^{+0.037}_{-0.043}$
 \\
\hline

\multirow{2}{*}{$n_s$} & 1.012 & 0.994 & 1.025 & 1.007 & 1.010 \\
& $0.985^{+0.038}_{-0.038}$ & $0.930_{-0.048}^{+0.053}$ & $0.976_{-0.043}^{+0.046}$ & $1.038_{-0.087}^{+0.050}$ & $0.985_{-0.039}^{+0.041}$\\
\hline

\multirow{2}{*}{$\Omega_k$} &\multirow{2}{*}{--}& $-0.021$ & \multirow{2}{*}{--} & \multirow{2}{*}{--} & \multirow{2}{*}{--} \\
&& $-0.063^{+0.036}_{-0.037}$ &&&  \\
\hline

\multirow{2}{*}{$w_{0}$} &\multirow{2}{*}{--}& \multirow{2}{*}{--} & $-0.975$ & \multirow{2}{*}{--} & \multirow{2}{*}{--} \\
&&& $-1.018^{+0.047}_{-0.043}$  &&  \\
\hline

\multirow{2}{*}{$\sum m_{\nu}$ [$e$V]} &\multirow{2}{*}{--}& \multirow{2}{*}{--} & \multirow{2}{*}{--} & $0.025$ & \multirow{2}{*}{--} \\
&&&& $<0.777$ &  \\
\hline

\multirow{2}{*}{$\Delta N_{\rm eff}$} &\multirow{2}{*}{--}& \multirow{2}{*}{--} & \multirow{2}{*}{--} & \multirow{2}{*}{--} & $-0.11$ \\
&&&&& $0.48^{+0.47}_{-0.86}$ \\
\hline
\hline
$\chi^2_{\rm min}$ & 1249.8 & 1249.6 & 1249.4 & 1249.7 & 1249.8 \\ 
\hline
$\Delta\chi^2_{\rm min}$ & -- & $-0.2$ & $-0.4$ & -0.1 & 0 \\
\hline
\end{tabular}}

\caption{\footnotesize Cosmological results (best-fit, posterior mean, and $68\%$ CL) from eBOSS + BOSS + ext-BAO + Pantheon for the $\Lambda$CDM model as well as several model extensions. In contrast to the previous analyses, we vary $n_s$ and $\omega_b$.
Note that we quote $95\%$ CL bound for $\sum m_{\nu}$. For each dataset we also report its best-fit $\chi^2$, and the $\Delta \chi^2$ with respect to the analogous $\Lambda$CDM best-fit model.}
\label{tab:parameter_extansion_ns_ob}
\end{table*}

Although these differences do not alter the conclusions of this paper, one may argue that to obtain LSS constraints that are truly independent of {\it Planck}, one should not include any prior on $n_s$. Therefore, we have redone, for all cosmological model considered in this work, the analysis of eBOSS + BOSS + ext-BAO + Pantheon, letting $n_s$ vary freely as well as restricting $\omega_b$ to the BBN prior. 
Those cosmological results are presented in table~\ref{tab:parameter_extansion_ns_ob}. 
From the first column of this table, we can compare the $n_s$-free analysis with the base-$\Lambda$CDM analysis of the last column of table~\ref{tab:parameter_BOSS_eBOSS} for the full combination of LSS datasets. 
We observe that the relative shifts between the mean values, the relative shifts between the best-fit values, and the degradation of the constraints are substantially similar to those observed in the above comparison on eBOSS + BOSS.
In addition, we list here the impact of varying $n_s$ (and $\omega_b$) on the constraints of the extension parameters to flat $\Lambda$CDM from the analysis of eBOSS + BOSS + ext-BAO + Pantheon: 
\begin{itemize}
\item For $\Omega_k$, the relative shift between the mean values is 0.74$\sigma$, the relative shift between the best-fit values is 0.40$\sigma$, while the constraint is degraded by $\sim 30\%$. 
\item For $w_0$, the relative shift between the mean values is 0.46$\sigma$, the relative shift between the best-fit values is 0.64$\sigma$, while the constraint is degraded by $\sim 10\%$. 
\item For $\sum m_{\nu}$, there is no relative shift between the best-fit values, while the $95\%$ CL bound becomes 2.8 times weaker.
\item For $\Delta N_{\rm eff}$, the relative shift between the mean values is 0.12$\sigma$, the relative shift between the best-fit values is 0.36$\sigma$, but the constraint is not degraded.
\end{itemize}

Finally, we perform a BOSS + ext-BAO + Pantheon analysis without eBOSS (\emph{i.e.}, the first column of table~\ref{tab:parameter_extansion_ns_ob} without eBOSS) to assess the impact of eBOSS on the determination of $n_s$.
In this case, we reconstruct $n_s = 0.943 \pm 0.043$ at $68\%$ CL (with a best-fit value at $0.957$). 
This shows that the inclusion of the eBOSS data allows us an improvement of $13\%$ on the $68\%$ CL error bar of $n_s$.


\section{Redshift resolution uncertainties}\label{app:redshift_error}
The broadness of the emission lines of QSOs, due to the rotating gas located around the black hole, increases the uncertainties in the determination of their redshift, see \emph{e.g.},~Ref.~\cite{Dawson:2015wdb}.  
Here, we show that the leading corrections coming from these redshift errors are, under minor assumptions concerning their distribution, degenerate with EFT counterterms. 
Although, nothing is really new in the following derivation, except for the inclusion of uncertainties in the redshift, we follow a pedagogic approach in order to make sure that we track correctly the corresponding modifications. 
We start with the case without error in the redshift determination, before introducing the uncertainty associated with the QSO broad emission lines.

\paragraph{Comoving coordinates in redshift space} The observed redshift $z_{\rm obs}$ of the light emitted from a distant object sitting on a cosmological background labelled with time $t$, or correspondingly scale factor $a$ or redshift $z$, with $a \equiv (1+z)^{-1}$, is given by: 
\begin{equation}\label{eq:zobs}
1+z_{\rm obs} = (1+z) (1+ \delta z_{\rm pec}) \ ,
\end{equation}
where $\delta z_{\rm pec}\simeq \v(z) \cdot \hat n/c$, with $c$ the speed of light, arises from the peculiar velocity $\v$ of the object in the direction of the line-of-sight $\hat n$ with respect to its local cosmological background (labelled with redshift $z$). 
eq.~\eqref{eq:zobs} follows straightforwardly from noticing that: \textit{i)} the peculiar velocity of the object with respect to the local comoving background induces a redshift $1+ \delta z_{\rm pec} \equiv \lambda_{c}/\lambda_{e}$, where $\lambda_{e}$ is the emitted physical wavelength while $\lambda_{c}$ is the wavelength seen in the local comoving frame; \textit{ii)}  the cosmological expansion induces another redshift $1+z \equiv \lambda_{o}/\lambda_{c}$ between the wavelength $\lambda_{o}$ that we observe today in our local inertial frame with respect to the wavelength $\lambda_{c}$; \textit{iii)}  the total redshift is the shift between the physical emitted wavelength $\lambda_{e}$ and the observed one $\lambda_{o}$: $1+z_{\rm obs} = \lambda_{o} / \lambda_{e}$. 
For what follows, it is convenient to rewrite eq.~\eqref{eq:zobs} as:
\begin{equation}
z_{\rm obs} = z + (1+z) \delta z_{\rm pec}  \ .
\end{equation}
The comoving distance $\chi(z)$ from us at redshift $z=0$ (accordingly with today's scale factor $a_0 \equiv 1$) to the observed object sitting on a cosmological background at redshift $z$, is given by:
\begin{equation}
\chi(z) \equiv c \int \frac{dt}{a} = c \int_{a}^{a_{0}} \frac{da'}{a'^2 \, H(a')} = c \int_0^z \frac{dz'}{H(z')} \ ,
\end{equation}
where $H$ the Hubble parameter. 
The line-of-sight distance associated to the volume distortion from the peculiar velocity $\v$ of the object measured in its local comoving frame at redshift $z$ is given by $\delta \chi'(\delta z_{\rm pec}) \simeq c \, \delta z_{\rm pec} /H(z)$, assuming that the change in the background $ d(H^{-1}(z))/dz$ around $z$ is small with respect to $\delta z_{\rm pec}$. 
Seen from our local comoving frame, this distance is rescaled by $a_0/a \equiv (1+z)$, then yielding:
\begin{equation}
\delta \chi \equiv (1+z) \, c \, \frac{\delta z_{\rm pec}}{H(z)} \simeq (1+z) \,  \frac{\v(z) \cdot \hat n}{H(z)}\ ,
\end{equation}
assuming that the velocity is non-relativistic such that $\delta z_{\rm pec}\simeq \v(z) \cdot \hat n/c$.
Thus, we get that the comoving coordinate in redshift space $\s \equiv \chi(z_{\rm obs})$ is related to the comoving coordinate in real space $\x \equiv \chi(z)$ through: 
\begin{equation}
\s(z) \simeq \x(z) + (1+z) \frac{\v(z) \cdot \hat n}{H(z)} \, \hat n = \x + \frac{\v \cdot \hat n}{\mathcal{H}} \hat n  \ .
\end{equation}
In the last equality, and from now on, we sometimes drop the time dependence from the notation and use the comoving Hubble parameter $\mathcal{H}(a)\equiv a H(a)$ for conciseness in the presentation. 

\paragraph{Density in redshift space.} Following Refs.~\cite{Matsubara:2007wj,Matsubara:2008wx} (see also Refs.~\cite{Senatore:2014vja,Vlah:2018ygt}), we can derive the relation between the density field in redshift space (that we denote with a subscript ``r'') at comoving redshift-space coordinate $\s$ with the density and velocity fields in real space at comoving real-space coordinate $\x$. 
Mass conservation of the infinitesimal volume $d^3 x$ in real space to $d^3 s$ in redshift space implies:
\begin{equation}\label{eq:mass_conservation}
\rho(\x) d^3 x = \rho_r(\s) d^3 s = \rho_r(\s) \left| \frac{\partial \s}{\partial \x} \right| d^3 x \, ,
\end{equation}
where $\left| \frac{\partial \s}{\partial \x} \right|$ denotes the Jacobian of the transformation. 
The overdensity field in redshift space thus transforms as:
\begin{equation}
1 + \delta_{r}(\s) = \left| \frac{\partial \s}{\partial \x} \right|^{-1} (1+\delta(\x)) \, .
\end{equation}
In Fourier space, the overdensity relation between redshift space and real space then reads:
\begin{align}
\delta_{r}(\k) &\equiv \int d^3s \, e^{-i \k \cdot \s} \delta_{r}(\s) \nonumber \\
& = \delta(\k) + \int d^3 x \, e^{-i \k \cdot \x} \left( e^{-i \k \cdot \hat n \, \tfrac{\v}{\mathcal{H}} \cdot \hat n} -1 \right) (1+ \delta(\x)) \, . \label{eq:delta_r}
\end{align}

Alternatively, we can derive the same relation from the Lagrangian description that relates both the density in real space and the density in redshift space to the density at some initial time. 
In real space, the position of the object in Lagrangian coordinate $\x$, at time $t$, is given by its initial position $\q$ and the displacement $\m (\q, t)$ from its initial position:
\begin{equation}
\x(\q, t) = \q + \m(\q, t) \, .
\end{equation}
For dark matter, the overdensity field is given by:
\begin{equation}
1+\delta(\x, t) = \int d^3q \, \delta_D(\x-\q-\m(\q, t)) \, .
\end{equation}
For biased tracers (that we denote with a subscript ``g''), this becomes: 
\begin{equation}\label{eq:lagrange_halo_conservation}
1+\delta_g(\x, t) = \int d^3q \, F_L \, \delta_D(\x-\q-\m(\q, t)) \, ,
\end{equation}
where $F_L \equiv F_L \left[\delta(\q, t), \partial^2\delta(\q, t), \dots \right]$ is the Lagrangian halo biasing function. 
Fourier transforming the above, the real-space galaxy overdensity reads:
\begin{equation}\label{eq:fourier_lagrange_halo_density}
(2\pi)^3 \delta_D(\k) + \delta_g(\k, t) = \int d^3q \, F_L  \, e^{- i \k \cdot \left(\q + \m(\q, t) \right)} \, .
\end{equation}
In redshift space, the position in Lagrangian coordinates is given by:
\begin{equation}
\s(\q, t) = \q + \m(\q, t) + \left(\frac{\dot{\m}(\q,t)}{\mathcal{H}} \cdot \hat n\right) \hat n \, .
\end{equation}
eq.~\eqref{eq:fourier_lagrange_halo_density} thus becomes:
\begin{align}
(2\pi)^3 \delta_D(\k) + \delta_{g,r}(\k, t) & = \int d^3q \, F_L \, e^{- i \k \cdot \left(\q + \m(\q, t) + \left(\tfrac{\dot{\m}(\q,t)}{\mathcal{H}} \cdot \hat n\right) \, \hat n \right)} \,  \nonumber \\
& = \int d^3 x \, (1+\delta_{g}(\x)) e^{-i \k \cdot \x}  e^{-i \k \cdot \hat n \, \tfrac{\v}{\mathcal{H}} \cdot \hat n} \ ,
\end{align}
where we have used eq.~\eqref{eq:lagrange_halo_conservation} to go to the second line. 
We thus find the same relation between the overdensities in redshift space and real space: 
\begin{equation}\label{eq:delta_g,r}
\delta_{g,r}(\k)  = \delta_g(\k) + \int d^3 x \, e^{-i \k \cdot \x} \left( e^{-i \k \cdot \hat n \, \tfrac{\v}{\mathcal{H}} \cdot \hat n} -1 \right) (1+ \delta_g(\x)) \, . 
\end{equation}

\paragraph{Introducing redshift errors.}
The uncertainties in the determination of the redshift by the experimentalist can be tracked by introducing an independent variable, $\delta z_{\rm sys}$, in eq.~\eqref{eq:zobs}: 
\begin{equation}\label{eq:zobs_zerr}
1+z_{\rm obs} = (1+z) (1+ \delta z_{\rm pec} + \delta z_{\rm sys}) \ .
\end{equation}
For convenience, we define an associated ``fake'' velocity variable given by $\delta z_{\rm sys} \equiv v_{\rm sys}(z) /c$. 
As such, it is easy to see that the derivation above proceeds similarly but with the replacement $\v \cdot \hat n \rightarrow \v \cdot \hat n+ v_{\rm sys}$. 
Therefore, in the presence of redshift errors, the relation between the overdensities in redshift space and real space, eq.~\eqref{eq:delta_g,r}, is modified to: 
\begin{equation}\label{eq:delta_g,r^sys}
\delta^{\rm sys}_{g,r}(\k)  = \delta_g(\k) + \int d^3 x \, e^{-i \k \cdot \x} \left( e^{-i \k \cdot \hat n \, \tfrac{\v}{\mathcal{H}} \cdot \hat n} e^{-i \k \cdot \hat n \, \tfrac{v_{\rm sys}}{\mathcal{H}}} -1 \right) (1+ \delta_g(\x)) \, . 
\end{equation}

We want to understand the leading corrections to our predictions from the presence of redshift errors. 
To see this, we can Taylor-expand $e^{-i \k \cdot \hat n \, \tfrac{v_{\rm sys}}{\mathcal{H}}} = 1 - i (\k \cdot \hat n) \, \mathcal{H}^{-1} \, v_{\rm sys} - \frac{1}{2} (\k \cdot \hat n)^2 \, \mathcal{H}^{-2} \, v_{\rm sys}^2 + \dots$ in the above equation. 
After some straightforward manipulations, eq.~\eqref{eq:delta_g,r^sys} becomes: 
\begin{equation}
\delta^{\rm sys}_{g,r}(\k) = \delta_{g,r}(\k) + \epsilon_{\rm sys}(\k) \, \delta_D(k) + \epsilon_{\rm sys}(\k) \, \delta_{g,r}(\k) + \dots \ ,
\end{equation}
where $\delta_{g,r}(\k)$ is the redshift-space galaxy density field without redshift error given by eq.~\eqref{eq:delta_g,r^sys}, and we have introduced the notation $\epsilon_{\rm sys}(\k) \equiv -i \mu k \mathcal{H}^{-1} \ v_{\rm sys} - \frac{1}{2}\mu^2 k^2 \mathcal{H}^{-2} \ v_{\rm sys}^2$, with $\mu \equiv (\k \cdot \hat n) / |\k|$. 
The $\dots$ represent higher-order corrections in powers of $\epsilon_{\rm sys}$. 
Assuming that, by definition, $v_{\rm sys}$ is a scale-independent variable that correlates only with itself, the power spectrum picks, at leading orders in derivatives, corrections going as:
\begin{align}\label{eq:Pzerr}
P_{g,r}^{\rm sys}(k, \mu) & = P_{g,r}(k, \mu) -2i\mu k \frac{\bar{v}_{\rm sys}}{\mathcal{H}} \sigma_0^2 - 2i\mu k \frac{\bar{v}_{\rm sys}}{\mathcal{H}} (b_1 + f \mu^2)^2 P_{11}(k) \\
& - \mu^2 k^2 \frac{\sigma_{v {\rm,sys}}^2}{\mathcal{H}^{2}} (\delta_D(k)+3\sigma_0^2) - 2 \mu^2 k^2 \frac{\sigma_{v {\rm,sys}}^2}{\mathcal{H}^{2}} (b_1 + f \mu^2)^2 P_{11}(k) + \dots \ , \nonumber
\end{align}
where we have introduced the following notation: $\bar{v}_{\rm sys} \equiv \braket{v_{\rm sys}}$, $\sigma_{v {\rm,sys}}^2 \equiv \braket{v_{\rm sys} v_{\rm sys}}$, and $\sigma_0^2 \equiv P_{g,r}(0)$. 

\paragraph{Lessons for eBOSS QSO full-shape analysis.}
From this derivation, we can make several observations:
\begin{itemize}
\item The first two correction terms at the first line of eq.~\eqref{eq:Pzerr}, are purely imaginary, and thus do not appear in the even multipoles that we use in our analysis. 
Moreover, those terms are significant only if the determination of the redshifts is biased on average, \textit{i.e.}, $\braket{v_{\rm sys}} \neq 0$.~\footnote{One could imagine measuring the redshift error bias by searching in the odd multipoles such signal, that may be clean from other known contributions such as relativistic or wide-angle effects that present a different scaling dependence, see \textit{e.g.}, Refs.~\cite{Beutler:2018vpe,Beutler:2020evf}. 
Besides, in principle, the odd window function multipoles that are imaginary also mix imaginary contributions in the power spectrum to the even multipoles of the power spectrum. 
We leave those explorations to future work. 
}
\item Therefore, as anticipated and according to eq.~\eqref{eq:powerspectrum}, we see that the leading corrections to uncertainties in the redshift determination are degenerate with EFT counterterms going like $\sim \mu^2 k^2$ (namely, the terms in $c_{\epsilon}^{\textrm{quad}}$) or $\sim \mu^{2} k^2 P_{11}(k)$ (namely, the terms in $c_{r,1}$). 
Therefore, albeit a potential adjustment in the prior for the coefficients associated to those counterterms to accomodate this new effect, our predictions are unchanged in the presence of redshift errors. 

\item Beyond the leading corrections, there could be a term going like $\sim k^3 P_{11}(k)$ from the contraction of $\braket{v_{\rm sys}^3}$, that is not degenerate with counterterms in the EFT. 
However, again, this term is purely imaginary, and thus does not appear in the even multipoles that we use in our analysis. 
Moreover, this term is significant only if the distribution of the redshift errors has some level of assymetry.

\item We can get insights on the size of the corrections by inspecting the distribution of the redshift errors. 
From Ref.~\cite{Zarrouk:2018vwy}, we learn that the variance of the estimated distribution of eBOSS QSO redshift errors corresponds roughly to $\sigma_{v,\textrm{sys}}^2 \sim (300 $km/s)${}^{2}$. 
This tells us that the size of the correction is about 
\begin{equation}\label{eq:vsys_estim}
\frac{\sigma_{v,\textrm{sys}}^2 }{\mathcal{H}^2(z)} \sim  \frac{(1+z_{\rm eBOSS})^2 (300 \ \textrm{km/s})^2}{\left(\Omega_m \cdot (1+z_{\rm eBOSS})^{3}+\Omega_\Lambda \right) \left(100 \ h \ \textrm{km/s/Mpc} \right)^2} \sim 10 \ (\Mpcinvh)^2 \ . 
\end{equation}
Despite this being slightly larger than the size of $c_{\epsilon}/k_{\rm M}^2 \sim 4 \ (\Mpcinvh)^2$, $\sigma_0^2$ is smaller than $1/\bar n_g \sim 2 \cdot 10^5 \ (\Mpcinvh)^3$ for eBOSS, and therefore, the first correction term at the second line of eq.~\eqref{eq:Pzerr} is smaller than the corresponding EFT counterterm going as $\sim \mu^2 k^2$. 
Furthermore, eq.~\eqref{eq:vsys_estim} also tells us that $b_1^2 \ \sigma_{v,\textrm{sys}}^2\, \mathcal{H}^{-2} \sim 40  \ (\Mpcinvh)^2$ is about $1.6$ times smaller than $b_1 \, c_{r,1} /k_{\rm R}^2  \sim 64 \ (\Mpcinvh)^2$, \textit{i.e.}, the typical size of the EFT counterterm going as $\sim \mu^2 k^2 P_{11}$ that is degenerate with the last correction in the second line of eq.~\eqref{eq:Pzerr}. 
Thus, corrections to redshift uncertainties are well accounted in our analysis given the prior we put on those counterterms.~\footnote{Note that here we have assumed that redshift uncertainties are thought to arise in the determination of the peculiar velocity of the objects, as given by Eq.~\eqref{eq:zobs_zerr}. If instead we attribute them as a global error, such as $z_{\rm obs} = z + (1+z)\delta z_{\rm pec} + \delta z_{\rm sys}$, the size of the corrections to our observables will be reduced by a factor $1+z$.} 

\item The next-to-leading correction to the even multipoles is going like $\sim \mu^4 k^4 P_{11}(k)$, which is degenerate with EFT counterterms at two loop, see eq.~\eqref{eq:nnlo}. 
From Ref.~\cite{Zarrouk:2018vwy} (see in particular figs.~4~and~5), we see that the redshift errors distribution of eBOSS QSO sample departs visibly from a Gaussian by the presence of fat tails. 
Naively, this indicates that higher moments of the distribution are suppressed, and in particular, $\braket{v_{\rm sys}^4} \ll \sigma_{v,\textrm{sys}}^4$. 
Thus, the higher-order corrections to redshift uncertainties, \textit {e.g.,} $\sim \mu^4 k^4 \frac{\braket{v_{\rm sys}^4}}{\mathcal{H}^4} P_{11}(k)$, become quickly negligible with respect to the EFT counterterms that share similar scale dependence, \textit{e.g.}, the NNLO counterterms of eq.~\eqref{eq:nnlo}, that are themselves already higher order and thus safely small with respect to the error bars of the data at the scales analyzed. 

\end{itemize}

We conclude that, given the presence of EFT counterterms that share similar scale dependence with corrections from redshift errors and large enough priors to encompass these effects, our analysis is unaffected by uncertainties in the determination of eBOSS QSO redshifts.

\bibliographystyle{JHEP}
\bibliography{references}

\end{document}